\documentclass[showpacs,preprintnumbers,superscriptaddress,amsmath,amssymb,floatfix]{revtex4}
\usepackage{tabularx}
\usepackage[usenames]{color}
\usepackage{epsfig}
\usepackage{amssymb}
\usepackage{graphicx}
\usepackage{subfigure}
\usepackage{float}

\usepackage{rotating,multirow}
\newcommand{\be}{\begin{equation}}
\newcommand{\ee}{\end{equation}}
\newcommand{\bee}{\begin{eqnarray}}
\newcommand{\eee}{\end{eqnarray}}

\newcommand{\piNpiN}{$\pi\!N\!\rightarrow \!\pi\!N$}
\newcommand{\piNetaN}{$\pi \! N \! \rightarrow \! \eta N \:$}

\definecolor{grey}{rgb}{0.5,0.5,0.5}

\definecolor{black}{rgb}{0,0,0}

\def \irbaddress{Rudjer Bo\v{s}kovi\'{c} Institute, Bijeni\v{c}ka cesta 54, P.O. Box 180, 10002 Zagreb, Croatia}
\def \Tuzlaaddress{University of Tuzla, Faculty of Science, Univerzitetska 4, 75000
 Tuzla, Bosnia and Herzegovina}

\begin{document}

\title{Stability of the Zagreb realization of the Carnegie-Mellon-Berkeley
    coupled-channels unitary model}

\author{ H. Osmanovi\'{c} }
\affiliation{\Tuzlaaddress \\}

\author{S. Ceci}
\affiliation{\irbaddress \\
*E-mail: alfred.svarc@irb.hr}

\author{A. \v{S}varc*}
\affiliation{\irbaddress \\
*E-mail: alfred.svarc@irb.hr}

\author{M. Had\v{z}imehmedovi\'{c} }
\affiliation{\Tuzlaaddress \\}

\author{J. Stahov }
\affiliation{\Tuzlaaddress \\}

\date{\today}

\begin{abstract}
In ref.~\cite{Mirza2011} we have used the Zagreb realization of Carnegie-Melon-Berkeley coupled-channel, unitary model as
 a tool for extracting pole positions from the world collection of partial wave data, with the aim  of eliminating model dependence in pole-search procedures. In order that the 
method is sensible, we in this paper discuss the stability of the method with respect to the strong variation of different model ingredients. We show that the Zagreb CMB procedure 
is very stable with strong variation of the model assumptions, and that it  can reliably predict the pole  positions of the fitted partial wave amplitudes.

\end{abstract}


\date{\today}

\maketitle

\section{Introduction}

{
Various analyses of meson-baryon scattering designed with the goal to extract
resonance properties from data are available in literature.  They fundamentally differ in technical approach how the model is implemented (effective Lagrangian approach, K matrix 
formalism, K matrix approximation, phenomenological coupled-channel formalism, t-channel dispersion relations, hyperbolic dispersion relations, e.t.c.), however most of them 
strongly insist in obeying fundamental physics principles like Lorentz invariance, crossing symmetry,  unitarity and analyticity. In ref.~\cite{Mirza2011}, we have extracted the 
pole positions from a ``world collection" of partial wave data and partial-wave amplitudes understanding them as partial-wave data, with the aim of eliminating  the model 
dependence from the individual pole-search procedures.
 We have analyzed some major PWA amplitudes like: i)~Karlsruhe-Helsinki (KH80) amplitudes \cite{KH80} for the  $\pi$N elastic scattering, where  fixed  t-channel dispersion 
relations are combined with Pietarinen expansion to impose and maintain point-to-point analyticity; ii) VPI/GWU single energy (GWU-SES) \cite{GWUWEB}, and energy dependent 
solutions (WI08) \cite{Arn04,GWUWEB} where Chew-Mandelstam K-matrix formalism together with dispersion relation constraints is applied; iii) dynamic coupled-channel Lagrangian 
approaches from EBAC model \cite{Diaz07, Dur08} which  is based on an energy independent
Hamiltonian which is derived from a set of Lagrangians by using a unitary transformation method \cite{Matsuyama2007}; iv)  J\"{u}lich model  \cite{Juelich} which is also a 
dynamical coupled-channels model of meson production reactions in the nucleon resonance region which includes $\pi$N, $\eta$N, $\pi \Delta$, $\sigma$N, and $\rho$N channels; v) 
Dubna-Mainz-Taipei (DMT) model \cite{Che03,Che07}, a meson-exchange model for pion-nucleon scattering which was constructed using a three-dimensional reduction scheme of the 
Bethe-Salpeter equation for a model Lagrangian involving $\pi$, $\eta$, N, $\Delta$, $\rho$, and $\sigma$ fields; and vi) and the Giessen K-matrix approximation (analyticity 
violating) model \cite{Giessen}.  Let us not fail to mention other, important theoretical approaches like Cutkosky CMU-LBL phenomenological coupled channel model \cite{Cut79}, 
Pittsburgh phenomenological coupled-channel model \cite{Vrana2000} and Bonn-Gatchina K-matrix approach without dispersive part~\cite{BonnGatchina}, which are for different reasons 
not included in analysis of ref. \cite{Mirza2011}. \\ }

 In this paper we  discuss how stable our recommended procedure is with respect to our  model assumptions in order to give an quantitative  estimate how confident the results of 
that report actually  are. \\

We have been triggered to start the whole enterprise  by the present uncertainty how to make the least model
 dependent comparison between  scattering theories which analyze experimental data on one side, and QCD models or lattice QCD calculations on the other. In other words, we have 
been dealing with the problem how to precisely define what a resonance actually is. And  it has to be done on both sides, on scattering theory side, as well as on the QCD side. \\

One, and the most intuitive way to define a resonance in a scattering process, is to define it as an intermediate state of two particles when they, because of attractive 
interaction, dwell in the vicinity of each other longer then in a standard scattering process. When this definition is transformed into the language of scattering theory, since 
1976 it seems that, as H\"{o}hler has said in \cite{Hoe2001}: \textit{``It is `noncontroversial among theorists' (see Chew \cite{Chew1976} and the references in my `pole-emics', 
p.697 in ref.~\cite{Hoe2000}) that in S-matrix theory the effects of resonances follow from first order poles in the 2nd sheet."}  However, if resonances are to be identified with 
the first order poles in the 2nd sheet, we can not but observe that characterizing such states, namely identifying and quantifying scattering matrix poles in the complex energy 
plane turns out to be quite a challenge.  \\

Different methods aimed onto identifying one of the possible repercussions of a resonant state onto measurable quantities have been 
developed. First it has been speculated that a Breit-Wigner function should be a good representation of the complex energy pole, 
so the Breit-Wigner formula for spin zero particles together with its generalization to the non-zero spin case has been developed in 1936 (see an illustrative discussion in 
Cottingham and Greenwood, p.241 in ref.~\cite{Breit-Wigner}). Then in \cite{DalitzMoorhouse} Dalitz and Moorehouse  considered the scattering matrix eigenphases, and their rapid 
increase through  $\pi$/2 have been taken as a signature of pinpointing a scattering matrix pole (namely, the corresponding eigenvalue of the reaction matrix \mbox{K = i 
(S-1)/(S+1)} then has a pole at this energy \cite{Dalitz61}). Finally, closely related to the afore two, the quick backwards looping of Argand diagram have been also extensively 
discussed by  Dalitz and Moorehouse \cite{DalitzMoorhouse}. \\

As the time passed by, the real meaning of these methods, namely the fact that they are devised only to reach the scattering 
matrix pole in the complex energy plane \textit{in a certain approximation}, tended to be overlooked. These methods alone have been
 identified as resonance definition procedures, and what they found has been proclaimed to be a resonant state.  { However, their inherent model dependence, like for instance 
unitary addition of non-resonant background, has been systematically neglected 
 (see the elaboration on Breit-Wigner model dependence in ref.~\cite{Hoe2000}).} The problems of eigenphases, linked to the manifest  violation of Neumann-Wigner no-crossing 
theorem \cite{Neumann-Wigner29} when rapidly going through $\pi/2$ even for the case of simply constant background terms, have been suppressed in spite of the extensive 
elaboration of the problem given by Dalitz and Moorhouse in ref.~\cite{DalitzMoorhouse}. They have explicitly said that: \textit{``With such a complexity of branch cuts without 
physical significance, we must conclude that  the eigenphase representation for the S matrix is not  generally a useful representation for the scattering in the neighbourhood of a 
resonance"}, but have anyhow allowed the possibility to use it in a more restricted situation: \textit{``However, we should emphasize also that the eigenphase representation is a 
perfectly acceptable  and economical representation of the S matrix in situations where the eigenphases vary slowly and do not cross".} The single-channel nature of pole search 
methods  (with the exception of the time delay which is in principle of multi-channel character) also did not represent much of a problem,because experimental data in inelastic 
channels were scarce. The fact that a single-channel procedure can not reveal much information about resonant states being `far away' from that channel, has also been suppressed. 
\\ 

{ 
In ref.~\cite{Mirza2011}  we have returned to the root of the problem --  we have fully respected that we are looking for the pole in the complex energy plane having at our 
disposal only data on the real axes, and instead of using different pole-extraction methods to extract a pole per resonance, we have suggested to use only one, Zagreb CMB method 
\cite{Bat98,Batinic2010} and extract poles from a selected collection of partial wave amplitudes simultaneously. { We have decided to use \emph{only one method} to extract pole 
positions from \emph{all published partial waves analyses}, and inspect the result { with the aim to distinguish which part of the disagreement in pole positions is coming from 
the different analytic structure, and which is coming from the different input (different PWA functions)}.  In other words, we have taken all sets of partial-wave amplitudes, 
treated them as nothing else but a good, energy dependent representations of all analyzed experimental data, and extracted the poles which are needed by the CMB method.} 
We observe that even when practically all analyzed PWA use very similar input data set,  their PWA solutions  do differ in spite of reporting the similar  quality of fit to the 
input data (similar reduced $\chi ^2$). So, from the theoretical point of view, they equally well describe the experiment.   And now these, however similar, but still different 
curves through different analytic continuations of each model generate corresponding and different sets of poles.  
So, in addition to the issue of slightly different input, the error of unknown analyticity is superimposed to it. It is important to realize that these poles,  even for the 
identical set of input amplitudes,  should not necessarily coincide, because the models used for analytic continuation are intrinsically different in their analytic form (for 
illustration see ref. \cite{Ceci2011}).  The idea of using only one model (Zagreb CMB) to extract the set of poles from different PWA treating them as partial wave data boils down 
to testing the internal agreement of input data sets. In this way, the difference between poles of various solutions is attributed only to the under-determinacy of input data and 
not to the analytic structure of the models in question. Simply, different poles obtained in this way quantify the difference in PWA solutions with respect to the similar input, 
and disregard the different analytic form used to obtain them.  In this manner, all  uncertainties originating  from different analytic { properties} of different models are 
avoided, and the only remaining errors { are the quality of the input and} the precision of CMB method itself. Therefore, averaging and error analysis of pole positions is 
sensible, and can be safely carried out. To answer the question of a correct choice of analytic form is a more complex problem and will be addressed elsewhere.  }
  \\ 
  
In this paper we discuss how sensitive this method is to { CMB} model assumptions when our input data set is well defined. We have first chosen the reliable input data set, a data 
set for which the minimization is fast, stable and unambiguous. Then we have identified our model assumptions, specific ones which we may check, and more general ones which are 
unfortunately beyond our reach.  The general ones, like the isobar character of the model representing the unstable channels with a quasi two body channel with 
 decay properly taken into account, could not be checked directly. Other ones, like  for instance the form of the channel propagator (channel-resonance vertex function), { 
dependence on the dispersion integral subtraction constant,}  background parameterization, number of channels,  mass of the effective channel, etc.~we  check directly. Instead of 
taking different recipes for each model assumption, we have simply modified the used functional form with an completely arbitrary function which is just changing the range of 
interest, and repeated the fit. In this way the stability with respect to different model assumption is performed.

\section{General idea}
If we want to apply a CMB formalism with the intention of extracting pole parameters from a ``world collection" of partial wave data and
 partial wave amplitudes, one has to answer a natural question how stable the formalism itself is with respect to model assumptions. 
 We are aware that we can test the sensitivity of the model upon its ingredients, but the reliability of the isobar model itself
  can not be tested in this way. Therefore, at this instant, we still can not say how strongly our answer depends on the fact that we assume
   that the interaction mechanism of isobar intermediate states is the only relevant one. \\

Our present analysis involves yet another approximation, and that is that  possible three body states are effectively represented by a
 group of two body ones. We shall not test this hypothesis either, even knowing that according to some considerations these effects can 
 build up to 30\%.  \\

Even before starting the procedure, we have to answer a nontrivial question: \\ \emph{``What is the optimal input data base so that our pole positions are unique { for the model 
without any modifications}?"} \\

Once this answer is given, we have to analyze how the final pole positions depend on model assumptions like (i) shape of the channel-resonance 
form factor; { (ii) choice of the dispersion integral subtraction constant;} (iii)  masses of the effective channel; (iv) number of channels; (v) type of the background 
parameterization.
\\ \noindent
The driving idea of this paper is: \\
\emph{``To {\textit{drastically}} modify only one part of the model at a time, and then make a fit to the chosen data base.
 The dissipation of pole positions known from the original fit will give us information how important certain part of the model is 
 for the stability of the solution."} \\

We will just add that these changes are intentionally artificial, aimed only at obtaining the effect, and not paying attention to the
 physical meaning. 
 
\section{Formalism}
The CMB model is  isobar, coupled-channel, analytic, and unitary  model, where the T matrix in a given channel is assumed 
to be a sum over the contributions from a number of intermediate particles (resonance and background contributions). The coupling of the channel asymptotic states to these
 intermediate particles determines the imaginary part of the channel function, and is represented effectively with a separable function. 
 The real part of the channel function is calculated by the dispersion relation technique, thus ensuring analyticity.  
 Besides the known resonance contributions, the background contributions are included via additional terms with poles below the $\pi$N threshold. 
 Due to the clear analytic and separable structure of the model, finding the pole positions in CMB model is trimmed down to the
  generalization of the dispersion integral for the channel propagator from real axes to the full complex energy plane, and this
   is a very well defined procedure. In practice, we instead use a very stable, and numerically much faster analytic continuation 
   method based on the Pietarienen expansion \cite{Pietarinen} in order to extrapolate the real valued channel propagator into the 
   complex energy plane. 

\subsection{Formulae}
Our current partial-wave analysis~\cite{Bat98} is based on the manifestly unitary, multichannel CMB approach of ref.~\cite{Cut79}. 
The most prominent property of this approach is  
analyticity of partial waves with respect to Mandelstam $s$ variable. In every discussion of partial-wave poles, analyticity plays a
 crucial role since the  
 poles are situated in a complex plane, away from physical region, and our measuring abilities are restricted to the real energy axis only.
  To gain any knowledge about the nature of partial-wave singularities would be impossible if partial waves were not analytic.
   Therefore, the ability to calculate pole positions is not just a benefit of the CMB model's  
analyticity but also a necessity for resonance extraction. In this approach, the resonance itself is considered to exist 
if there is an  associated partial-wave pole in the ``unphysical'' sheet. 
\\ \\ \noindent
We use the  multichannel $T$ matrix related to the scattering matrix $S$ as:
\begin{equation}
S_{ab}(s)= \delta_{ab} + 2\,i\,T_{ab}(s), 
\nonumber
\end{equation}
 where $\delta_{ab}$ is Kronecker delta symbol.  
The T-matrix matrix element is in the CMB model given as: 
\begin{eqnarray}
\label{eq:Tmatrix}
 T_{ab}^{JL}(s) 
= \sum_{i,j=1}^{N^{JL}}  f_a^{JL}(s)  \sqrt{\rho_a (s)}  \gamma_{ai}^{JL}G_{ij}^{JL}(s)\gamma_{jb}^{JL}\sqrt{\rho_b(s)}f_b^{JL}(s) & & 
\end{eqnarray}
where $a(b)$ represents the outgoing (incoming) channel. In our analyses we use $a,b=\pi N,\eta N,\pi^2 N$. The initial and final channel $b(a)$ 
couple through intermediate particles labeled $i$ and $j$. The factors $\gamma_{ia}$  are energy-independent parameters 
occurring graphically at the vertex between channel $a$ and intermediate particle $i$ and are determined by fitting procedure. Also occurring at
 each initial or final vertex is form factor $f_a^{JL}(s)$
\begin{equation}
\label{eq:ch-resvfun}
 f_a^{JL}(s)=\left(\frac{q_a}{Q_{1a}+\sqrt{Q_{2a}^{2}+q_a^2}}\right)^L,
\end{equation}
and phase-space factor $\rho_a(s)$
\begin{equation}
 \rho_a(s)=\frac{q_a(s)}{\sqrt{s}},
\end{equation}
where $s=W^2$ is a Mandelstam variable, and  $q_a(s)$ is the meson momentum for any of the three channels given as 
\begin{eqnarray}
 q_a(s) =  \frac{\sqrt{(s-(m+m_a)^2)(s-(m-m_a)^2)}}{2\sqrt{s}}. 
\end{eqnarray}

Furthermore, $L$ is the angular momentum in channel $a$, and $Q_{1a}$, $Q_{2a }$ are constants. The factor $f_a^{JL}(s)$ provides appropriate threshold behavior 
on the right-hand cut, and also produces a left-hand branch cut in the $s$ plane. Parameters $Q_{1a}$ and $Q_{2a}$ are chosen to determine the
 branch point and strength of the left-hand branch cut. In our analyses they have been taken to be the same, and are fixed to the mass of
 the channel meson $a$. \\

$G_{ij}^{JL}$ is a dressed propagator for partial wave $JL$ and particles $i$ and $j$, and may be written in terms of a diagonal bare propagator 
$G_{ij}^{0JL}$ and a self-energy matrix $\Sigma_{kl}^{JL}$ using Dyson equation
\begin{eqnarray}\label{Dyson}
 G_{ij}^{JL}(s) & = & G_{ij}^{0JL}(s)+\sum_{k,l=1}^{N^{JL}}G_{ik}^{0JL}(s)\Sigma_{kl}^{JL}(s)G_{ij}^{JL}(s).   
\end{eqnarray}
\\

The bare propagator 
\begin{equation}\label{bare}
 G_{ij}^{0JL}(s)=\frac{e_i\delta_{ij}}{s_i-s}
\end{equation}
has a pole at the real value $s_i$. The sign $e_i=\pm 1$ must be chosen to be positive for poles above the elastic threshold which correspond 
to resonance.  \\

The nonresonant background is described by a meromorphic function, in most of the cases consisting of two terms of the form 
(\ref{bare}) with pole positions below $\pi N$ threshold. For that case, the signs of the terms are opposite. The positive sign correspond to the
 repulsive and the negative sign to the attractive potential. In principle the number of poles can be increased arbitrarily 
 (see the next subsection on background representation), but in reality the number is never larger than three. \\

$\Sigma_{kl}^{JL}$  is the self-energy term for the particle propagator
\begin{equation}
 \Sigma_{kl}^{JL}(s)=\sum_a \gamma_{ka}^{JL} \cdot \Phi_a^{JL}(s) \cdot \gamma_{la}^{JL}
\end{equation}
The $\Phi_{a}^{JL}(s)$ are called ``channel propagators''. They are constructed in an approximation that { treats} each channel as containing
only two particles. We require that $T_{ab}^{JL}$ have, in all channels, correct unitarity and analyticity properties consistent with a 
 quasi-two-body approximation. \\
 
 The imaginary part of $\Phi_a^{JL}(s)$ is the effective phase-space factor for channel $a$:
\begin{equation}
\label{eq:F}
 \mathrm{Im} \ \Phi_a^{JL}(s)=[f_a^{JL}(s)]^2\rho_a(s).
\end{equation}

\begin{figure*}[!htb]\begin{center}
 \vspace*{-6.cm}
\includegraphics[width=8.cm]{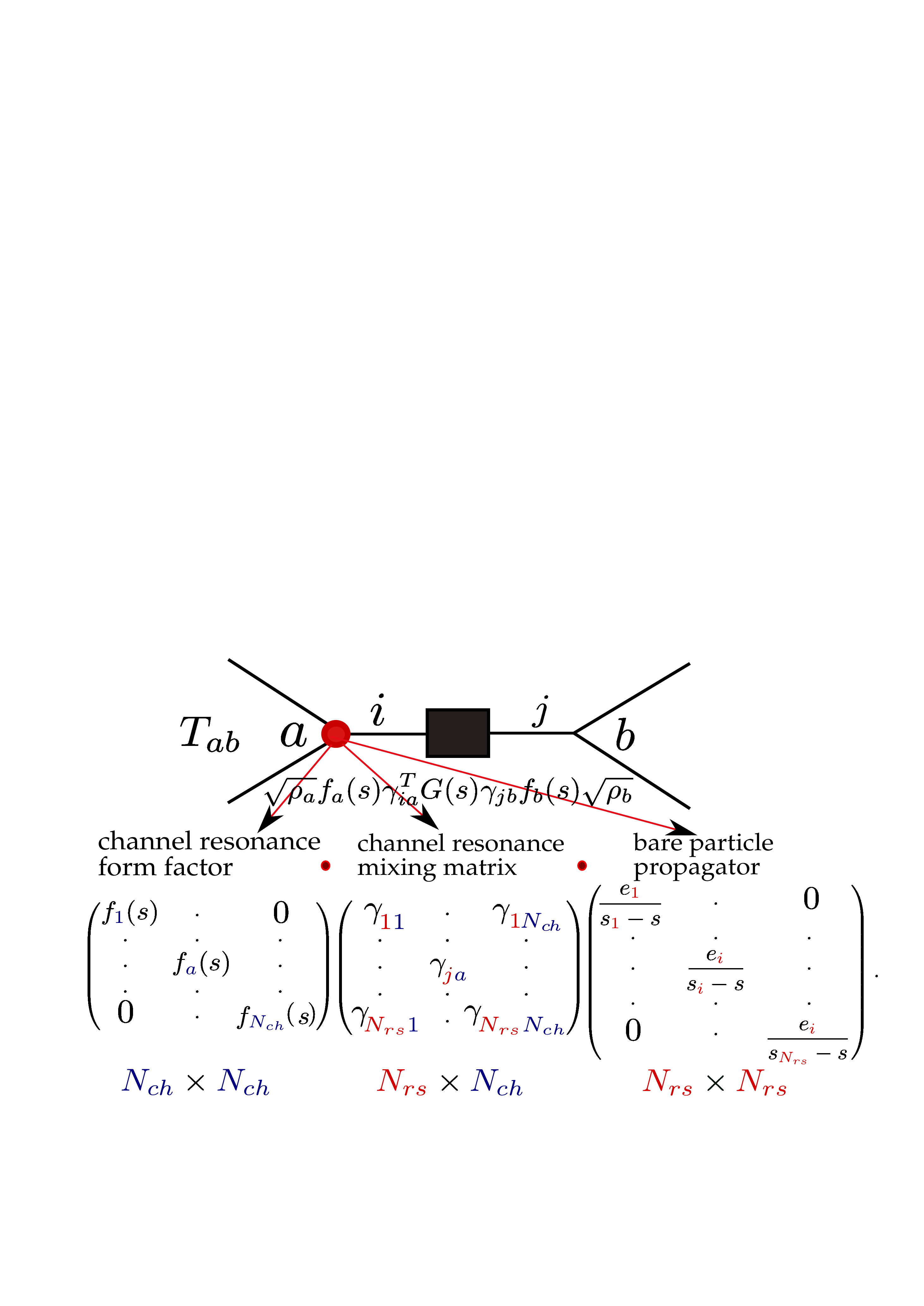}\\
\end{center} 
\vspace{-2.cm}
 \caption{(Color online) Parameterization of channel-intermediate particle vertex in CMB model.}
 \label{Figure1}
\end{figure*}

The channel propagator is evaluated on the real axes only
\begin{equation}\label{eq:imphi}
\mathrm{Im}\, \Phi(x)=\frac{\left[q(x)\right]^{2L+1}}{\sqrt{x}\,\left\{Q_1+\sqrt{Q_2^2+\left[q(x)\right]^2}\right\}^{2L}}, \vspace*{0.1cm}
\end{equation}
where by $x$ we stress the fact that values are on the real axes. 
The real part of $\Phi_a^{JL}(x)$ is calculated using a subtracted dispersion relation

\begin{equation}\label{eq:DR}
\mathrm{Re} \, \Phi(x)= { \mathrm{Re} \, \Phi(x_0)}+ \frac{x-x_0}{\pi}\,\,\mathrm{P}\int_{x_a}^{\infty}\frac{\mathrm{Im}\,\Phi(x')\,dx'}{(x'-x)(x'-x_0)}.
\end{equation}
where $x_a=(m+m_a)^2$ { and $\mathrm{Re} \, \Phi(x_0)$ is a subtraction constant at the subtraction point $x_0$. In all publications \cite{Bat98,Batinic2010} we have chosen the 
convention that $x_0=x_a$ and $\mathrm{Re} \, \Phi(x_0)$ = 0, but this is not necessarily so. }
For better understanding, the structure of the channel-intermediate particle form factor and its relation to the full T-matrix is given in Fig.~\ref{Figure1}.
 \\
We give a matrix form of the final T-matrix as defined in Eq.~(\ref{eq:Tmatrix}): 
\begin{widetext}
 \begin{eqnarray}
      \hat{T}(s) = 
	  \sqrt{\mathrm{Im}  \hat{\Phi}(s)} \cdot
    \hat{\gamma}^{\rm T} \cdot  \frac{\hat{G}_{0}(s)}{I- \left[\hat{\gamma} \cdot \hat{\Phi}(s) \cdot \hat{\gamma}{\rm ^T}\right] \cdot
	 \hat{G}_{0}(s)} \cdot  \hat{\gamma}  \cdot \sqrt{\mathrm{Im}  \hat{\Phi} (s) } 
\label{eq:final}
\end{eqnarray}
\end{widetext}
\subsection{Background contributions}
Theoretically, background term is in principle a free function of energy. However, an approximation is very often used that one 
may decompose the background term into a finite number of  unphysical pole terms, i.e., the background is a meromorphic function. This approach is very convenient for the CMB type 
calculation because integral equations  for the background poles are solved in the identical manner as for physical ones, so has it been applied and  strongly defended by Cutkosky 
et al \cite{Cut79} in CMB PWA.   Zagreb group has taken it over in \cite{Bat98}. \\

Let us elaborate on what we have actually done when we assumed that the background {\it indeed is} a meromorphic function. In that case, instead of solving a nonlinear equation 
for the T-matrix pole positions when the background has a general energy dependent form, we have  actually introduced a new type of  poles  which mimic the background. Here we 
rely on the approximation that the solutions of a nonlinear equation with general type background will be close to the solutions of ``pole-type" equation, the equation when the 
meromorphic background  is very close to the original function. Consequently, we do not have one, but two types of T-matrix poles: i) poles  which correspond to the physical 
internal singularities; and ii) poles which correspond to the background. We call the first 
	  type  genuine poles, and the second ones we call dynamic poles. \\ \\ \noindent
The speculative question is: {\it ``How realistic the meromorphic background approximation  actually is?"} 
\subsection{Pole extraction procedure}
The poles of the T matrix given in Eq.~(\ref{eq:Tmatrix}) are found by solving the equation:
\be
\label{eq:pole}
\mathrm{det} \, \, G^{-1}(s)=\mathrm{det} \, \left[ e_{ij} \delta_{ij}(s_i-s)-\Sigma_{ij} \right]=0.
\ee
This ia a complex equation, and can be solved only by {  analytic} continuation 
of the physical, experimentally accessible T-matrix values which of course lie on the real energy axes into the complex energy plane.
 Let us observe that the only part of the Eq.~(\ref{eq:pole}) which is not  defined in the complex energy
  plane is the channel propagator which is defined on the real axes only. \\

Therefore, the problem of finding the poles of Eq.~(\ref{eq:Tmatrix}) is reduced to analytic continuation of the channel propagator. \\

This analytic continuation can be done either by numeric principle value integration, or by Pietarinen expansion. 
The numeric integration of the channel propagator, used in ref.~\cite{Bat98}, is a straightforward procedure, but requires a lot of
computer time. Instead, we have used a more elegant way, a \textit{Pietarinen expansion} used by Karlsruhe-Helsinki group \cite{KH80}. 
\\ \indent
From Eq.~(\ref{eq:imphi}) it is evident that $\Phi(s)$ has a square-root type singularity in cm momentum, and analytic continuation
 into the complex energy plane should take it into account. Instead of calculating the dispersion integral  
(\ref{eq:DR}) for each point in complex plane, we decided to expend the function $\Phi(s)$ in power series of a new variable, which 
accounts for all analyticity requirements. We use the expansion (similar to Pietarinen's in Ref.~\cite{Pietarinen}  or Ciuli's~\cite{Ciu62})
\begin{equation}\label{eq:expansion}
\Phi _{\mathrm{I}}(s)=\sum_{n=0}^{N}C_n\,\left[Z_{\mathrm{I}}(s)\right]^n,
\end{equation}
where $C_n$ are coefficients of expansion. The new channel dependent variable is given by its principal branch
\begin{equation}
Z_{\mathrm{I}}(s)=\frac{\alpha-\sqrt{x_a-s}}{\alpha+\sqrt{x_a-s}},
\end{equation}
with the tuning parameter $\alpha$. This function is fitted to a data set consisting of imaginary parts of $\Phi(x)$ from Eq.~(\ref{eq:imphi}) and  
real parts of $\Phi(x)$ calculated from dispersion relation (\ref{eq:DR}), both of them evaluated at real axis (hence $x$). \\

The general idea is  
that the $\Phi(s)$ inherits analytic structure from $Z(s)$. We obtained parameters $\alpha$ and coefficients $C_n$ for each channel, and for all  
analyzed partial waves. The least-square fit is considered to be good if it meets following conditions: (i) small number of coefficients $C_n$  
needed (7 or 8, at most), (ii) the function fitted to the part of data set, when extrapolated outside of the fitted region is consistent with the  
rest of data, and (iii) fitting just imaginary part of $\Phi(x)$ produces real part that is in agreement with values obtained from (\ref{eq:DR}). \\

The channel propagator given by expansion (\ref{eq:expansion}) is obtained quite accurately { on the real axes}, and { as the expansion possess a correct analytic structure of 
2$\rightarrow$2 scattering processes, the deviation between expansion (\ref{eq:expansion}) and the correct value given by Eq. (\ref{eq:DR}) should be fairly small for the resonant 
region in the  
vicinity of physical axis.} \\

Every channel opening is responsible for two Riemann sheets: the first (physical) sheet with physical partial waves, and the secondary 
(unphysical) sheet with resonant poles. To get to the unphysical sheet it is enough to use the second branch of $Z(z)$
\begin{equation}\label{eq:simplerecipe}
Z_{\mathrm{II}}(s)=\frac{\alpha+\sqrt{x_a-s}}{\alpha-\sqrt{x_a-s}}.
\end{equation}

Finally, it is evident from Eq.~(\ref{eq:Tmatrix}) that all poles of each partial wave must by construction be the same in all channels and, in  
fact, equal to the poles of the resolvent $\mathbf{G}(s)$. \\

\begin{figure*}[!htb]
\centering
\vspace*{-6.cm}
{\includegraphics[width=10cm]{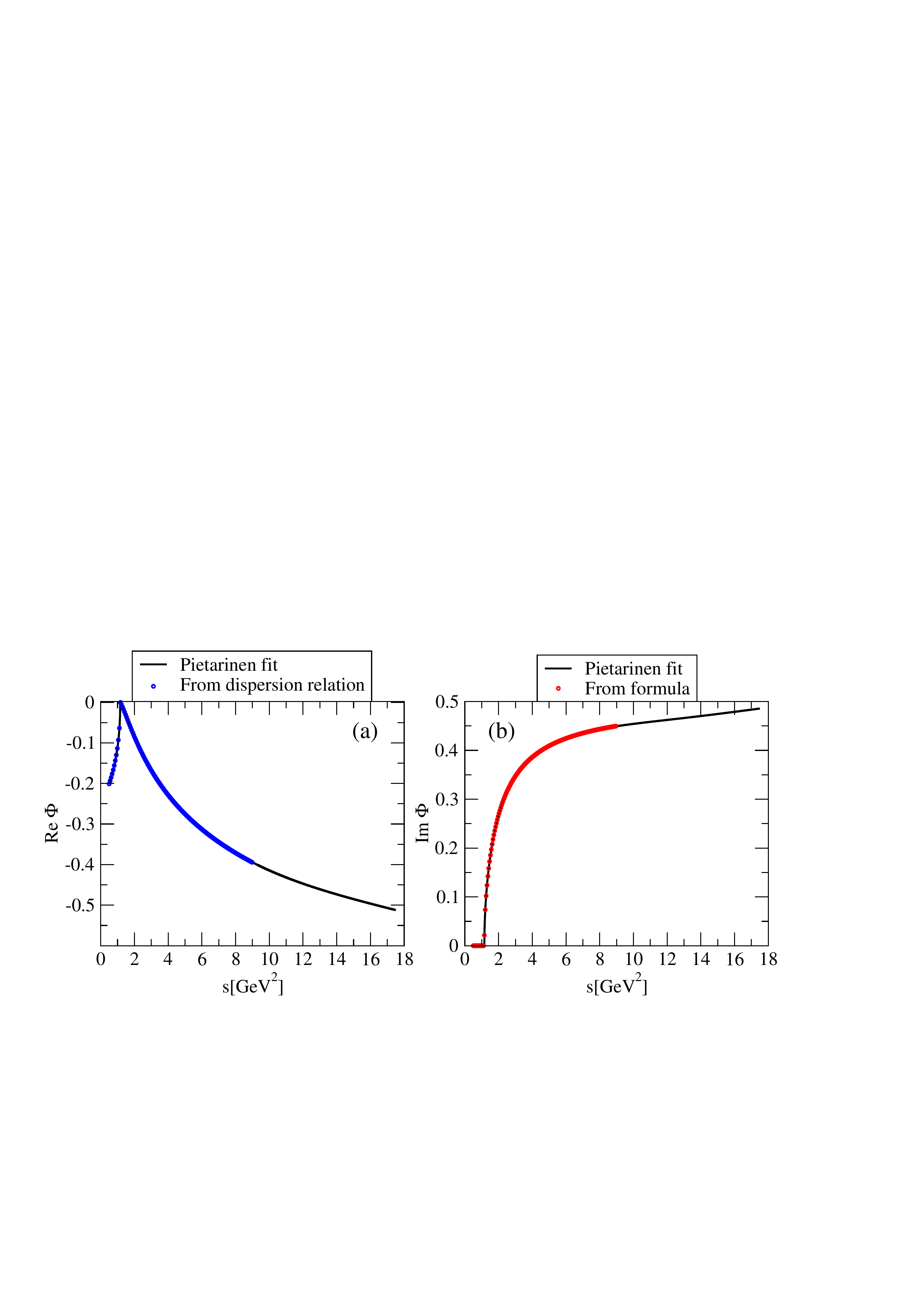}}
\vspace*{-2.cm}
\caption{\footnotesize (Color online) Pietarinen expansion for the S$_{11}$ partial wave.}\label{Fig2}
\end{figure*}

In practice, for the S-wave $\pi$N channel propagator, the realization of the Pietarinen expansion { on the real axes} is shown in Fig. \ref{Fig2}. 
Red points represent the imaginary part of the channel propagator $\Phi$ on the real axes, and are given by Eq.~(\ref{eq:imphi}), 
blue points represent the real part of the channel propagator $\Phi$ on the real axes, and are calculated by the dispersion relation 
(\ref{eq:DR}), and the full line represent the function $\Phi_{\mathrm{I}}$(x), a complex function defined on the full complex energy plane,
 and having a proper branch points and corresponding cuts for each channel. 

As the channel propagator is now known in the whole complex energy plane, the equation Eq.~(\ref{eq:pole}) { was solved numerically as the complex function of the  complex 
arguments.} In reality, instead of looking for zeroes of Eq.~(\ref{eq:pole}) we have 
looked for the poles of $\mathrm{det} \, G (s)^{-1}$.
In practice we have calculated $|\mathrm{det} \, G|$ as a function of Re(s) and Im(s) - (3D plot) specifying exactly the Riemann 
sheet in question (we have defined the appropriate momentum square roots in the channel propagator). Then we have looked  for the values 
of Real(s) and Imag(s) when $|\mathrm{det} \, G|$ becomes  bigger then an arbitrary number.  \\

\begin{figure}[!htb]
\begin{center}
\vspace*{-6.cm}
\includegraphics[width=6.cm]{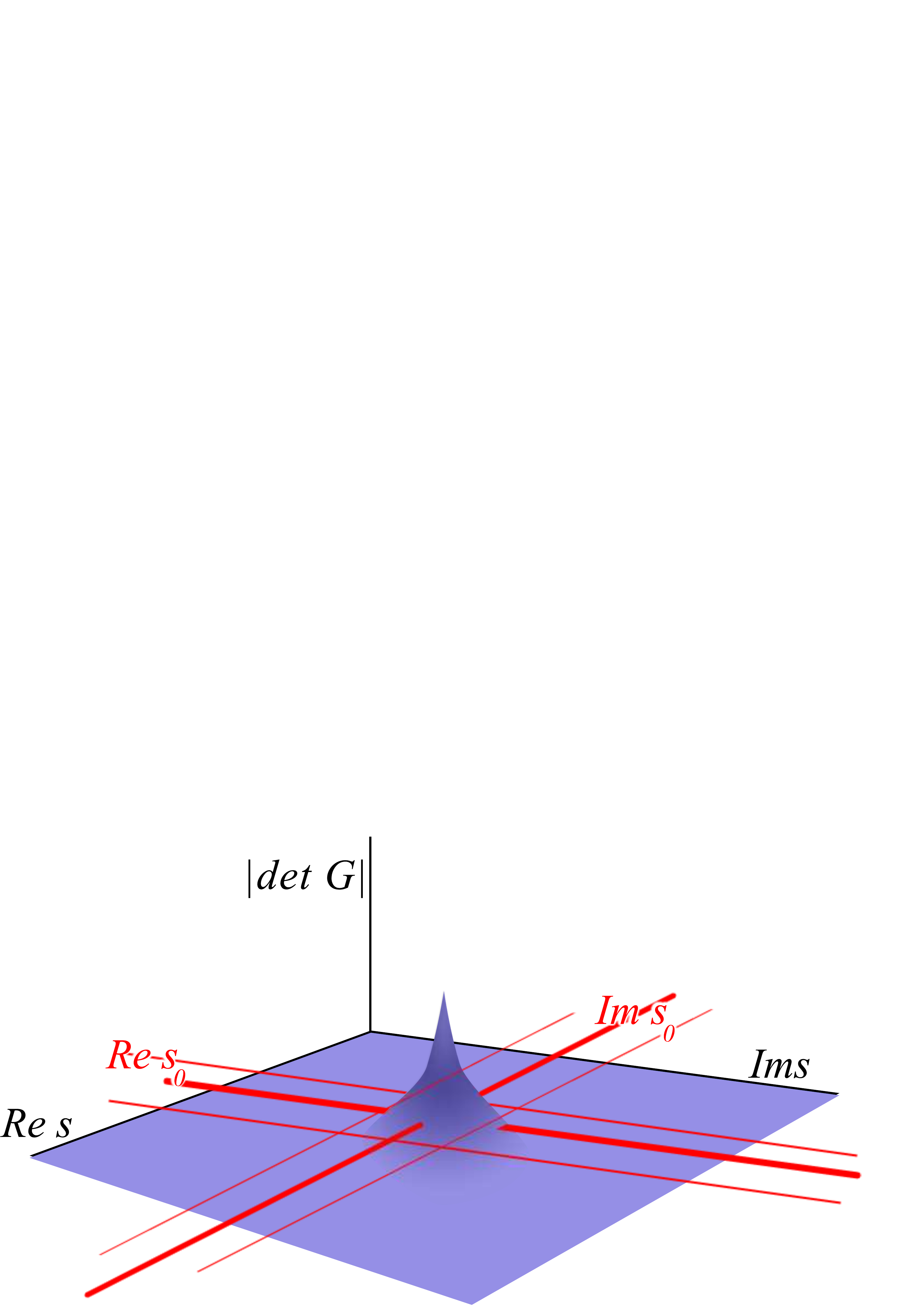}\\
\end{center} 
 \caption{(Color online) The principle of numerical finding the pole positions using Eq.~(\ref{eq:pole}) with the Pietarinen expansion of the channel propagator.}
 \label{Fig3}
\end{figure}

We regarded the obtained values of Re(s$_0$) and Im(s$_0$) as pole positions, and illustrate the procedure in  Fig.~\ref{Fig3}. 
\section{Sources of model dependence}
Understanding that the general assumptions of the model like its isobar, two body character in which the three and more body channels 
are identified with a quasi two body channels are properly taken into account,  we now identify other sources of model dependence which
 at the present moment represent additional,  systematic and uncontrollable uncertainties:  
\begin{enumerate}
\item the form of the channel propagator (meson-resonance vertex function)
       \begin{itemize}
	        \item its asymptotic part
			\item its intermediate part
			\item  threshold behavior
		\end{itemize}	
 { \item dispersion integral subtraction constant}
 \item background parameterization
 \item number of channels 
 \item mass of the effective channel
 \end{enumerate} 
 
The idea of present work is to allow for the extensive change of the chosen model features, and see at which level the pole positions
 are starting to move. The extensive change of the chosen model feature is realized by multiplying it with a freely chosen function 
 (no physical meaning whatsoever) which influences only the tested part of the model assumption, and parameters are varied in such a way that the change in the analyzed part 
in/decreases from $\sim$ 10\% to more than 100\% of the original value. \\

Prerequisite of such a procedure is that the input data base is such that it ensures a confident and reliable fit. Namely, we have to be positive that the variation of the 
extracting pole positions is coming from changing the ingredients of the model drastically, and that the fit to the input data base converges fast and uniquely. Therefore, we 
first describe the choice of the input data base, and then we test the ingredients of the model.

\section{Testing the model dependence}
\subsection{{ Defining the input data set}}
If we want to test the stability of the { model with respect to its ingredients}, we first have to choose such an input data base { (set of input data our modified model will 
fit)} for which the fitting procedure with original 
model is fast, unique and reproducible. { In other words, we want to choose such an input data base for which the original model reproduces its own, input values safely, quickly 
and unambiguously.} In that case, we can vary the { ingredients} of the model, and claim that the resulting shift in pole 
positions is the result of change in the model { itself}, and not the  consequence of inadequate input data. {  This is in a way  a test of internal consistency of the model, the 
proof that the model when fitting its own result obtains the parameters that were used to generate the input (logistic tautology).  However, when we start changing the ingredients 
of the model, the analytic structure of the model itself is changed, so we necessarily expect to obtain a different set of resulting pole parameters. The deviations of these  
parameters from the original input values are now the measure of sensitivity of the model to different model assumptions. }\\ 

As the input for this work we have chosen  partial wave amplitudes of Zagreb model \cite{Bat98}. Once the input parameters are chosen 
and fixed (bare poles and channel-resonance coupling constants $\gamma$), we have all partial waves at our disposal. 
In this section we test how stable the fit is for the first three lowest angular momentum partial waves S$_{11}$, P$_{11}$ and D$_{13}$ 
with respect to how many channel data sets we have used.  \\

In Zagreb model we have three set of partial wave amplitudes at our disposal: \piNpiN, \piNetaN and $\pi \! N \! \rightarrow \! \pi^2 N$.
In principle, all Zagreb PWA are smooth energy dependent functions. We could have fitted these solutions directly. However, we have been of an 
opinion that by doing so it would introduce too many analytic constraints coming from the fact that these 
solutions ``memorize" the analytic form of CMB formalism used to obtain them. We wanted to eliminate these additional constraints, 
so we have distributed the energy continuous Zagreb solutions by the recipe given in ref.~\cite{Ceci06}. \\

\begin{figure}[!htb]
\centering
\includegraphics[width=11.5cm]{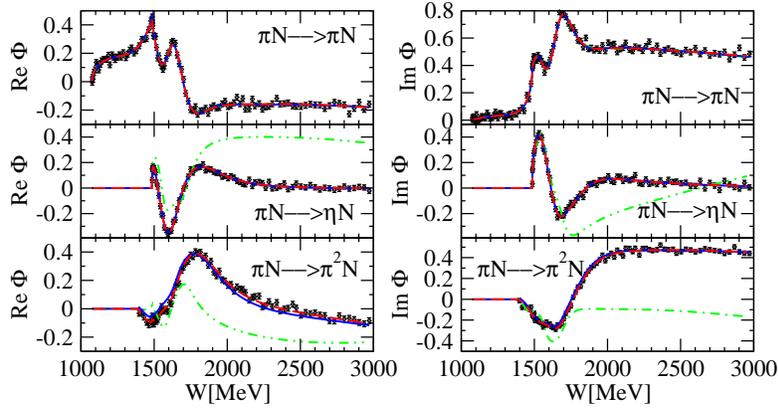} \\ 
\vspace{-9.5cm}   (a) S11 wave \vspace*{0.7cm} \\
\includegraphics[width=11.5cm]{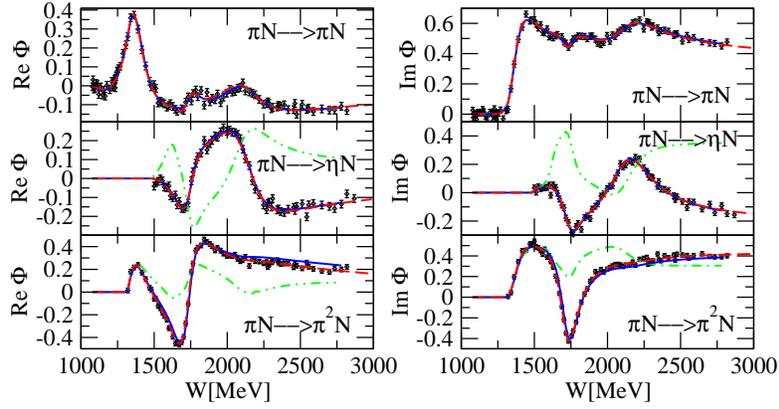} \\
\vspace{-9.5cm}  (b) P11 wave \vspace*{0.7cm}  \\
\includegraphics[width=11.5cm]{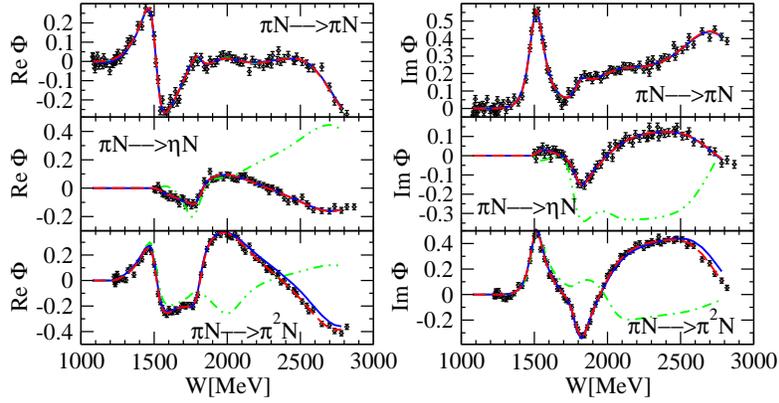} \\
\vspace{-9.5cm}  (c) D13 wave 
\caption{\footnotesize{(Color online) The input data sets together with our fit to them with the standard model are shown for the S$_{11}$, $P_{11}$
 and $D_{13}$ partial waves T-matrices.}}
\label{Fig:input_data}
\end{figure}

 Instead of using smooth theoretical curves { and avoid pathologically small $\chi_R^2$} , we constructed  pseudo data points by normally distributing the model input in order to
simulate the statistical nature of really measured data.  The standard deviation $\sigma$  was set to 0.02,  similar to the average error
 value  of GWU data { \cite{Arn04,GWUWEB}}. By using this procedure we were able to produce a set of \piNpiN, \piNetaN and $\pi \! N \! \rightarrow \! \pi^2 N \:$
  partial-wave data which reproduced the experiment,
  and when fitted, they gave realistic $\chi^2$ values comparable with those obtained by GWU/VPI~SES fits.
The pole parameters of the input data set are given in Table  \ref{Table:tab1}.  The input data sets together with our fit are shown in Fig.~\ref{Fig:input_data}. \\

To simplify the discussion, we introduce the generic abbreviation L$_{2I  2J}($mass$)$ for each pole position 
produced in Zagreb model. \\

To illustrate, the S$_ {11}$ (1518) pole denotes the first, 1518 MeV S$_{11}$ pole, and is identified with the N(1535) S$_{11}$ PDG resonance.  
Observe that all Zagreb solutions correspond to an existing PDG resonant state, with the exception of two P$_{11}$ intermediate energy solutions.
 In Zagreb model we get two P$_{11}$ solutions: P$_{11}$ (1708) pole which is in \cite{Bat98,Batinic2010} identified with N(1710) P$_{11}$ state, 
 and a new solution P$_{11}$ (1728) which is not confirmed by PDG. \\

{ In Tables \ref{tab2}, \ref{tab3}, \ref{tab4} we show the results of the test in which we have looked for the optimal number of fitted channels required for the stable fit. 
Results obtained by fitting only elastic channel are denoted by Fit~1. Fit~2  stands for a case when the $\pi N \rightarrow \eta N$ channel is included in a fit in addition to 
elastic channel. Results denoted by Standard fit are obtained by fitting all three channel simultaneously.  The reduced chi-squared for the limited data input (one channel for Fit 
1 and two channels for Fit 2) is denoted with $\chi^2$. We have calculated the new chi-square parameter which is obtained when solutions for Fit 1 and Fit 2 are used to calculate 
the agreement with all three channels  without any fitting of the third channel, and denoted it  as $^{tot}\chi_{R}^2$.  This new parameter now describes how well single and two 
channel fits reproduce all channels at the same time. We are aware that fitting two out of three channels should in principle fully predict the third one for the manifestly 
unitary model as Zagreb CMB,  but this is true for the ideal, theoretical and smooth input.  However, when we have decided to distribute the smooth input by adding realistic 
errors, in spite of the fact that the statement is still in principle valid, the reality makes the fit rather complicated and slightly unstable if only two channels are fitted. 
That can be seen when analyzing the tables containing the $\chi^2$  and $^{tot}\chi_{R}^2$ for the two and three channel fits. Both parameters are qualitatively similar, but for 
the three channel fits, $\chi^2$ is still always slightly better then the  $^{tot}\chi_{R}^2$ parameter for the two channel ones. \\

Hence, we have shown that reduced $\chi^2$ indeed is sufficiently  stable if we fit two out of three channels, but in spite of the fact that fitting three out of three channels 
effectively is over-constraining the problem, we have decided to use it thought this paper because  this technically simplifies the fitting procedure when generated pseudo-data 
are fitted, and also opens the possibility to increase the number of channels. } \newpage

\clearpage
\begin{table*}[!htb]
\caption{The original parameters of the Zagreb solution  for the S$_{11}$, P$_{11}$ and D$_{13}$ partial  waves. }
\vspace{-0.1cm}
\begin{center}
\begin{tabular}{|c|cccc|cccc|}
\hline \hline 
 Partial&  \multicolumn{4}{c|}{ Bare poles} & \multicolumn{4}{c|}{ Dressed poles } \\ \cline{2-9}
\rule{0cm}{0.5cm}waves &$\mathrm{W_{s_1}}$ & $\mathrm{W_{s_2}}$ & $\mathrm{W_{s_3}}$   &$\mathrm{W_{s_4}}$ & {\scriptsize { $\begin{pmatrix}
\mathrm{Re W}\\ \mathrm{-2Im W} 
\end{pmatrix}$}} & {\scriptsize { $\begin{pmatrix}
\mathrm{Re W}\\ \mathrm{-2Im W} 
\end{pmatrix}$}} & {\scriptsize { $\begin{pmatrix}
\mathrm{Re W}\\\mathrm{ -2Im W} 
\end{pmatrix}$}}   & {\scriptsize { $\begin{pmatrix}
\mathrm{Re W}\\ \mathrm{-2Im W} 
\end{pmatrix}$}}     \\ 
& & \multicolumn{2}{c}{$\mathrm{MeV}$}  &  & & \multicolumn{2}{c}{$\mathrm{MeV}$}  &  \\ \hline \hline
\rule{0cm}{0.4cm}S$_{11}$ &1523 & 1640 & 1837 &  &{\scriptsize { $\begin{pmatrix}
1518\\ 191
\end{pmatrix}$}} &  {\scriptsize { $\begin{pmatrix}
1642\\ 205 
\end{pmatrix}$}} &	 {\scriptsize { $\begin{pmatrix}
1784\\ 424
\end{pmatrix}$}}& {\scriptsize { $\begin{array}{c}
\\ 
\end{array}$}} \\ [1.2ex] \hline 
\rule{0cm}{0.4cm} P$_{11}$  &1607 & 1771 & 2186 & 2852&{\scriptsize { $\begin{pmatrix}
1360\\ 161
\end{pmatrix}$}} & {\scriptsize { $\begin{pmatrix}
1708\\ 175 
\end{pmatrix}$}} &	{\scriptsize { $\begin{pmatrix}
1728\\ 138
\end{pmatrix}$}}& {\scriptsize { $\begin{pmatrix}
2114\\ 345 
\end{pmatrix}$}}  \\[1.2ex]  \hline 
\rule{0cm}{0.4cm} D$_{13}$ &1582 & 1880 & 2497 & &{\scriptsize { $\begin{pmatrix}
1507\\ 121
\end{pmatrix}$}} & {\scriptsize { $\begin{pmatrix}
1805\\ 128 
\end{pmatrix}$}} &	{\scriptsize { $\begin{pmatrix}
1942\\ 471
\end{pmatrix}$}}& {\scriptsize { $\begin{pmatrix}
2699\\ 558
\end{pmatrix}$}}  \\ [1.2ex] \hline \hline
 \end{tabular}
\end{center}
\label{Table:tab1}
\end{table*}
\begin{table*}[!ht]
\caption{The stability of the fitting result for the S$_{11}$ partial  wave.}~
\vspace{-0.3cm}
\begin{center}
\begin{tabular}{|c|c|c|c|ccc|ccc|c|c|}
\hline \hline 
 &\hspace{0.05cm}{\multirow{5}{*}{\begin{sideways}Number of \end{sideways}\hspace{0.1cm}\begin{sideways}channels\end{sideways}}}
 \hspace{0.1cm}&\hspace{0.05cm}{\multirow{5}{*}{\begin{sideways}Number of \end{sideways}\hspace{0.2cm}\begin{sideways}resonances\end{sideways}}}
 \hspace{0.1cm}&\hspace{0.05cm}{\multirow{6}{*}{\begin{sideways}Fitted \end{sideways}\hspace{0.1cm}\begin{sideways}channels\end{sideways}}}
 \hspace{0.1cm} & \multicolumn{3}{c|}{ Bare poles} & \multicolumn{3}{c|}{ Dressed poles } & &\\ \cline{5-10}
 \rule{0cm}{0.5cm}& &  & & $\mathrm{W_{s_1}}$ & $\mathrm{W_{s_2}}$ &  $\mathrm{W_{s_3}}$ & {\scriptsize { $\begin{pmatrix}
\mathrm{Re W}\\ \mathrm{-2Im W} 
\end{pmatrix}$}} & {\scriptsize { $\begin{pmatrix}
\mathrm{Re W}\\ \mathrm{-2Im W} 
\end{pmatrix}$}} & {\scriptsize { $\begin{pmatrix}
\mathrm{Re W}\\ \mathrm{-2Im W} 
\end{pmatrix}$}}   & $\mathrm{\chi_R^2}$& $\mathrm{^{tot}\chi_{R}^2}$  \\ [1.3ex]
 &&& & &$\mathrm{MeV}$   & & & $\mathrm{MeV}$&   & & \\[1.3ex] \hline \hline
\rule{0cm}{0.4cm}\textit{Fit 1}&3& 3& 1&1455 & 1654 & 2860 &{\scriptsize { $\begin{pmatrix}
1562\\ 100
\end{pmatrix}$}} & {\scriptsize { $\begin{pmatrix}
1684\\ 179 
\end{pmatrix}$}} &	{\scriptsize { $\begin{pmatrix}
4090\\ 2450
\end{pmatrix}$}}&  1.094  &117.186\\ [1.2ex] \hline 
\rule{0cm}{0.4cm}\textit{Fit 2}&3& 3& 2&1522 & 1635 & 1823 &{\scriptsize { $\begin{pmatrix}
1516\\ 185
\end{pmatrix}$}} & {\scriptsize { $\begin{pmatrix}
1640\\ 207 
\end{pmatrix}$}} &	{\scriptsize { $\begin{pmatrix}
1790\\ 435
\end{pmatrix}$}}&  1.004 & 1.711\\ [1.2ex] \hline 
 {\footnotesize { $\begin{array}{l}
Standard\\ fit 
\end{array}$}}&3& 3& 3&1523 & 1640 & 1834 &{\scriptsize { $\begin{pmatrix}
1518\\ 189
\end{pmatrix}$}} & {\scriptsize { $\begin{pmatrix}
1642\\ 208 
\end{pmatrix}$}} &	{\scriptsize { $\begin{pmatrix}
1779\\ 427
\end{pmatrix}$}}&  1.016  &1.016\\ [1.ex] \hline \hline 
\end{tabular}
\end{center}

\label{tab2}
\end{table*}
\begin{table*}[!ht]
\caption{The stability of the fitting result for the P$_{11}$ partial \vspace*{0.1cm} wave.}~
\vspace{-0.3cm}
\begin{center}
\begin{tabular}{|c|c|c|c|cccc|cccc|c|c|}
\hline \hline 
\rule{0cm}{0.3cm}
&\hspace{0.05cm}{\multirow{5}{*}{\begin{sideways}Number of \end{sideways}\hspace{0.1cm}\begin{sideways}channels\end{sideways}}}
\hspace{0.1cm}&\hspace{0.05cm}{\multirow{5}{*}{\begin{sideways}Number of \end{sideways}\hspace{0.2cm}\begin{sideways}resonances\end{sideways}}}
\hspace{0.05cm}&\hspace{0.05cm}{\multirow{6}{*}{\begin{sideways}Fitted \end{sideways}\hspace{0.1cm}\begin{sideways}channels\end{sideways}}}
\hspace{0.05cm}&  \multicolumn{4}{c|}{ Bare poles} & \multicolumn{4}{c|}{ Dressed poles } & & \\ \cline{5-12}
\rule{0cm}{0.5cm}&  & & &$\mathrm{W_{s_1}}$ & $\mathrm{W_{s_2}}$ &  $\mathrm{W_{s_3}}$   & $\mathrm{W_{s_4}}$ & {\scriptsize {
 $\begin{pmatrix}
\mathrm{Re W}\\ \mathrm{-2Im W} 
\end{pmatrix}$}} & {\scriptsize { $\begin{pmatrix}
\mathrm{Re W}\\ \mathrm{-2Im W} 
\end{pmatrix}$}} & {\scriptsize { $\begin{pmatrix}
\mathrm{Re W}\\ \mathrm{-2Im W} 
\end{pmatrix}$}}  & {\scriptsize { $\begin{pmatrix}
\mathrm{Re W}\\ \mathrm{-2Im W }
\end{pmatrix}$}}   & $\mathrm{\chi_R^2}$ & $\mathrm{^{tot}\chi_{R}^2}$ \\ [1.3ex]
 &&  && & \multicolumn{2}{c}{$\mathrm{MeV}$}  &  & & \multicolumn{2}{c}{$\mathrm{MeV}$}  & & & \\ [1.3ex]\hline \hline
\rule{0cm}{0.4cm}\textit{Fit 1}&3& 4& 1&1712 & 1781 & 2172 & 3162&{\scriptsize { $\begin{pmatrix}
1416\\ 69
\end{pmatrix}$}} & {\scriptsize { $\begin{pmatrix}
1701\\ 30 
\end{pmatrix}$}} &	{\scriptsize { $\begin{pmatrix}
1788\\ 128
\end{pmatrix}$}}& {\scriptsize { $\begin{pmatrix}
2118\\ 321
\end{pmatrix}$}}&  0.872 &99.866 \\ [1.2ex] \hline 
\rule{0cm}{0.4cm}
\textit{Fit 2}&3& 4& 2&1734 & 1923 & 2191 & 2330&{\scriptsize { $\begin{pmatrix}
1360\\ 169
\end{pmatrix}$}} & {\scriptsize { $\begin{pmatrix}
1695\\ 216 
\end{pmatrix}$}} &	{\scriptsize { $\begin{pmatrix}
1745\\ 112
\end{pmatrix}$}}& {\scriptsize { $\begin{pmatrix}
2120 \\ 350 
\end{pmatrix}$}}&  1.048 &1.739 \\[1.2ex]  \hline 
{\footnotesize { $\begin{array}{l}
Standard\\ fit 
\end{array}$}}&3& 4& 3&1607 & 1772 & 2182 & 2841&{\scriptsize { $\begin{pmatrix}
1365\\ 157
\end{pmatrix}$}} & {\scriptsize { $\begin{pmatrix}
1708\\ 174 
\end{pmatrix}$}} &	{\scriptsize { $\begin{pmatrix}
1731\\ 136
\end{pmatrix}$}}& {\scriptsize { $\begin{pmatrix}
2117\\  345
\end{pmatrix}$}}&  0.958  & 0.958\\ [1.ex] \hline \hline
\end{tabular}
\end{center}

\label{tab3}
\end{table*}
~
\begin{table*}[!ht]
\caption{The stability of the fitting result for the D$_{13}$ partial   wave.}~
\vspace{-0.3cm}
\begin{center}
\begin{tabular}{|c|c|c|c|ccc|cccc|c|c|}
\hline \hline 
\rule{0cm}{0.3cm} & \hspace{0.05cm}{\multirow{5}{*}{\begin{sideways}Number of \end{sideways}\hspace{0.1cm}\begin{sideways}channels\end{sideways}}}
\hspace{0.1cm}&\hspace{0.05cm}{\multirow{5}{*}{\begin{sideways}Number of \end{sideways}\hspace{0.2cm}\begin{sideways}resonances\end{sideways}}}
\hspace{0.05cm} &\hspace{0.05cm}{\multirow{6}{*}{\begin{sideways}Fitted \end{sideways}\hspace{0.1cm}\begin{sideways}channels\end{sideways}}}
\hspace{0.05cm} &  \multicolumn{3}{c|}{ Bare poles} & \multicolumn{4}{c|}{ Dressed poles } & & \\ \cline{5-12}
\rule{0cm}{0.5cm}& &   & & $\mathrm{W_{s_1}}$ & $\mathrm{W_{s_2}}$ &  $\mathrm{W_{s_3}}$ & {\scriptsize { $\begin{pmatrix}
\mathrm{Re W}\\ \mathrm{-2Im W} 
\end{pmatrix}$}} & {\scriptsize { $\begin{pmatrix}
\mathrm{Re W}\\ \mathrm{-2Im W} 
\end{pmatrix}$}} & {\scriptsize { $\begin{pmatrix}
\mathrm{Re W}\\ \mathrm{-2Im W} 
\end{pmatrix}$}}   &{\scriptsize { $\begin{pmatrix}
\mathrm{Re W}\\ \mathrm{-2Im W} 
\end{pmatrix}$}}   & $\mathrm{\chi_R^2}$ &$\mathrm{^{tot}\chi_{R}^2}$ \\ [1.3ex]
 &&  &&&  $\mathrm{MeV}$&  & & \multicolumn{2}{c}{$\mathrm{MeV}$}  & & &\\ [1.3ex] \hline \hline
\rule{0cm}{0.4cm}\textit{Fit 1}&3& 3& 1&1589& 1998 & 2537 &{\scriptsize { $\begin{pmatrix}
1500\\ 127
\end{pmatrix}$}} & {\scriptsize { $\begin{pmatrix}
1765\\ 200
\end{pmatrix}$}} &	{\scriptsize { $\begin{pmatrix}
1980\\ 232
\end{pmatrix}$}}&  {\scriptsize { $\begin{pmatrix}
2720\\ 597
\end{pmatrix}$}}&1.295 &137.071 \\  [1.2ex]\hline 
 \rule{0cm}{0.4cm}\textit{Fit 2}&3& 3& 2&1589 & 1880 & 2557 &{\scriptsize { $\begin{pmatrix}
1507\\ 127
\end{pmatrix}$}} & {\scriptsize { $\begin{pmatrix}
1800\\ 126 
\end{pmatrix}$}} &	{\scriptsize { $\begin{pmatrix}
1932\\ 508
\end{pmatrix}$}}&  {\scriptsize { $\begin{pmatrix}
2700\\ 595
\end{pmatrix}$}}& 1.070 &1.555 \\[1.2ex]  \hline 
{\footnotesize { $\begin{array}{l}
Standard\\ fit 
\end{array}$}}&3 &3 &3 &1582 & 1880 & 2499 &{\scriptsize { $\begin{pmatrix}
1506\\ 121
\end{pmatrix}$}} & {\scriptsize { $\begin{pmatrix}
1807\\ 127 
\end{pmatrix}$}} &	{\scriptsize { $\begin{pmatrix}
1939\\ 485
\end{pmatrix}$}}&  {\scriptsize { $\begin{pmatrix}
2691\\ 583
\end{pmatrix}$}}&1.027 & 1.027 \\ [1.ex] \hline \hline 
\end{tabular}
\end{center}
\label{tab4}
\end{table*}
\clearpage
\subsection{The form of the channel propagator $\Phi(s)$}
The { major} assumption of the CMB model is done about the functional form of the channel propagator. We first define the channel-resonance 
vertex function in Eq.~(\ref{eq:ch-resvfun}), and then combine it with the phase space factors into the analytic channel propagator, see 
Eq.~(\ref{eq:imphi}). We take over a recipe extensively used by CMB group \cite{Cut79}, and the full definitions are given by Eqs.
 (\ref{eq:ch-resvfun}, \ref{eq:imphi} and \ref{eq:DR}).

However, it is very difficult to justify the correctness of these assumptions. So, instead of defending its physical reasonability, and discussing 
the dependence of the vertex function upon the channel resonance vertex constants $Q_{1a}$ and $Q_{2a}$, we shall
 test how sensitive our model is substantial {\textit{but artificial}} changes of its constituent parts: asymptotic part, intermediate part,
  threshold behavior. \\
\noindent
\textit{Procedure} \\
\textit{(i)} We have chosen the modification function completely artificially, requiring that the change of the channel propagator is 
localized only to its constituent part, and that the transition from the modified to non-modified part is smooth. \\
\textit{(ii)} The imaginary part of the channel propagator is evaluated. \\
\textit{(iii)} The real part of the channel propagator is calculated using the dispersion relation given by Eq. (\ref{eq:DR}). \\ 
{ \textit{(iv)} The input data set is refitted with modified channel propagator. }
\\ \\
\noindent
\textit{Important:} \\
We have modified all three channel propagators {\textit{simultaneously}} using {\textit{the same modification function}}. 

\subsubsection{Asymptotic part}
\begin{figure}[!ht]
\vspace*{-8.cm}
\centering
{\includegraphics[width=12cm]{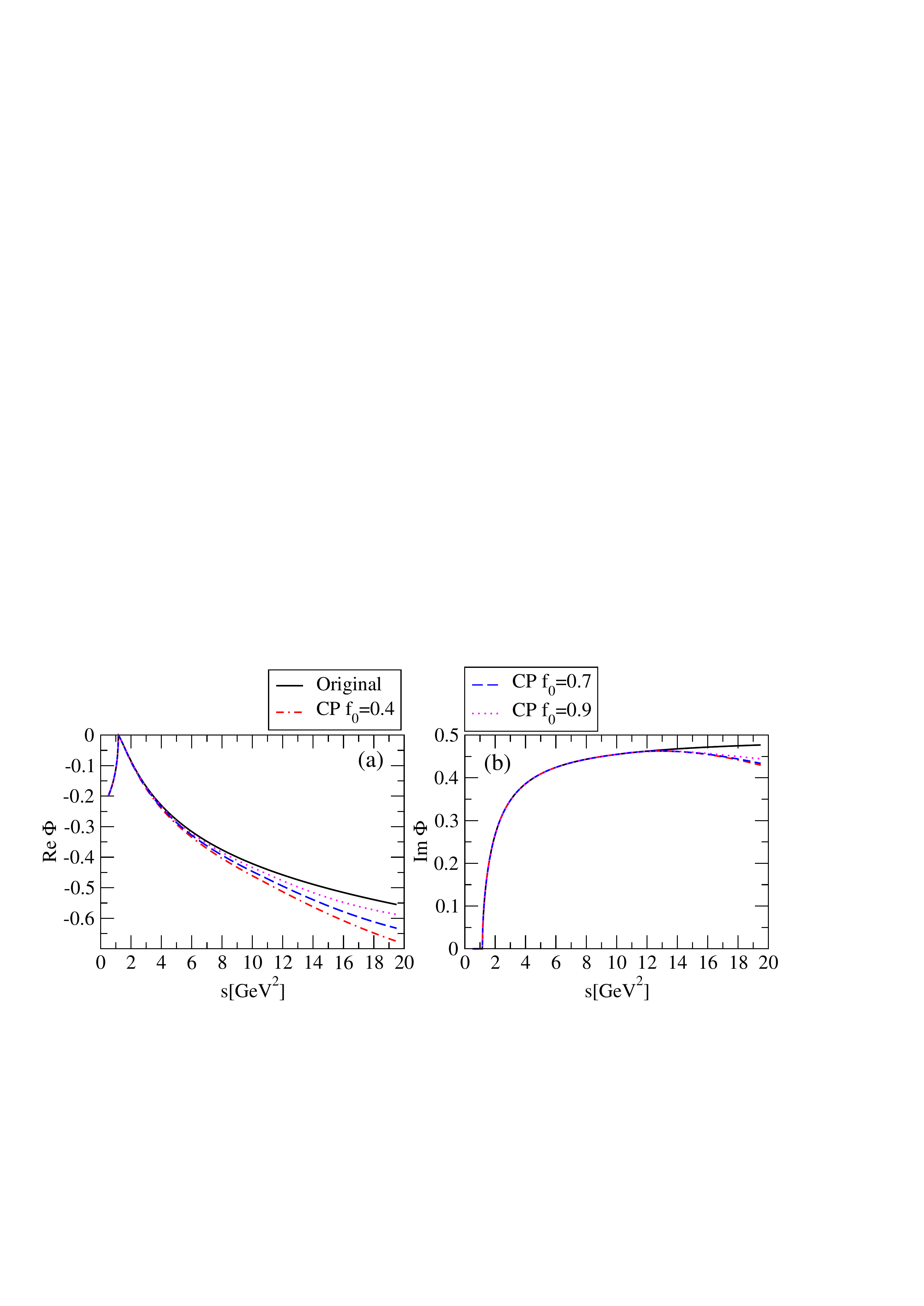}}
\vspace*{-4.cm}
\caption[]{\footnotesize (Color online) Real and imaginary part of channel propagator. Several values of parameter $f_0$ are used: $0.4$, $0.7$ and $0.9$.}
\label{asim}
\end{figure}

The imaginary part of the channel propagator was kept unchanged at $s$ below 10 GeV$^2$.
Above 10 GeV$^2$ the correction function g(s) which changes the asymptotic behavior of the channel propagator is introduced: 
\be \label{gs}
f(s)=f_0+ f_1 \cdot e^{-\lambda \cdot (s-s_0)^2},
\ee
 { Modified} imaginary part of the channel propagator (omitting channel indices) now reads:
\be
{\mathrm Im} \Phi_{corr}= \left\{^{{\mathrm Im} \Phi (s) {\rm \ \ \ for \ s \leq 10 \ GeV^2}}_{ {\mathrm Im} \Phi(s) \cdot f(s) {\rm \ \ \ for \ s>10 \ GeV^2 }} \right.
\ee 
Parameters $f_0$, $f_1$ and $\lambda$ are determined in such a way to ensure  continuity of channel propagator and its first derivative at $s=10$  GeV$^2$, and $s=10.5$ GeV$^2$.
Imaginary and real parts of channel propagator are shown in the Fig. (\ref{asim}) for several values of parameter $f_0$. 

{ Let us briefly discuss the analyticity and unitarity of afore used piecewise defined functions. When analyticity of the channel propagator is imposed via dispersion relations, 
the analyticity of real and imaginary parts themselves is not required \cite{Analyticity-piecewise}. Piecewise defined imaginary part of channel propagator is by construction a 
continuous function having continuous first derivative and not necessarily analytic, but when we use dispersion relations to obtain the corresponding real part, we automatically  
construct the analytic function having the same cut. The idea of the paper is to analyze how this change influences pole position of the T-matrix.}  \\

{ Such a change does not influence unitarity, because in ref. \cite{Cut79} it has been explicitly shown that the CMB formalism by the very construction retains unitary if all used 
functions are analytic.} \\ 

\noindent
\underline{\textit{Result:}} 

As it is seen in Fig. (\ref{asim}) neither imaginary nor real part of the channel propagator are changed in the relevant energy range s 
$\leq$ 10 GeV$^2$. \\

\noindent
\underline{\textit{Conclusion:}} 
 
Pole positions are unaffected. 

\subsubsection{Intermediate part}
As the intermediate part of the channel propagator, we define the energy domain which is notably above threshold, but not exceeding the high energy
 part above  10 GeV$^2$. In this  energy range, \mbox{1 GeV$^2 <$ $s$ $<$ 10 GeV$^2$},   most of the known nucleon resonances are situated.
   \\

In this part of our analysis channel propagator is multiplied by a function 
\be 
g(s)=\left[\frac{1+a \cdot s + s^2}{1+s^2}\right]^2
\label{gs1}
\ee 
which changes a shape of channel propagator without changing its asymptotic or threshold behavior. Depending on the size and the sign of the 
free parameter $a$, the value of the channel propagator in the intermediate range 2 GeV$^2<$ $s$ $<$ 5 GeV$^2$ can be enhanced, or reduced for as
 much as 100\%.  \\

\noindent
The imaginary part of channel propagator is given by
\be 
\mathrm{Im} \ \Phi_{corr}=\mathrm{Im}  \ \Phi \cdot g(s),
\ee 
and the real part is calculated using the dispersion relation (\ref{eq:DR}). \\

\begin{figure}[!t]
 \centering
 \vspace*{-6.cm}
\includegraphics[width=12cm]{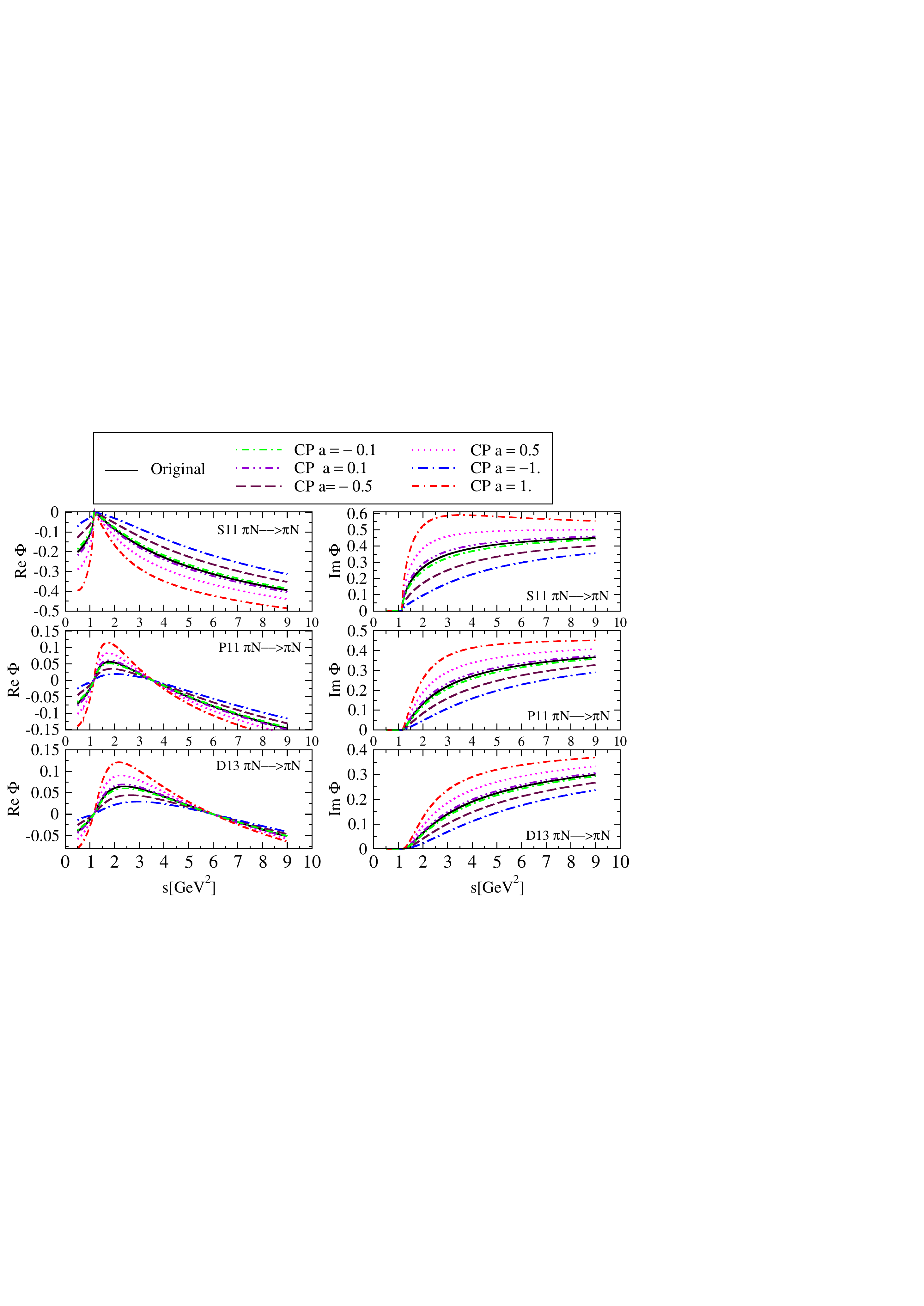}
\vspace*{-5.cm}
\caption[]{\footnotesize {(Color online) Real and imaginary parts of S$_{11}$, $P_{11}$ and $D_{13}$ elastic channel propagator. We change intermediate part  of channel  
propagator.}}
\label{Fig:innpartS}
\end{figure}

The change of the channel propagator for the S$_{11}$, P$_{11}$ and D$_{13}$ partial wave is given in  \vspace*{0.5cm} Fig. (\ref{Fig:innpartS}).

\noindent
\underline{\textit{Result:}}

\begin{figure}[!tb]
\centering
\vspace*{-6.cm}
{\includegraphics[width=12.cm]{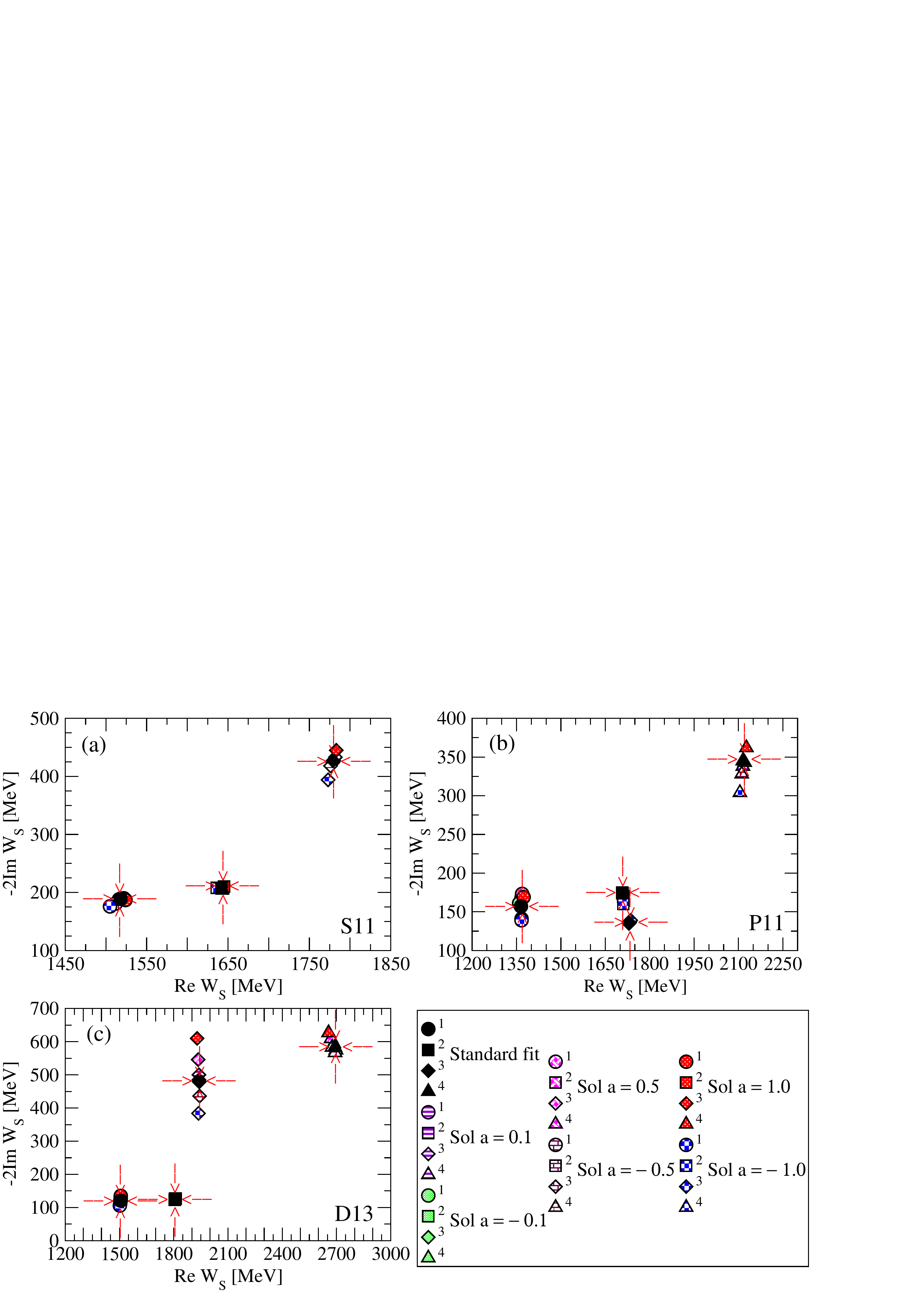}}
\vspace{-0.2cm}
\caption[]{\footnotesize (Color online) The extracted S$_{11}$, $P_{11}$ and $D_{13}$ partial waves T-matrix poles using different solutions of
 channel propagator. We change intermediate part  of channel propagator and the modification is the same for all three used channels.
  { Red arrows indicate the position of Standard solution poles.}}
\label{S11_inner_poles}
\end{figure}

The shift in pole positions for partial waves S$_{11}$, P$_{11}$ and D$_{13}$ are presented in Tables \ref{S11_inner}- \ref{D13_inner} and  depicted in 
Fig.(\ref{S11_inner_poles}). \\

\begin{table*}[!h]
\caption{\footnotesize The extracted S$_{11}$ partial wave T-matrix poles using different solutions of channel propagator. We change intermediate
part of channel propagator and the modification is the same for all three used channels}~
\begin{center}
\begin{tabular}{|l|ccc|ccc|c|}
\hline \hline 
\rule{0cm}{0.3cm} &  \multicolumn{3}{c|}{ Bare poles} & \multicolumn{3}{c|}{ Dressed poles } &  \\ \cline{2-7}
\rule{0cm}{0.4cm}Solutions  &   $\mathrm{W_{s_1}}$ & $\mathrm{W_{s_2}}$ &  $\mathrm{W_{s_3}}$   & {\scriptsize { $\begin{pmatrix}
\mathrm{Re W}\\ \mathrm{-2Im W} 
\end{pmatrix}$}} & {\scriptsize { $\begin{pmatrix}
\mathrm{Re W}\\ \mathrm{-2Im W} 
\end{pmatrix}$}} & {\scriptsize { $\begin{pmatrix}
\mathrm{Re W}\\ \mathrm{-2Im W} 
\end{pmatrix}$}}   & $\mathrm{\chi_R^2}$  \\ [1.1ex]
 &  & $\mathrm{MeV}$  &  &   &  $\mathrm{MeV}$&  &  \\ \hline \hline
\rule{0cm}{0.4cm}Sol $a=-1.0$ & 1522 & 1640 & 1837 &{\scriptsize { $\begin{pmatrix}
1505\\ 176
\end{pmatrix}$}} & {\scriptsize { $\begin{pmatrix}
1636\\ 208
\end{pmatrix}$}} &	{\scriptsize { $\begin{pmatrix}
1773\\  394
\end{pmatrix}$}} &  1.198 \\[1.1ex] 
Sol $a=-0.5$ & 1523 & 1641 & 1835 &{\scriptsize { $\begin{pmatrix}
1512\\ 181
\end{pmatrix}$}} & {\scriptsize { $\begin{pmatrix}
1640\\ 208
\end{pmatrix}$}} &	{\scriptsize { $\begin{pmatrix}
1776\\ 418
\end{pmatrix}$}} &  1.048  \\[1.2ex]
Sol $a=-0.1$ & 1523 & 1640 & 1834 &{\scriptsize { $\begin{pmatrix}
1516\\ 188
\end{pmatrix}$}} & {\scriptsize { $\begin{pmatrix}
1642\\  207
\end{pmatrix}$}} &	{\scriptsize { $\begin{pmatrix}
1779\\   424
\end{pmatrix}$}} &  1.031 \\ [1.1ex]  \hline

\rule{0cm}{0.4cm}Standard fit& 1523 & 1640 & 1834 &{\scriptsize { $\begin{pmatrix}
1518\\  189
\end{pmatrix}$}} & {\scriptsize { $\begin{pmatrix}
1642\\  208 
\end{pmatrix}$}} &	{\scriptsize { $\begin{pmatrix}
1779\\  427
\end{pmatrix}$}} &  1.016  \\ [1.1ex] \hline 

\rule{0cm}{0.4cm}Sol $a=0.1$ & 1523 & 1640 & 1834 &{\scriptsize { $\begin{pmatrix}
1518\\   189
\end{pmatrix}$}} & {\scriptsize { $\begin{pmatrix}
1642\\ 207
\end{pmatrix}$}} &	\centering {\scriptsize { $\begin{pmatrix}
1780\\  427
\end{pmatrix}$}} &  1.037 \\ [1.1ex]
Sol $a=0.5$ & 1523 & 1639 & 1833 &{\scriptsize { $\begin{pmatrix}
1522\\ 190
\end{pmatrix}$}} & {\scriptsize { $\begin{pmatrix}
1644\\ 207
\end{pmatrix}$}} &	{\scriptsize { $\begin{pmatrix}
1782\\ 433
\end{pmatrix}$}} &  1.062 \\ [1.1ex]
Sol $a=1.0$ & 1523 & 1638 & 1831 &{\scriptsize { $\begin{pmatrix}
1524\\ 187
\end{pmatrix}$}} & {\scriptsize { $\begin{pmatrix}
1645\\ 210
\end{pmatrix}$}} &	{\scriptsize { $\begin{pmatrix}
1783\\ 445
\end{pmatrix}$}} &  1.105  \\ [1.1ex]\hline \hline
\end{tabular}
\end{center}
\label{S11_inner}
\end{table*}
\begin{table*}[!h]
\caption{\footnotesize The extracted P$_{11}$ partial wave T-matrix poles using different solutions of channel propagator. We change 
intermediate part of channel propagator and the modification is the same for all three used channels}
\begin{center}
\begin{tabular}{|l|cccc|cccc|c|}
\hline \hline 
\rule{0cm}{0.3cm} &  \multicolumn{4}{c|}{ Bare poles} & \multicolumn{4}{c|}{ Dressed poles } & \\ \cline{2-9}
\rule{0cm}{0.4cm}Solutions  &   $\mathrm{W_{s_1}}$ & $\mathrm{W_{s_2}}$ &  $\mathrm{W_{s_3}}$   & $\mathrm{W_{s_4}}$ & {\scriptsize {
 $\begin{pmatrix}
\mathrm{Re W}\\ \mathrm{-2Im W} 
\end{pmatrix}$}} & {\scriptsize { $\begin{pmatrix}
\mathrm{Re W}\\ \mathrm{-2Im W} 
\end{pmatrix}$}} & {\scriptsize { $\begin{pmatrix}
\mathrm{Re W}\\ \mathrm{-2Im W} 
\end{pmatrix}$}}  & {\scriptsize { $\begin{pmatrix}
\mathrm{Re W}\\ \mathrm{-2Im W} 
\end{pmatrix}$}}  & $\mathrm{\chi_R^2}$  \\ [1.1ex]
  &  & \multicolumn{2}{c}{$\mathrm{MeV}$}  &  & & \multicolumn{2}{c}{$\mathrm{MeV}$}  & &  \\ \hline \hline
\rule{0cm}{0.4cm}Sol $a=-1.0$ & 1603 & 1772 & 2181 &2929 &{\scriptsize { $\begin{pmatrix}
1367\\ 139
\end{pmatrix}$}} & {\scriptsize { $\begin{pmatrix}
1711\\ 160
\end{pmatrix}$}} &	{\scriptsize { $\begin{pmatrix}
1736\\ 139
\end{pmatrix}$}} &{\scriptsize { $\begin{pmatrix}
2105\\ 304
\end{pmatrix}$}} & 1.137\\[1.1ex]
Sol $a=-0.5$ & 1606 & 1772 & 2182 &2874&{\scriptsize { $\begin{pmatrix}
1367\\ 141
\end{pmatrix}$}} & {\scriptsize { $\begin{pmatrix}
1710\\ 170
\end{pmatrix}$}} &	{\scriptsize { $\begin{pmatrix}
1733\\ 137
\end{pmatrix}$}}&	{\scriptsize { $\begin{pmatrix}
2111\\ 328
\end{pmatrix}$}} &  0.971  \\[1.1ex]
Sol $a=-0.1$ & 1606 & 1772 & 2182 &2849& {\scriptsize { $\begin{pmatrix}
1359\\ 161
\end{pmatrix}$}} & {\scriptsize { $\begin{pmatrix}
1708\\ 174
\end{pmatrix}$}} &	{\scriptsize { $\begin{pmatrix}
1731\\ 137
\end{pmatrix}$}} &	{\scriptsize { $\begin{pmatrix}
2115\\ 345
\end{pmatrix}$}}&  0.955 \\[1.1ex] \hline
\rule{0cm}{0.4cm}Standard fit& 1607 & 1772 & 2182 & 2841&{\scriptsize { $\begin{pmatrix}
1365\\ 157
\end{pmatrix}$}} & {\scriptsize { $\begin{pmatrix}
1708\\ 174 
\end{pmatrix}$}} &	{\scriptsize { $\begin{pmatrix}
1731\\ 136
\end{pmatrix}$}}& {\scriptsize { $\begin{pmatrix}
2117\\ 345
\end{pmatrix}$}}&  0.958  \\ [1.1ex] \hline 

\rule{0cm}{0.4cm}Sol $a=0.1$ & 1606 & 1772 & 2182 & 2796 &{\scriptsize { $\begin{pmatrix}
1367\\ 168
\end{pmatrix}$}} & {\scriptsize { $\begin{pmatrix}
1711\\ 168
\end{pmatrix}$}} &	\centering {\scriptsize { $\begin{pmatrix}
1732\\ 136
\end{pmatrix}$}} & \centering {\scriptsize { $\begin{pmatrix}
2115\\ 338
\end{pmatrix}$}}& 1.001 \\[1.1ex]
Sol $a=0.5$ & 1607 & 1772 & 2184 &2811&{\scriptsize { $\begin{pmatrix}
1369\\ 173
\end{pmatrix}$}} & {\scriptsize { $\begin{pmatrix}
1707\\ 174
\end{pmatrix}$}} &	{\scriptsize { $\begin{pmatrix}
1730\\ 135
\end{pmatrix}$}} &	{\scriptsize { $\begin{pmatrix}
2121\\ 343
\end{pmatrix}$}}&  0.973 \\[1.1ex]

Sol $a=1.0$ & 1607 & 1772 & 2186 &2784&{\scriptsize { $\begin{pmatrix}
1374\\ 169
\end{pmatrix}$}} & {\scriptsize { $\begin{pmatrix}
1707\\ 175
\end{pmatrix}$}} &	{\scriptsize { $\begin{pmatrix}
1730\\ 135
\end{pmatrix}$}} &	{\scriptsize { $\begin{pmatrix}
2127\\ 362
\end{pmatrix}$}}&  0.995  \\[1.1ex] \hline \hline
\end{tabular}
\end{center}
\label{P11_inner}
\end{table*}
\begin{table*}[!h]
\caption{\footnotesize The extracted D$_{13}$ partial wave T-matrix poles using different solutions of channel propagator. 
We change intermediate part of channel propagator and the modification is the same for all three used channels}
\begin{center}
\begin{tabular}{|l|ccc|cccc|c|}
\hline \hline 
\rule{0cm}{0.25cm} &  \multicolumn{3}{c|}{ Bare poles} & \multicolumn{4}{c|}{ Dressed poles } &  \\ \cline{2-8}
\rule{0cm}{0.4cm}Solutions  &   $\mathrm{W_{s_1}}$ & $\mathrm{W_{s_2}}$ &  $\mathrm{W_{s_3}}$   & {\scriptsize { $\begin{pmatrix}
\mathrm{Re W}\\ \mathrm{-2Im W} 
\end{pmatrix}$}} & {\scriptsize { $\begin{pmatrix}
\mathrm{Re W}\\ \mathrm{-2Im W} 
\end{pmatrix}$}} & {\scriptsize { $\begin{pmatrix}
\mathrm{Re W}\\ \mathrm{-2Im W} 
\end{pmatrix}$}}   & {\scriptsize { $\begin{pmatrix}
\mathrm{Re W}\\ \mathrm{-2Im W} 
\end{pmatrix}$}}   &{$\chi_R^2$}  \\ 
 &&  $\mathrm{MeV}$&  & & \multicolumn{2}{c}{$\mathrm{MeV}$}  & & \\ \hline \hline 
\rule{0cm}{0.4cm}Sol $a=-1.0$ & 1578 & 1881 & 2503 &{\scriptsize { $\begin{pmatrix}
1502\\ 106
\end{pmatrix}$}} & {\scriptsize { $\begin{pmatrix}
1806\\ 123
\end{pmatrix}$}} &	{\scriptsize { $\begin{pmatrix}
1937\\ 384
\end{pmatrix}$}} &  {\scriptsize { $\begin{pmatrix}
2675\\ 583
\end{pmatrix}$}} &1.396\\ [1.1ex]
Sol $a=-0.5$ & 1579 & 1881 & 2500 &{\scriptsize { $\begin{pmatrix}
1504\\ 113
\end{pmatrix}$}} & {\scriptsize { $\begin{pmatrix}
1807\\ 125
\end{pmatrix}$}} &	{\scriptsize { $\begin{pmatrix}
1943\\ 436
\end{pmatrix}$}} &  {\scriptsize { $\begin{pmatrix}
2692\\ 567
\end{pmatrix}$}} &1.087  \\[1.1ex]
Sol $a=-0.1$ & 1581 & 1880 & 2498 &{\scriptsize { $\begin{pmatrix}
1505\\ 120
\end{pmatrix}$}} & {\scriptsize { $\begin{pmatrix}
1808\\ 127
\end{pmatrix}$}} &	{\scriptsize { $\begin{pmatrix}
1941\\ 479
\end{pmatrix}$}} &{\scriptsize { $\begin{pmatrix}
2697\\ 575
\end{pmatrix}$}}  &1.027 \\[1.1ex] \hline
\rule{0cm}{0.4cm}Standard fit & 1582 & 1880 & 2499 &{\scriptsize { $\begin{pmatrix}
1506\\ 121
\end{pmatrix}$}} & {\scriptsize { $\begin{pmatrix}
1807\\ 127 
\end{pmatrix}$}} &	{\scriptsize { $\begin{pmatrix}
1939\\ 485
\end{pmatrix}$}} & {\scriptsize { $\begin{pmatrix}
2691\\ 583
\end{pmatrix}$}}& 1.027  \\[1.1ex]  \hline 

\rule{0cm}{0.4cm}Sol $a=0.1$ & 1582 & 1879 & 2497 &{\scriptsize { $\begin{pmatrix}
1505\\ 121
\end{pmatrix}$}} & {\scriptsize { $\begin{pmatrix}
1808\\ 127
\end{pmatrix}$}} &	{\scriptsize { $\begin{pmatrix}
1939\\ 500
\end{pmatrix}$}} &  {\scriptsize { $\begin{pmatrix}
2685\\ 594
\end{pmatrix}$}} &1.032 \\[1.1ex]

Sol $a=0.5$& 1584 & 1878 & 2495 &{\scriptsize { $\begin{pmatrix}
1506\\ 127
\end{pmatrix}$}} & {\scriptsize { $\begin{pmatrix}
1808\\ 127
\end{pmatrix}$}} &	{\scriptsize { $\begin{pmatrix}
1935\\ 546
\end{pmatrix}$}} & {\scriptsize { $\begin{pmatrix}
2670\\ 609
\end{pmatrix}$}} & 1.084 \\[1.1ex]

Sol $a=1.0$& 1585 & 1876 & 2493 &{\scriptsize { $\begin{pmatrix}
1507\\ 135
\end{pmatrix}$}} & {\scriptsize { $\begin{pmatrix}
1811\\ 127
\end{pmatrix}$}} &	{\scriptsize { $\begin{pmatrix}
1929\\ 610
\end{pmatrix}$}} &  {\scriptsize { $\begin{pmatrix}
2656\\ 627
\end{pmatrix}$}} & 1.183   \\[1.1ex] \hline \hline
\end{tabular}
\label{D13_inner}
\end{center}
\end{table*}

\noindent
\underline{\textit{Conclusions:}} \\

As seen in Fig.~(\ref{Fig:innpartS}) changing the $a$ parameter in ad hoc modification function given by Eq.~(\ref{gs1}) within the range -1 $<$ $a$ $<$ 1 
changes the imaginary part of the channel propagator for up to 100\% in the intermediate range 2  $GeV^2$ $<$ $s$ $<$ 5 GeV$^2$. \\

\vbox{
We observe:
\begin{enumerate}
   \item The strong modification of the imaginary part of the channel propagator is via the dispersion relation  Eq.~(\ref{eq:DR})
    transmitted to the real part, causing opposite effect (reduction) for the S-wave, and a combined effect for other partial waves.
   \item From Tables \ref{S11_inner}- \ref{D13_inner} we learn that bare poles are fairly stable with respect with change of the intermediate part of the channel propagator, while 
dressed poles experience quite a change. 
    \item The dressed poles can, however, be classified in two categories:
	        \begin{description}
			     \item[a.] Poles weekly dependent upon the strong changes of the intermediate part - stable poles \\
				          S$_{11}$ (1518), S$_{11}$ (1642), P$_{11}$ (1731), D$_{13}$ (1506) and D$_{13}$ (1807)
	             \item[b.] Poles moderately and notably dependent upon the strong changes of the intermediate part - running poles \\
				          S$_{11}$ (1779), $P_{11}(1365)$, { P$_{11}$ (1731}), P$_{11}$ (2117), D$_{13}$ (1939) and D$_{13}$ (2691) 
			\end{description}
	\item We may connect the category of the pole to its self energy content:
	      \begin{description}
			     \item[a.] For the stable poles  the self energy correction to the bare mass is small 
	             \item[b.] For the running poles, the self energy correction to the bare mass is either strong 
				 (as for  S$_{11}$ (1779), { P$_{11}$ (1731)}, P$_{11}$ (2117) and D$_{13}$ (2691) ),  or poles are entirely dynamically
				  generated meaning that they do not at all have a corresponding, nearby bare poles (as for  { $P_{11}(1365)$} Roper resonance,
				   and D$_{13}$ (1939) )
			\end{description}
			\item Consequently, ``difference"  of bare and dressed pole position can be taken as the criterion for the stability
			 of the pole position with respect to the strong modification of the intermediate part: ``The lower the self energy correction is, the more stable the pole position 
is." 
            \item As seen in Tables \ref{S11_inner} - \ref{D13_inner}, the reduced $\chi^2_R$, the quantity representing the measure of
			 ability of our model to reproduce the input data, is systematically smallest for the non-modified solution.  {In other words, our fit to the input data worsens 
whenever any change to the intermediate part of channel propagator is introduced.}  
\end{enumerate}
}
\subsubsection{Threshold behavior}
\begin{figure}[!htb]
\centering
\vspace*{-6.cm}
\includegraphics[width=12cm]{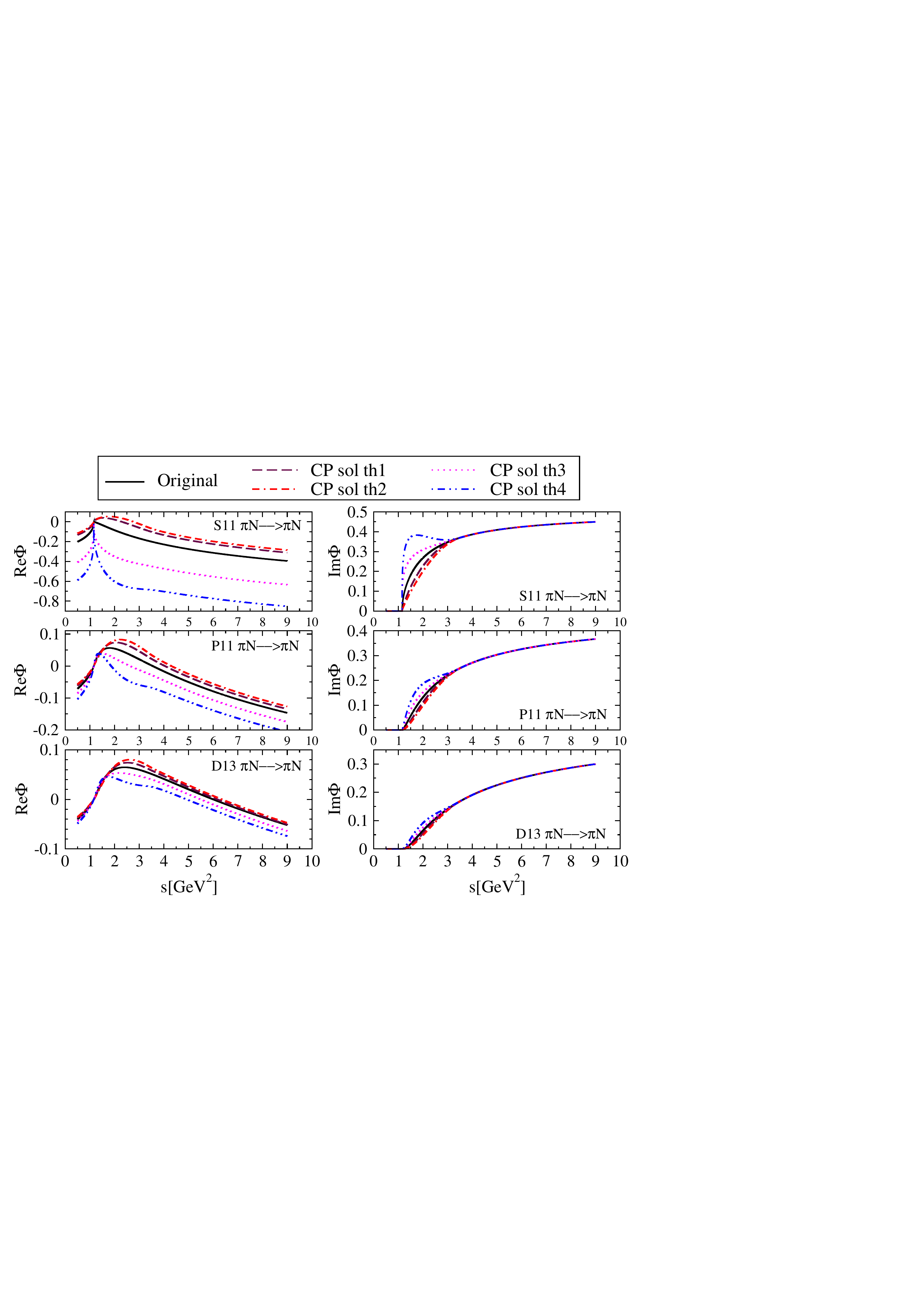}
\vspace*{-5.cm}
\caption[]
{\footnotesize (Color online) Real and imaginary parts of the $S_{11}$, $P_{11}$ and $D_{13}$ elastic channel propagator. Threshold behavior. Sol th1:
 $\mathrm{Im}\Phi_{corr}=\mathrm{Im}\Phi\cdot\sqrt{q}(a+cs^2)$, Sol th2: $\mathrm{Im}\Phi_{corr}=\mathrm{Im}\Phi\cdot \sqrt{q}(a+0.5s+cs^2)$, Sol th3: 
 $\mathrm{Im}\Phi_{corr}=\mathrm{Im}\Phi\cdot(\frac{1}{\sqrt{q}})(a+cs^2)$, Sol th4: $\mathrm{Im}\Phi_{corr}=\mathrm{Im}\Phi\cdot(\frac{1}{\sqrt{q}})(a-0.5s+cs^2)$.}
\label{Fig:thpartS} 
\end{figure}

In this part of our analysis, behavior of channel propagator's imaginary part close to the threshold a given channel is changed. For values 
$s$ $>$ 3.5 GeV$^2$ channel propagator is kept unchanged. The threshold behavior of imaginary part of the channel propagator is given by 
$q^{2L+1}$. It is  changed by multiplying the channel propagator with  $q^{\lambda}$. The continuity in energy is ensured by the matching function $h(s)=(a+b\cdot s+c\cdot s^2)$. 
{ Modified} imaginary part is given by:
\be 
\mathrm{Im} \Phi_{corr}=\mathrm{Im} \Phi \cdot q_a^{\lambda}\cdot h(s).
\label{eq:thr}
\ee 
with $\lambda=\pm 0.5$. 
Parameters $a$, $b$ and $c$ are adjusted in such a way to ensure continuity of channel propagator and first derivative at $s$ =3.5 GeV$^2$ .\\
The change of the channel propagator for the S$_{11}$, P$_{11}$ and D$_{13}$ partial wave is given  in Fig.  (\ref{Fig:thpartS}).\\

\noindent
\underline{\textit{Result:}} \\

\begin{figure}[!htb]
\vspace{-6cm}
\centering
{\includegraphics[width=12.8cm]{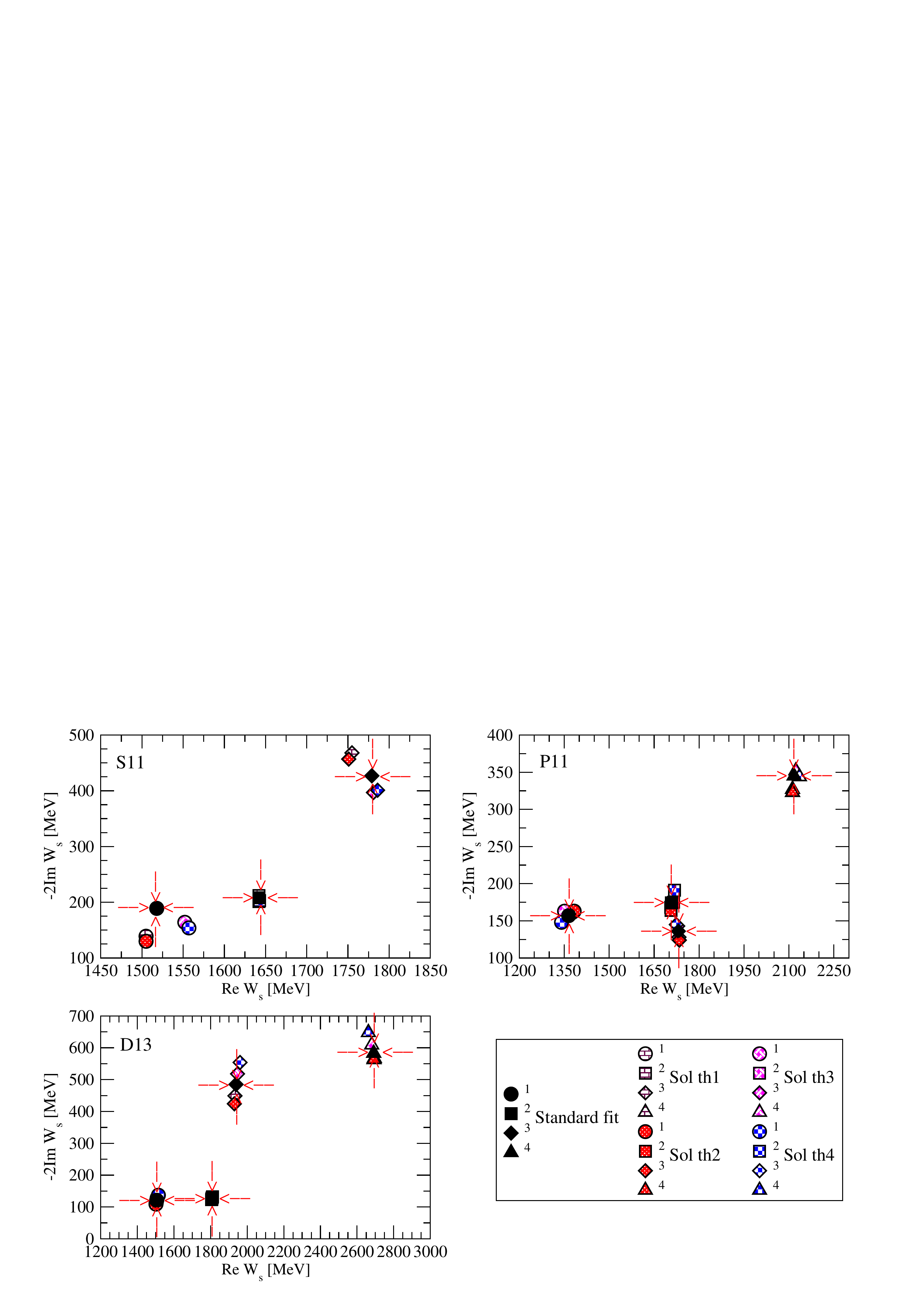}}
\caption[]{\footnotesize (Color online) The extracted S$_{11}$, $P_{11}$ and D$_{13}$ partial waves T-matrix poles using different solutions
 of channel propagator. We change threshold behavior  of channel propagator and the modification is the same for all three used channels.
  { Red arrows indicate the position of Standard solution poles.}}
\label{S11_th_poles}
\end{figure}

The shift in pole positions for partial waves S$_{11}$, P$_{11}$ and D$_{13}$ are presented in Tables \ref{S11_th_table}- \ref{D13_th_table} 
and depicted in Fig. (\ref{S11_th_poles}).~ 
{ We personally do not believe that such strong variations in the threshold behavior are physics-wise permissible, but we have investigated this aspect of our model just to check 
the how quickly our results deteriorate when even uphysical changes are introduced. We are happy to conclude that the results are only moderately sensitive to even huge changes.  
}
\begin{table*}[!hbt]
\vspace{-0.4cm}
\caption{\footnotesize  The extracted $S_{11}$ partial wave T-matrix poles using different solutions of channel propagator. We change
 threshold behavior of channel propagator and the modification is the same for all three used channels. }
\begin{center}
\begin{tabular}{|l|ccc|ccc|c|}
\hline \hline 
\rule{0cm}{0.3cm} &  \multicolumn{3}{c|}{ Bare poles} & \multicolumn{3}{c|}{ Dressed poles } & \\  \cline{2-7}
\rule{0cm}{0.4cm} Solutions &   $\mathrm{W_{s_1}}$ & $\mathrm{W_{s_2}}$ &  $\mathrm{W_{s_3}}$   & {\scriptsize { $\begin{pmatrix}
\mathrm{Re W}\\ \mathrm{-2Im W} 
\end{pmatrix}$}} & {\scriptsize { $\begin{pmatrix}
\mathrm{Re W}\\ \mathrm{-2Im W} 
\end{pmatrix}$}} & {\scriptsize { $\begin{pmatrix}
\mathrm{Re W}\\ \mathrm{-2Im W} 
\end{pmatrix}$}}   & $\mathrm{\chi_R^2}$  \\ 
 &  & $\mathrm{MeV}$  &  &   & $\mathrm{MeV}$ &  &  \\ \hline \hline 
\rule{0cm}{0.4cm}Standard fit& 1523 & 1640 & 1834 &{\scriptsize { $\begin{pmatrix}
1518\\ 189
\end{pmatrix}$}} & {\scriptsize { $\begin{pmatrix}
1642\\ 208
\end{pmatrix}$}} &	{\scriptsize { $\begin{pmatrix}
1779\\ 427
\end{pmatrix}$}} &  1.016  \\[1.1ex] \hline 

\rule{0cm}{0.4cm}Sol th1 & 1551 & 1673 & 1899 &{\scriptsize { $\begin{pmatrix}
1505\\ 139
\end{pmatrix}$}} & {\scriptsize { $\begin{pmatrix}
1642\\ 212
\end{pmatrix}$}} &	{\scriptsize { $\begin{pmatrix}
1755\\ 468
\end{pmatrix}$}} &  1.289 \\ [1.1ex]

Sol th2 & 1557 & 1683 & 1915 &{\scriptsize { $\begin{pmatrix}
1505\\ 130
\end{pmatrix}$}} & {\scriptsize { $\begin{pmatrix}
1642\\ 208
\end{pmatrix}$}} &	{\scriptsize { $\begin{pmatrix}
1751\\ 457
\end{pmatrix}$}} &  1.372 \\[1.1ex]

Sol th3 & 1477 & 1575 & 1728 &{\scriptsize { $\begin{pmatrix}
1552\\ 164
\end{pmatrix}$}} & {\scriptsize { $\begin{pmatrix}
1642\\ 212
\end{pmatrix}$}} &	{\scriptsize { $\begin{pmatrix}
1781\\ 397
\end{pmatrix}$}} &  1.409 \\[1.1ex]

Sol th4 & 1453 & 1540& 1678 &{\scriptsize { $\begin{pmatrix}
1557\\ 154
\end{pmatrix}$}} & {\scriptsize { $\begin{pmatrix}
1643\\ 203
\end{pmatrix}$}} &	{\scriptsize { $\begin{pmatrix}
1786\\ 401
\end{pmatrix}$}} &  1.587 \\ [1.1ex]\hline \hline 
\end{tabular}\label{S11_th_table}
\end{center}
\end{table*}
\begin{table*}[!ht]
\vspace{-0.1cm}
\caption{\footnotesize The extracted $P_{11}$ partial wave T-matrix poles using different solutions of channel propagator.
 We change threshold behavior of channel propagator and the modification is the same for all three used channels}
\begin{center}
\begin{tabular}{|l|cccc|cccc|c|}
\hline \hline 
\rule{0cm}{0.3cm} &  \multicolumn{4}{c|}{ Bare poles} & \multicolumn{4}{c|}{ Dressed poles } &  \\ \cline{2-9}
\rule{0cm}{0.4cm}Solutions  &   $\mathrm{W_{s_1}}$ & $\mathrm{W_{s_2}}$ &  $\mathrm{W_{s_3}}$   & $\mathrm{W_{s_4}}$ & {\scriptsize {
 $\begin{pmatrix}
\mathrm{Re W}\\ \mathrm{-2Im W} 
\end{pmatrix}$}} & {\scriptsize { $\begin{pmatrix}
\mathrm{Re W}\\ \mathrm{-2Im W} 
\end{pmatrix}$}} & {\scriptsize { $\begin{pmatrix}
\mathrm{Re W}\\ \mathrm{-2Im W} 
\end{pmatrix}$}}&{\scriptsize { $\begin{pmatrix}
\mathrm{Re W}\\ \mathrm{-2Im W} 
\end{pmatrix}$}}& $\mathrm{\chi_R^2}$  \\ 
 &   & \multicolumn{2}{c}{$\mathrm{MeV}$}  &  & & \multicolumn{2}{c}{$\mathrm{MeV}$}  & &  \\ \hline \hline 
\rule{0cm}{0.4cm}Standard fit& 1607 & 1772 & 2182 &2841&{\scriptsize { $\begin{pmatrix}
1365\\ 157
\end{pmatrix}$}} & {\scriptsize { $\begin{pmatrix}
1708\\ 174 
\end{pmatrix}$}} &	{\scriptsize { $\begin{pmatrix}
1731\\ 136
\end{pmatrix}$}} &	{\scriptsize { $\begin{pmatrix}
2117\\ 345
\end{pmatrix}$}}&  0.957 \\[1.2ex] \hline 

\rule{0cm}{0.4cm}Sol th1 & 1629 & 1780 & 2192 &2802&{\scriptsize { $\begin{pmatrix}
1376\\ 161
\end{pmatrix}$}} & {\scriptsize { $\begin{pmatrix}
1705\\ 168
\end{pmatrix}$}} &	{\scriptsize { $\begin{pmatrix}
1733\\ 128
\end{pmatrix}$}} &	{\scriptsize { $\begin{pmatrix}
2112\\ 327
\end{pmatrix}$}}&  1.157 \\ [1.2ex]

Sol th2 & 1639 & 1784 & 2196 &2801&{\scriptsize { $\begin{pmatrix}
1383\\ 163
\end{pmatrix}$}} & {\scriptsize { $\begin{pmatrix}
1706\\ 164
\end{pmatrix}$}} &	{\scriptsize { $\begin{pmatrix}
1734\\ 124
\end{pmatrix}$}} &	{\scriptsize { $\begin{pmatrix}
2112\\ 323
\end{pmatrix}$}}&  1.335 \\[1.2ex]

Sol th3 & 1572 & 1763 & 2174 &2712 &{\scriptsize { $\begin{pmatrix}
1351\\ 163
\end{pmatrix}$}} & {\scriptsize { $\begin{pmatrix}
1715\\ 175
\end{pmatrix}$}} &	{\scriptsize { $\begin{pmatrix}
1729\\ 143
\end{pmatrix}$}} &	{\scriptsize { $\begin{pmatrix}
2123\\ 352
\end{pmatrix}$}}&  1.010 \\[1.2ex]

Sol th4 & 1534 & 1753 & 2168 &2546&{\scriptsize { $\begin{pmatrix}
1341\\ 148
\end{pmatrix}$}} & {\scriptsize { $\begin{pmatrix}
1718\\ 191
\end{pmatrix}$}} &	{\scriptsize { $\begin{pmatrix}
1724\\ 145
\end{pmatrix}$}} &	{\scriptsize { $\begin{pmatrix}
2135\\ 344
\end{pmatrix}$}}&  1.274 \\ [1.2ex]\hline \hline 
\end{tabular}
\label{P11_th_table}
\end{center}
\end{table*}
\begin{table*}[!htb]
\vspace{-0.4cm}
\caption{\footnotesize The extracted $D_{13}$ partial wave T-matrix poles using different solution
 of channel propagator. We change threshold behavior of channel propagator and the modification is the same for all three used channels}
\begin{center}
\begin{tabular}{|l|ccc|cccc|c|}
\hline \hline 
\rule{0cm}{0.3cm} &  \multicolumn{3}{c|}{ Bare poles} & \multicolumn{4}{c|}{ Dressed poles } &  \\ \cline{2-8}
\rule{0cm}{0.4cm}Solutions  &   $\mathrm{W_{s_1}}$ & $\mathrm{W_{s_2}}$ &  $\mathrm{W_{s_3}}$   & {\scriptsize { $\begin{pmatrix}
\mathrm{Re W}\\ \mathrm{-2Im W} 
\end{pmatrix}$}} & {\scriptsize { $\begin{pmatrix}
\mathrm{Re W}\\ \mathrm{-2Im W} 
\end{pmatrix}$}} & {\scriptsize { $\begin{pmatrix}
\mathrm{Re W}\\ \mathrm{-2Im W} 
\end{pmatrix}$}}   &{\scriptsize { $\begin{pmatrix}
\mathrm{Re W}\\ \mathrm{-2Im W} 
\end{pmatrix}$}}& $\chi_R^2$  \\ 
  &&  $\mathrm{MeV}$&  & & \multicolumn{2}{c}{$\mathrm{MeV}$}  & & \\ \hline \hline 
\rule{0cm}{0.4cm}Standard fit& 1582 & 1880 & 2499 &{\scriptsize { $\begin{pmatrix}
1506\\ 121
\end{pmatrix}$}} & {\scriptsize { $\begin{pmatrix}
1807\\ 127 
\end{pmatrix}$}} &	{\scriptsize { $\begin{pmatrix}
1939\\ 485
\end{pmatrix}$}} & {\scriptsize { $\begin{pmatrix}
2691\\ 583
\end{pmatrix}$}} & 1.027  \\ [1.2ex]\hline 

Sol th1 & 1589 & 1881 & 2500 &{\scriptsize { $\begin{pmatrix}
1503\\ 114
\end{pmatrix}$}} & {\scriptsize { $\begin{pmatrix}
1807\\ 124
\end{pmatrix}$}} &	{\scriptsize { $\begin{pmatrix}
1933\\ 449
\end{pmatrix}$}} & {\scriptsize { $\begin{pmatrix}
2694\\ 567
\end{pmatrix}$}} & 1.057 \\ [1.2ex]

Sol th2 & 1593 & 1882 & 2505 &{\scriptsize { $\begin{pmatrix}
1502\\ 109
\end{pmatrix}$}} & {\scriptsize { $\begin{pmatrix}
1806\\ 123
\end{pmatrix}$}} &	{\scriptsize { $\begin{pmatrix}
1930\\ 424
\end{pmatrix}$}} & {\scriptsize { $\begin{pmatrix}
2692\\ 563
\end{pmatrix}$}} &1.148 \\[1.2ex]

Sol th3 & 1573 & 1875 & 2481 &{\scriptsize { $\begin{pmatrix}
1508\\ 128
\end{pmatrix}$}} & {\scriptsize { $\begin{pmatrix}
1808\\ 129
\end{pmatrix}$}} &	{\scriptsize { $\begin{pmatrix}
1947\\ 518
\end{pmatrix}$}} & {\scriptsize { $\begin{pmatrix}
2680\\ 609
\end{pmatrix}$}} & 1.119 \\[1.2ex]

Sol th4 & 1563 & 1871 & 2468 &{\scriptsize { $\begin{pmatrix}
1515\\ 137
\end{pmatrix}$}} & {\scriptsize { $\begin{pmatrix}
1810\\ 132
\end{pmatrix}$}} &	{\scriptsize { $\begin{pmatrix}
1961\\ 554
\end{pmatrix}$}} &	{\scriptsize { $\begin{pmatrix}
2662\\ 648
\end{pmatrix}$}} &  1.363 \\[1.2ex] \hline \hline 
\end{tabular}\label{D13_th_table}
\end{center}
\end{table*}

\noindent
{\vbox{
\underline{\textit{Conclusion:}} \\

As seen in Fig.~(\ref{Fig:thpartS}) the threshold behavior was modified by the continuous and arbitrary function given by Eq.~(\ref{eq:thr}). 
We have chosen 4 possibilities, in two of them the threshold behavior has been made shallower (th1 and th2), and for the two of them 
(th3 and th4) much steeper. \\

We observe:
\begin{enumerate}
   \item The strong modification of the imaginary part threshold behavior of the channel propagator is via the dispersion relation  Eq.~(\ref{eq:DR})
    transmitted to the real part, causing significant changes of the real part {\textit{in complete energy range}}.
   \item Contrary to the former case when intermediate part of the channel propagator has been analyzed, bare poles now experience significant change
    for all energies, and all partial waves. 
   \item Dressing tries to compensate for the strong shift of the bare poles, but not always very successfully. 
   \item As a result, we conclude that the correct threshold behavior of the channel propagator {\textit{is essential}}
    for the applicability of the model. 
   \item With some possible coincidental exceptions, all dressed poles experience a notable shift.  
   \item As seen in Tables \ref{S11_th_table} - \ref{D13_th_table}, the reduced $\chi^2_R$, the quantity representing the measure of ability 
   of our model to reproduce the input data, is systematically smallest for the non-modified solution.   { In other words, our fit to the input data worsen whenever any change to 
the threshold behavior of channel propagator is introduced.}
\end{enumerate}
}

{ \subsection{Dispersion integral subtraction constant}
As positions of bare poles are in principle not too well defined (they must depend on the renormalization scheme, see Capstick et.al \cite{CapstickEPJA}), the fact that they can 
only be significantly moved by changing the threshold behavior troubled us a lot. As our model is an effective model, changing its ingredients should enable us to "mimic" at least 
some of other models, and that means {  we should observe a} significant shift of  bare poles. So, having the bare pole model dependence hidden only in fairly well defined 
threshold behavior was non-satisfactory. \\

The answer  to this issue was  already hinted by Vrana et al in \cite{Vrana2000} where they have shown that position of bare  poles significantly depends on the dispersion 
integral subtraction constant, and its importance for the position of the bare {  and dressed} poles was briefly discussed. 
We have, therefore,  extensively tested this freedom within our model, and shifted the subtraction constant $\mathrm{Re}\Phi(s_0)$ defined in Eq. (\ref{eq:DR}) from its {  
original} zero value. We have chosen SC Sol 1: $\mathrm{Re}\Phi(s_0)=m_{\pi}$ for all three used  channels; SC Sol 2: $\mathrm{Re}\Phi(s_0)=m_{\pi}$ for $\pi N$ channel only, 
while in other two channels $\mathrm{Re}\Phi(s_0)=0$; SC Sol 3: $\mathrm{Re}\Phi(s_0)=m_{\eta}$ for $\eta N$ channel only, while in other two channels $\mathrm{Re}\Phi(s_0)=0$; SC 
Sol 4: $\mathrm{Re}\Phi(s_0)=m_{\pi^2}$ for $\pi^2 N$ channel only, while in other two channels $\mathrm{Re}\Phi(s_0)=0$; SC Sol 5: $\mathrm{Re}\Phi(s_0)=-m_{\pi}$ for all three 
used channels, SC Sol 6: $\mathrm{Re}\Phi(s_0)=m_a$, where $m_a$ is the mass of the meson in corresponding  channel. \\

As changing the dispersion integral subtraction constant only shifts the real part of the  channel propagator into positive or negative y-direction for a certain constant value, 
we show the modified real part of the S$_{11}$, P$_{11}$ and D$_{13}$ partial wave channel propagator in  Fig. (\ref{Fig:dispersion constant}) for only two representative values 
of subtraction constant $\mathrm{Re}\Phi(s_0)= \pm m_{\pi}$ in all three channels simultaneously. {  For other modifications related plots look qualitatively the same, so we do 
not show them explicitly.}\\

\noindent
\emph{\underline{Results:}} \\  

Results are given in Tables \ref{S11_SC_table} -\ref{D13_SC_table} and in Fig. (\ref{Fig:dcp}). \\

\begin{figure}[!ht]
 \centering
 \vspace*{-6.cm}
\includegraphics[width=12cm]{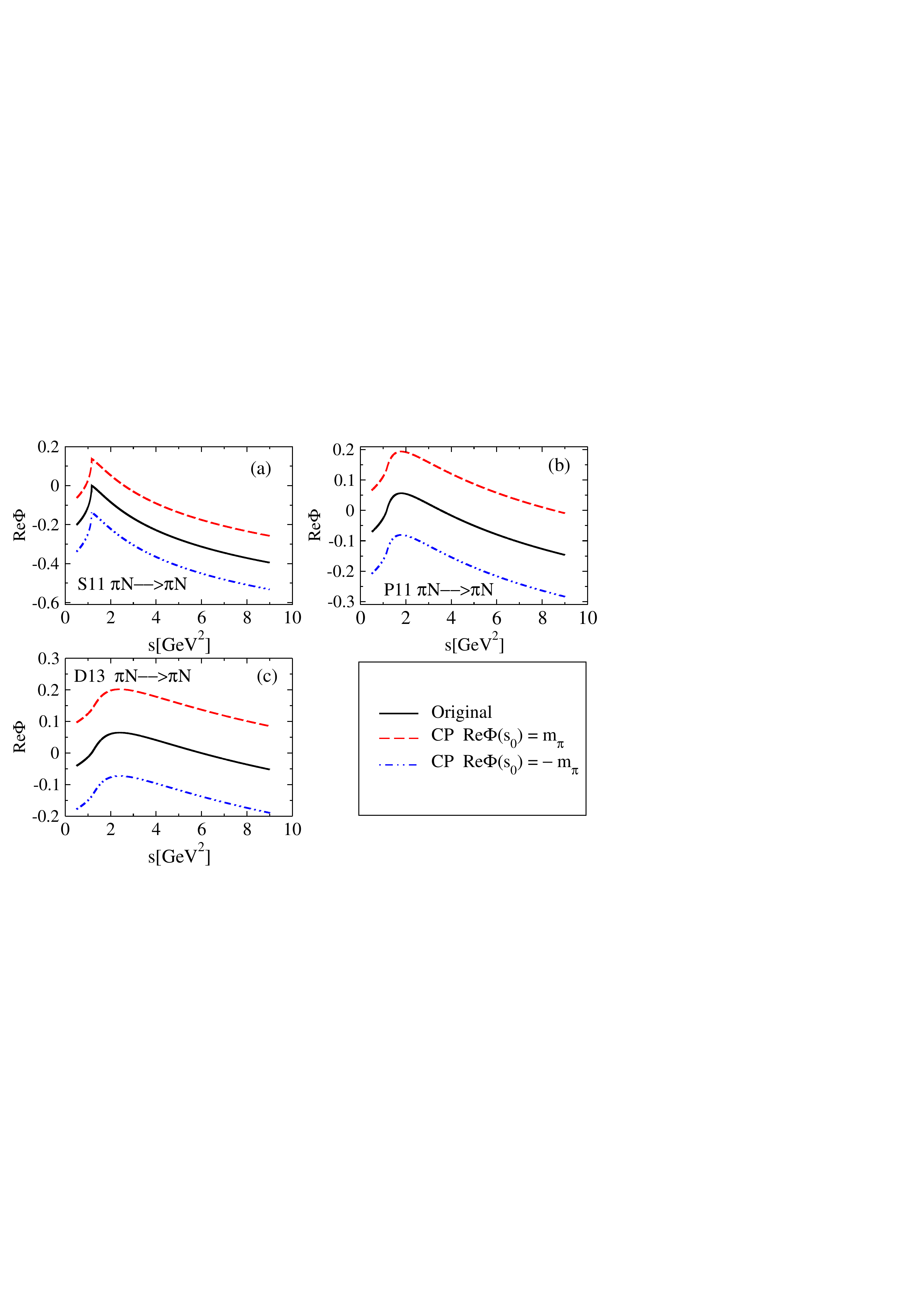}
\vspace*{-6.cm}
\caption[]{\footnotesize {(Color online) Real part of S$_{11}$, $P_{11}$ and $D_{13}$ elastic channel propagator. We change the dispersion integral subtraction constant.}}
\label{Fig:dispersion constant}
\end{figure}

\begin{figure}[!htb]
\vspace*{-7.cm}
\centering
{\includegraphics[width=12.8cm]{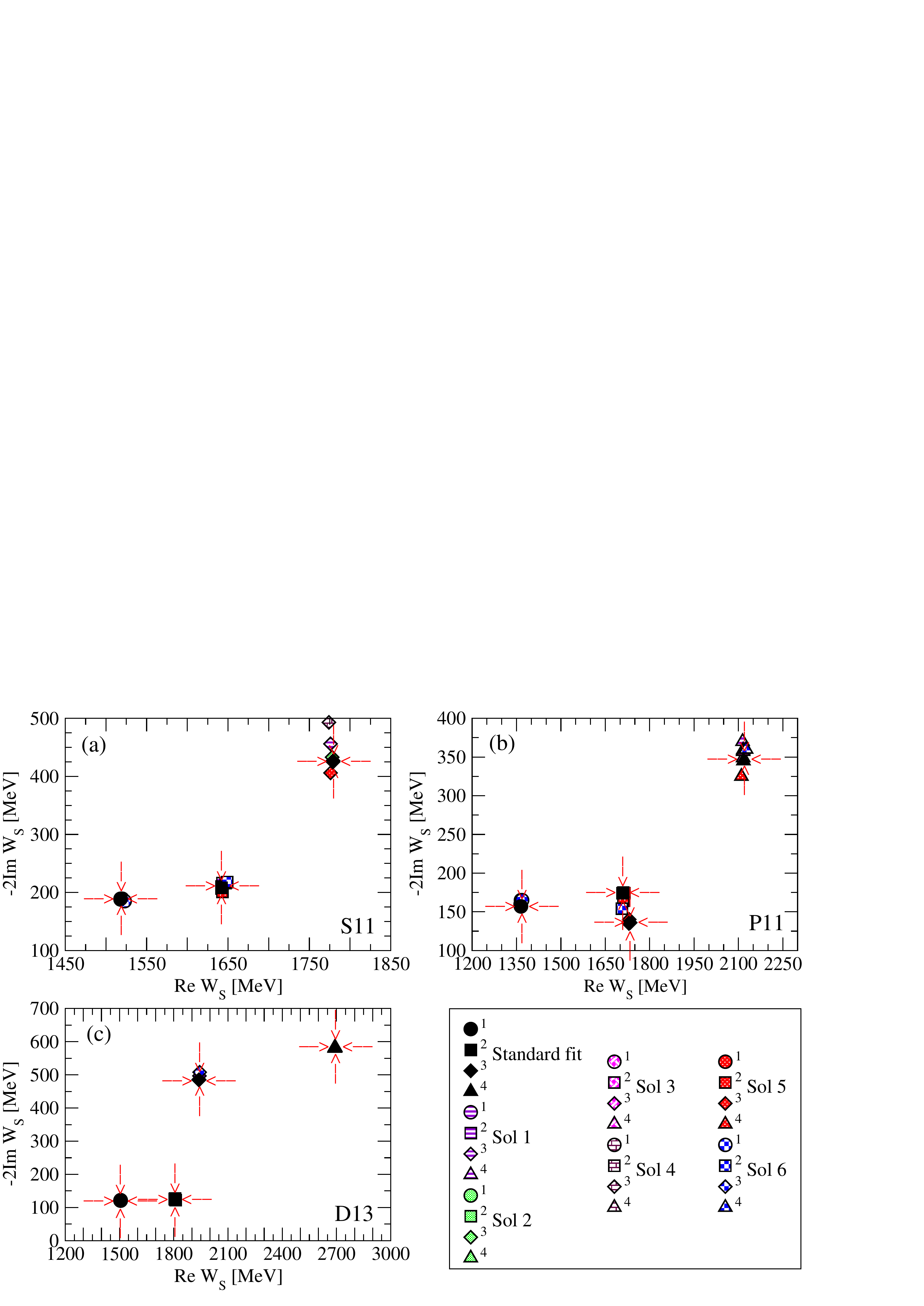}}
\caption[]{\footnotesize (Color online) The extracted S$_{11}$, $P_{11}$ and D$_{13}$ partial waves T-matrix poles using different solutions of channel propagator. We change 
dispersion integral subtraction constant for all three used channels. Red arrows indicate the position of Standard solution poles.}
\label{Fig:dcp}
\end{figure}

\begin{table}[!h] 
\caption{\footnotesize { The extracted $S_{11}$ partial wave $T$-matrix poles using different values of subtraction constant. SC Sol 1: $\mathrm{Re}\Phi(s_0)=m_{\pi}$ for all 
three used  channels; SC Sol 2: $\mathrm{Re}\Phi(s_0)=m_{\pi}$ for $\pi N$ channel only, while in other two channels $\mathrm{Re}\Phi(s_0)=0$; SC Sol 3: 
$\mathrm{Re}\Phi(s_0)=m_{\eta}$ for $\eta N$ channel only, while in other two channels $\mathrm{Re}\Phi(s_0)=0$; SC Sol 4: $\mathrm{Re}\Phi(s_0)=m_{\pi^2}$ for $\pi^2 N$ channel 
only, while in other two channels $\mathrm{Re}\Phi(s_0)=0$; SC Sol 5: $\mathrm{Re}\Phi(s_0)=-m_{\pi}$ for all three used channels, SC Sol 6: $\mathrm{Re}\Phi(s_0)=m_a$, where 
$m_a$ is the mass of the meson in corresponding  channel.}}
\begin{center}
\begin{tabular}{|l|ccc|ccc|c|}
\hline \hline 
\rule{0cm}{0.3cm}
 &  \multicolumn{3}{c|}{ Bare poles} & \multicolumn{3}{c|}{ Dressed poles } &\\ [1.1ex] \cline{2-7}
\rule{0cm}{0.4cm}&   $W_{s_1}$ & $W_{s_2}$ &  $W_{s_3}$   & {\scriptsize { $\begin{pmatrix}
\mathrm{Re W}\\ \mathrm{-2Im W} 
\end{pmatrix}$}} & {\scriptsize { $\begin{pmatrix}
\mathrm{Re W}\\ \mathrm{-2Im W} 
\end{pmatrix}$}} & {\scriptsize { $\begin{pmatrix}
\mathrm{Re W}\\ \mathrm{-2Im W} 
\end{pmatrix}$}}   & $\mathrm{\chi_R^2}$  \\ [1.1ex]
\rule{0cm}{0.4cm} &  & $MeV$  &  &   & $MeV$ &  &  \\ [1.1ex]\hline \hline 
\rule{0cm}{0.4cm}Standard fit& 1523 & 1640 & 1834 &{\scriptsize { $\begin{pmatrix}
1518\\ 189
\end{pmatrix}$}} & {\scriptsize { $\begin{pmatrix}
1642\\ 208
\end{pmatrix}$}} &	{\scriptsize { $\begin{pmatrix}
1779\\ 427
\end{pmatrix}$}} &  1.016  \\[1.1ex] \hline 
\rule{0cm}{0.4cm}SC Sol 1& 1579 & 1686 & 1953 &{\scriptsize { $\begin{pmatrix}
1520\\ 188
\end{pmatrix}$}} & {\scriptsize { $\begin{pmatrix}
1643\\ 216
\end{pmatrix}$}} &	{\scriptsize { $\begin{pmatrix}
1776\\ 456
\end{pmatrix}$}} &  1.054 \\ [1.1ex]\hline
\rule{0cm}{0.4cm}SC Sol 2& 1537 & 1662 & 1868 &{\scriptsize { $\begin{pmatrix}
1520\\ 189
\end{pmatrix}$}} & {\scriptsize { $\begin{pmatrix}
1642\\ 210
\end{pmatrix}$}} &	{\scriptsize { $\begin{pmatrix}
1778\\ 433
\end{pmatrix}$}} &  1.019 \\ [1.1ex]\hline
\rule{0cm}{0.4cm}SC Sol 3& 1587 & 1730 & 1907 &{\scriptsize { $\begin{pmatrix}
1518\\ 188
\end{pmatrix}$}} & {\scriptsize { $\begin{pmatrix}
1642\\ 207
\end{pmatrix}$}} &	{\scriptsize { $\begin{pmatrix}
1779\\ 425
\end{pmatrix}$}} &  1.016 \\ [1.1ex]\hline
\rule{0cm}{0.4cm}SC Sol 4& 1529 & 1659 & 2150 &{\scriptsize { $\begin{pmatrix}
1522\\ 186
\end{pmatrix}$}} & {\scriptsize { $\begin{pmatrix}
1647\\ 216
\end{pmatrix}$}} &	{\scriptsize { $\begin{pmatrix}
1774\\ 493
\end{pmatrix}$}} &  1.141 \\ [1.1ex]\hline
\rule{0cm}{0.4cm}SC Sol 5& 1462 & 1593 & 1756 &{\scriptsize { $\begin{pmatrix}
1518\\ 188
\end{pmatrix}$}} & {\scriptsize { $\begin{pmatrix}
1643\\ 201
\end{pmatrix}$}} &	{\scriptsize { $\begin{pmatrix}
1776\\ 406
\end{pmatrix}$}} &  1.052 \\ [1.1ex]\hline
\rule{0cm}{0.4cm}SC Sol 6& 1663 & 1761 & 2252 &{\scriptsize { $\begin{pmatrix}
1523\\ 185
\end{pmatrix}$}} & {\scriptsize { $\begin{pmatrix}
1649\\ 218
\end{pmatrix}$}} &	{\scriptsize { $\begin{pmatrix}
1775\\ 510
\end{pmatrix}$}} &  1.216 \\ [1.1ex]\hline\hline
\end{tabular}\label{S11_SC_table}
\end{center}
\end{table}

\begin{table*}[!h]
\caption{\footnotesize { The extracted $P_{11}$ partial wave $T$-matrix poles using different values of subtraction constant. SC Sol 1: $\mathrm{Re}\Phi(s_0)=m_{\pi}$ for all 
three used  channels; SC Sol 2: $\mathrm{Re}\Phi(s_0)=m_{\pi}$ for $\pi N$ channel only, while in other two channels $\mathrm{Re}\Phi(s_0)=0$; SC Sol 3: 
$\mathrm{Re}\Phi(s_0)=m_{\eta}$ for $\eta N$ channel only, while in other two channels $\mathrm{Re}\Phi(s_0)=0$; SC Sol 4: $\mathrm{Re}\Phi(s_0)=m_{\pi^2}$ for $\pi^2 N$ channel 
only, while in other two channels $\mathrm{Re}\Phi(s_0)=0$; SC Sol 5: $\mathrm{Re}\Phi(s_0)=-m_{\pi}$ for all three used channels, SC Sol 6: $\mathrm{Re}\Phi(s_0)=m_a$, where 
$m_a$ is the mass of the meson in corresponding  channel.}}
\begin{center}
\begin{tabular}{|l|cccc|cccc|c|}
\hline \hline 
\rule{0cm}{0.3cm} &  \multicolumn{4}{c|}{ Bare poles} & \multicolumn{4}{c|}{ Dressed poles } & \\ \cline{2-9}
\rule{0cm}{0.4cm}Solutions  &   $\mathrm{W_{s_1}}$ & $\mathrm{W_{s_2}}$ &  $\mathrm{W_{s_3}}$   & $\mathrm{W_{s_4}}$ & {\scriptsize {
 $\begin{pmatrix}
\mathrm{Re W}\\ \mathrm{-2Im W} 
\end{pmatrix}$}} & {\scriptsize { $\begin{pmatrix}
\mathrm{Re W}\\ \mathrm{-2Im W} 
\end{pmatrix}$}} & {\scriptsize { $\begin{pmatrix}
\mathrm{Re W}\\ \mathrm{-2Im W} 
\end{pmatrix}$}}  & {\scriptsize { $\begin{pmatrix}
\mathrm{Re W}\\ \mathrm{-2Im W} 
\end{pmatrix}$}}  & $\mathrm{\chi_R^2}$  \\ [1.1ex]
  &  & \multicolumn{2}{c}{$\mathrm{MeV}$}  &  & & \multicolumn{2}{c}{$\mathrm{MeV}$}  & &  \\ \hline \hline
\rule{0cm}{0.4cm}Standard fit& 1607 & 1772 & 2182 & 2841&{\scriptsize { $\begin{pmatrix}
1365\\ 157
\end{pmatrix}$}} & {\scriptsize { $\begin{pmatrix}
1708\\ 174 
\end{pmatrix}$}} &	{\scriptsize { $\begin{pmatrix}
1731\\ 136
\end{pmatrix}$}}& {\scriptsize { $\begin{pmatrix}
2117\\ 345
\end{pmatrix}$}}&  0.958  \\ [1.1ex] \hline 
\rule{0cm}{0.4cm}SC Sol 1 & 1686 & 1858 & 2323 &4216&{\scriptsize { $\begin{pmatrix}
1365\\ 157
\end{pmatrix}$}} & {\scriptsize { $\begin{pmatrix}
1715\\ 174
\end{pmatrix}$}} &	{\scriptsize { $\begin{pmatrix}
1730\\ 135
\end{pmatrix}$}} &{\scriptsize { $\begin{pmatrix}
2114\\ 370
\end{pmatrix}$}} & 1.011\\[1.1ex]\hline
\rule{0cm}{0.4cm}SC Sol 2 & 1660 & 1801 & 2191 &3279&{\scriptsize { $\begin{pmatrix}
1364\\ 160
\end{pmatrix}$}} & {\scriptsize { $\begin{pmatrix}
1713\\ 173
\end{pmatrix}$}} &	{\scriptsize { $\begin{pmatrix}
1730\\ 136
\end{pmatrix}$}}&	{\scriptsize { $\begin{pmatrix}
2116\\ 359
\end{pmatrix}$}} &  0.965  \\[1.1ex]\hline
\rule{0cm}{0.4cm}SC Sol 3 & 1611 & 1796 & 2376 &3586& {\scriptsize { $\begin{pmatrix}
1364\\ 159
\end{pmatrix}$}} & {\scriptsize { $\begin{pmatrix}
1711\\ 175
\end{pmatrix}$}} &	{\scriptsize { $\begin{pmatrix}
1730\\ 135
\end{pmatrix}$}} &	{\scriptsize { $\begin{pmatrix}
2116\\ 349
\end{pmatrix}$}}&  0.953 \\[1.1ex] \hline
\rule{0cm}{0.4cm}SC Sol 4& 1662 & 1809 & 2181 & 4842&{\scriptsize { $\begin{pmatrix}
1364\\ 160
\end{pmatrix}$}} & {\scriptsize { $\begin{pmatrix}
1712\\ 173 
\end{pmatrix}$}} &	{\scriptsize { $\begin{pmatrix}
1730\\ 136
\end{pmatrix}$}}& {\scriptsize { $\begin{pmatrix}
2116\\ 357
\end{pmatrix}$}}&  0.959  \\ [1.1ex] \hline 

\rule{0cm}{0.4cm}SC Sol 5 & 1461 & 1634 & 1966 &  2412 &{\scriptsize { $\begin{pmatrix}
1365 \\ 165
\end{pmatrix}$}} & {\scriptsize { $\begin{pmatrix}
1711 \\ 164
\end{pmatrix}$}} &	\centering {\scriptsize { $\begin{pmatrix}
 1731 \\ 140
\end{pmatrix}$}} & \centering {\scriptsize { $\begin{pmatrix}
 2110\\ 325
\end{pmatrix}$}}& 1.027 \\[1.1ex]\hline
SC Sol 6 & 1696 & 1876 & 2433 & 7115 &{\scriptsize { $\begin{pmatrix}
1369\\ 165
\end{pmatrix}$}} & {\scriptsize { $\begin{pmatrix}
1706\\ 154
\end{pmatrix}$}} &{\scriptsize { $\begin{pmatrix}
1734\\ 136
\end{pmatrix}$}} &{\scriptsize { $\begin{pmatrix}
2126\\ 360
\end{pmatrix}$}}&  1.303 \\[1.1ex] \hline \hline
\end{tabular}
\end{center}
\label{P11_SC_table}
\end{table*}

\begin{table*}[!hb]
\caption{\footnotesize { The extracted $D_{13}$ partial wave $T$-matrix poles using different values of subtraction constant. SC Sol 1: $\mathrm{Re}\Phi(s_0)=m_{\pi}$ for all 
three used  channels; SC Sol 2: $\mathrm{Re}\Phi(s_0)=m_{\pi}$ for $\pi N$ channel only, while in other two channels $\mathrm{Re}\Phi(s_0)=0$; SC Sol 3: 
$\mathrm{Re}\Phi(s_0)=m_{\eta}$ for $\eta N$ channel only, while in other two channels $\mathrm{Re}\Phi(s_0)=0$; SC Sol 4: $\mathrm{Re}\Phi(s_0)=m_{\pi^2}$ for $\pi^2 N$ channel 
only, while in other two channels $\mathrm{Re}\Phi(s_0)=0$; SC Sol 5: $\mathrm{Re}\Phi(s_0)=-m_{\pi}$ for all three used channels, SC Sol 6: $\mathrm{Re}\Phi(s_0)=m_a$, where 
$m_a$ is the mass of the meson in corresponding  channel.}}
\begin{center}
\begin{tabular}{|l|ccc|cccc|c|}
\hline \hline 
\rule{0cm}{0.25cm} &  \multicolumn{3}{c|}{ Bare poles} & \multicolumn{4}{c|}{ Dressed poles } &  \\ \cline{2-8}
\rule{0cm}{0.4cm}Solutions  &   $\mathrm{W_{s_1}}$ & $\mathrm{W_{s_2}}$ &  $\mathrm{W_{s_3}}$   & {\scriptsize { $\begin{pmatrix}
\mathrm{Re W}\\ \mathrm{-2Im W} 
\end{pmatrix}$}} & {\scriptsize { $\begin{pmatrix}
\mathrm{Re W}\\ \mathrm{-2Im W} 
\end{pmatrix}$}} & {\scriptsize { $\begin{pmatrix}
\mathrm{Re W}\\ \mathrm{-2Im W} 
\end{pmatrix}$}}   & {\scriptsize { $\begin{pmatrix}
\mathrm{Re W}\\ \mathrm{-2Im W} 
\end{pmatrix}$}}   &{$\chi_R^2$}  \\ 
 &&  $\mathrm{MeV}$&  & & \multicolumn{2}{c}{$\mathrm{MeV}$}  & & \\ \hline \hline 
\rule{0cm}{0.4cm}Standard fit & 1582 & 1880 & 2499 &{\scriptsize { $\begin{pmatrix}
1506\\ 121
\end{pmatrix}$}} & {\scriptsize { $\begin{pmatrix}
1807\\ 121
\end{pmatrix}$}} &	{\scriptsize { $\begin{pmatrix}
1939\\ 485
\end{pmatrix}$}} &  {\scriptsize { $\begin{pmatrix}
2675\\ 583
\end{pmatrix}$}} &1.027\\ [1.1ex]\hline
SC Sol 1 & 1666 & 1988 & 2683&{\scriptsize { $\begin{pmatrix}
1506\\ 121
\end{pmatrix}$}} & {\scriptsize { $\begin{pmatrix}
1808\\ 124
\end{pmatrix}$}} &	{\scriptsize { $\begin{pmatrix}
1942\\ 497
\end{pmatrix}$}} &  {\scriptsize { $\begin{pmatrix}
2691\\ 581
\end{pmatrix}$}} &1.041  \\[1.1ex] \hline
SC Sol 2 & 1619 & 1891 & 2547 &{\scriptsize { $\begin{pmatrix}
1506\\ 121
\end{pmatrix}$}} & {\scriptsize { $\begin{pmatrix}
1807\\ 127
\end{pmatrix}$}} &	{\scriptsize { $\begin{pmatrix}
1939\\ 485
\end{pmatrix}$}} &{\scriptsize { $\begin{pmatrix}
2691\\ 583
\end{pmatrix}$}}  &1.031 \\[1.1ex] \hline
\rule{0cm}{0.4cm}SC Sol 3& 1628 & 2071 & 2599 &{\scriptsize { $\begin{pmatrix}
1506\\ 122
\end{pmatrix}$}} & {\scriptsize { $\begin{pmatrix}
1807\\ 127 
\end{pmatrix}$}} &	{\scriptsize { $\begin{pmatrix}
1939\\ 486
\end{pmatrix}$}} & {\scriptsize { $\begin{pmatrix}
2692\\ 583
\end{pmatrix}$}}& 1.031  \\[1.1ex]  \hline 

\rule{0cm}{0.4cm}SC Sol 4 & 1661 & 1930 & 2661 &{\scriptsize { $\begin{pmatrix}
1506\\ 122
\end{pmatrix}$}} & {\scriptsize { $\begin{pmatrix}
1807\\ 127
\end{pmatrix}$}} &	{\scriptsize { $\begin{pmatrix}
1939\\ 486
\end{pmatrix}$}} &  {\scriptsize { $\begin{pmatrix}
2692\\ 583
\end{pmatrix}$}} &1.031 \\[1.1ex] \hline

SC Sol 5 & 1317 & 1697 & 2376 &{\scriptsize { $\begin{pmatrix}
1505 \\ 121
\end{pmatrix}$}} & {\scriptsize { $\begin{pmatrix}
 1805 \\ 125
\end{pmatrix}$}} &	{\scriptsize { $\begin{pmatrix}
 1939 \\ 485
\end{pmatrix}$}} & {\scriptsize { $\begin{pmatrix}
 2692 \\ 583
\end{pmatrix}$}} & 1.032 \\[1.1ex]\hline

SC Sol 6& 1678 & 2071 & 2722 &{\scriptsize { $\begin{pmatrix}
1506\\ 121
\end{pmatrix}$}} & {\scriptsize { $\begin{pmatrix}
1810\\ 122
\end{pmatrix}$}} &	{\scriptsize { $\begin{pmatrix}
1943\\ 508
\end{pmatrix}$}} &  {\scriptsize { $\begin{pmatrix}
2690\\ 580
\end{pmatrix}$}} & 1.183   \\[1.1ex] \hline \hline
\end{tabular}
\label{D13_SC_table}
\end{center}
\end{table*}

We see that bare pole positions are notably changed with the variation of the dispersion relation subtraction  point value. For the shift into the time-like direction the bare 
poles are shifted upwards; and when the subtraction point is shifted into the negative energy region the value of bare poles is reduced. The change is notable, and for inspected 
values varies for 150 - 200 MeV.  It is important to notice that the dressed pole positions are extremely stable, see Fig. (\ref{Fig:dcp}), and not influenced by the subtraction 
constant values. \\

\clearpage

\vbox{
{ We have also tested the freedom in choosing the dispersion integral subtraction point point $s_0$. In Fig. (\ref{Fig:spvariation}) where the resulting real part is depicted, we 
show that the results are qualitatively and quantitatively similar to changing the size of the subtraction constant. We see that changing the subtraction point value shifts the 
real part upwards and downwards depending on the direction of the shift. So, in practice, this freedom is equivalent to retaining the subtraction point at threshold and varying 
the subtraction constant. Therefore, the similar conclusions  as in case of dispersion constant variation are drawn that the bare poles are strongly influenced, and dressed poles 
remain virtually untouched. }}

\begin{figure}[!hb]
\vspace*{-5.cm}
 \centering
\includegraphics[width=12cm]{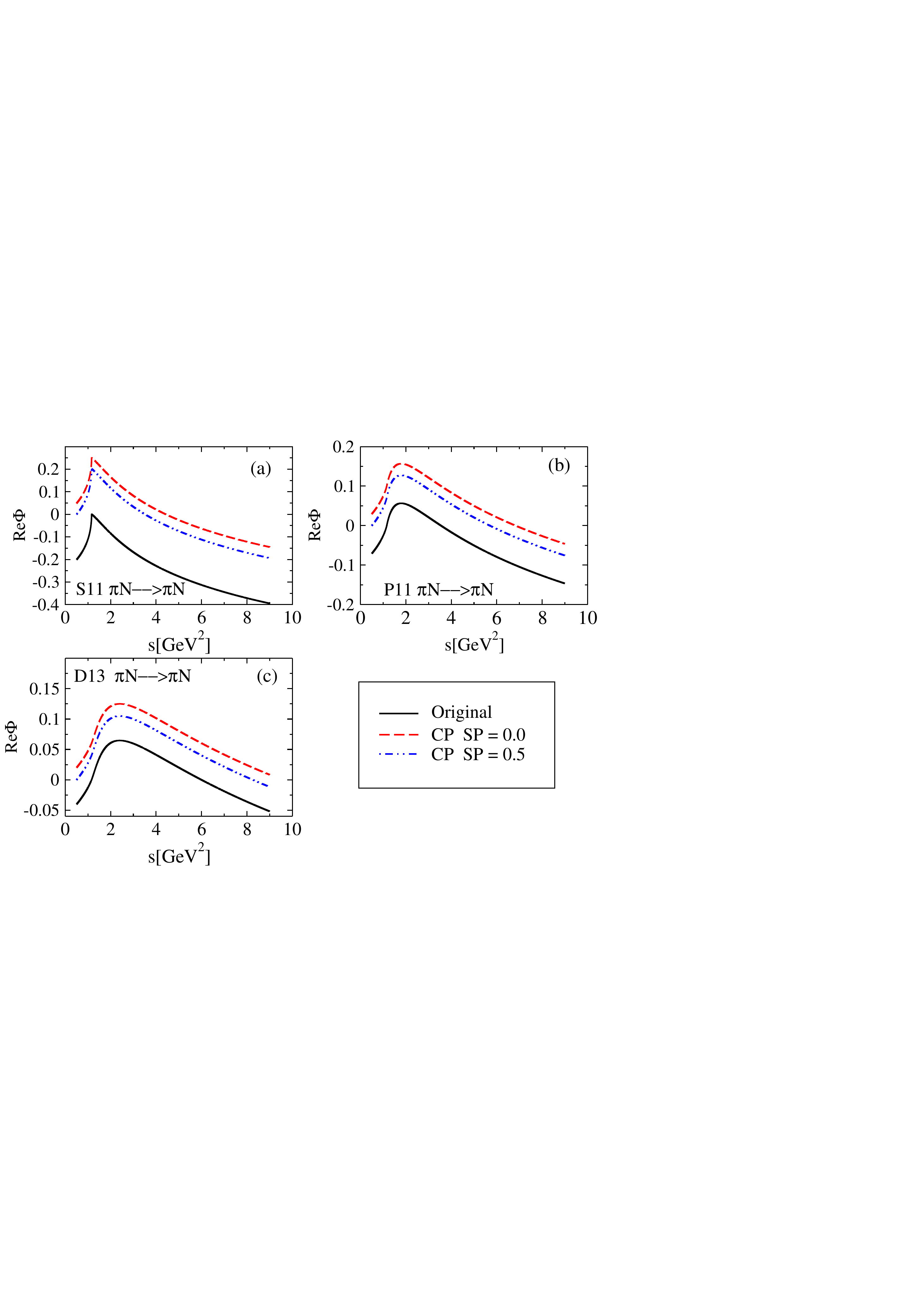}
\vspace*{-6.cm}
\caption[]{\footnotesize {(Color online) Real  part of S$_{11}$, $P_{11}$ and $D_{13}$ elastic channel propagator. We change the dispersion integral subtraction point value.}}
\label{Fig:spvariation}
\end{figure}

\noindent
\underline{\textit{Conclusion:}} \\

{ Results  shown in Tables \ref{S11_SC_table} - \ref{D13_SC_table}} prove that the value of subtraction constant affect bare pole positions significantly. Dressed poles remain 
practically unchanged. }

 \subsection{Background parameterization}

Additional background poles below and above physical region are added.  Our model was based on the premises that two subthreshold unphysical 
poles are enough to represent the smooth energy dependent background, and we test this assumption. We test two aspects:
\begin{enumerate}
   \item  Are two poles absolutely essential?
   \item   Does adding new subthreshold, or distant, positive energy poles change the pole positions?
\end{enumerate}   

Therefore, we have added first one, and then two poles, below, and far above threshold, and with different signs representing attraction or 
repulsion. All possible combinations for three analyzed partial waves are analyzed. \\

\noindent
\underline{\textit{Result:}} \\

The shift in pole positions for partial waves S$_{11}$, P$_{11}$ and D$_{13}$ are presented in Tables \ref{S11_BG_table}- \ref{D13_BG_table}
 and depicted in Fig. (\ref{S11_BG_poles}).

\begin{table*}[!h]
\caption{\footnotesize The extracted $S_{11}$ partial wave T-matrix poles. Additional background poles below and above physical region are added.
}
\begin{center}
\begin{tabular}{|l|cc|ccc|ccc|c|}
\hline \hline 
 & \multicolumn{5}{c}{ Bare poles} & \multicolumn{3}{c|}{ Dressed poles } & \\ \cline{2-9}
Solutions  & $\begin{array}{c}\mathrm{B1}\\\mathrm{B2}\end{array}$ &$\begin{array}{c}\mathrm{B3}\\\mathrm{B4}\end{array}$& $\mathrm{W_{s_1}}
$ & $\mathrm{W_{s_2}}$ &  $\mathrm{W_{s_3}}$ & {\scriptsize { $\begin{pmatrix}
\mathrm{Re W}\\ \mathrm{-2Im W} 
\end{pmatrix}$}} & {\scriptsize { $\begin{pmatrix}
\mathrm{Re W}\\ \mathrm{-2Im W} 
\end{pmatrix}$}} & {\scriptsize { $\begin{pmatrix}
\mathrm{Re W}\\ \mathrm{-2Im W} 
\end{pmatrix}$}}   & $\mathrm{\chi_R^2}$  \\ 

  & $\mathrm{GeV^2}$&$\mathrm{GeV^2}$ &  & $\mathrm{MeV}$ & &  &  $\mathrm{MeV}$& &  \\ \hline \hline
Standard fit&$\begin{array}{c}-1.256 \\ 0.210\end{array}$ & &1523 & 1640 & 1834 &{\scriptsize { $\begin{pmatrix}
1518\\ 189
\end{pmatrix}$}} & {\scriptsize { $\begin{pmatrix}
1642\\ 208
\end{pmatrix}$}} &	{\scriptsize { $\begin{pmatrix}
1779\\ 427
\end{pmatrix}$}}&  1.016  \\  \hline 
{ Sol 1}{\scriptsize$\begin{array}{c} e1=+1(Below)\\e2=-1(Below)\\e3=+1(Below) \end{array}$}&\scriptsize{$\begin{array}{c}-1.382 \\ 
0.460 \end{array}$ }&\scriptsize{ -1.851}&1523 & 1640 & 1835 &{\scriptsize { $\begin{pmatrix}
1518\\ 189
\end{pmatrix}$}} & {\scriptsize { $\begin{pmatrix}
1642\\ 208
\end{pmatrix}$}} &	{\scriptsize { $\begin{pmatrix}
1779\\ 427
\end{pmatrix}$}}&  1.016  \\  \hline 
{ Sol 2}{\scriptsize$\begin{array}{c} e1=+1(Below)\\e2=-1(Below)\\e3=-1(Below) \end{array}$} &\scriptsize{$\begin{array}{c}-2.508 \\ 
-6.794 \end{array}$ }&\scriptsize{0.419} &1523 & 1640 & 1834 &{\scriptsize { $\begin{pmatrix}
1518\\ 189
\end{pmatrix}$}} & {\scriptsize { $\begin{pmatrix}
1642\\ 208
\end{pmatrix}$}} &	{\scriptsize { $\begin{pmatrix}
1780\\ 427
\end{pmatrix}$}}&  1.019  \\  \hline 
{ Sol 3}{\scriptsize$\begin{array}{c} e1=+1(Below)\\e2=-1(Below)\\e3=-1(Below)\\e4=+1(Below) \end{array}$}&\scriptsize{$\begin{array}{c}-0.0049
\\ 0.103 \end{array}$} &\scriptsize{ $\begin{array}{c} -1.078 \\ -0.0024 \end{array}$ }&1523 & 1640 & 1835 & {\scriptsize 
{ $\begin{pmatrix}
1520\\ 191
\end{pmatrix}$}} & {\scriptsize { $\begin{pmatrix}
1641\\ 207 
\end{pmatrix}$}} &	{\scriptsize { $\begin{pmatrix}
1781\\ 420
\end{pmatrix}$}}&   1.019 \\  \hline 

\rule{0cm}{0.4cm}{ Sol 4}{\scriptsize$\begin{array}{c} e1=+1(Below)\\e2=-1(Above) \end{array}$}&\scriptsize{$\begin{array}{c}-46594\\ 36.17
\end{array}$}&   &1525 & 1646 & 1844 &{\scriptsize { $\begin{pmatrix}
1529\\ 171
\end{pmatrix}$}} & {\scriptsize { $\begin{pmatrix}
1659\\ 223 
\end{pmatrix}$}} &	{\scriptsize { $\begin{pmatrix}
1713\\ 681
\end{pmatrix}$}}&  2.461  \\ [1.2ex] \hline 

{ Sol 5}{\scriptsize$\begin{array}{c} e1=+1(Below)\\e2=-1(Below)\\e3=+1(Above) \end{array}$}&\scriptsize{$\begin{array}{c}-1.694\\ 0.344
\end{array}$}& \scriptsize{7536}  &1523 & 1640 & 1834 &{\scriptsize { $\begin{pmatrix}
1518\\ 193
\end{pmatrix}$}} & {\scriptsize { $\begin{pmatrix}
1642\\ 208 
\end{pmatrix}$}} &	{\scriptsize { $\begin{pmatrix}
1779\\428
\end{pmatrix}$}}&  1.019  \\  \hline 

{ Sol 6}{\scriptsize$\begin{array}{c} e1=+1(Below)\\e2=-1(Above)\\e3=+1(Above) \end{array}$}&\scriptsize{$\begin{array}{c}-7725\\ 12.85
\end{array}$}& \scriptsize{12.80}  &1525 & 1645 & 1844 &{\scriptsize { $\begin{pmatrix}
1528\\ 177
\end{pmatrix}$}} & {\scriptsize { $\begin{pmatrix}
1657\\ 220 
\end{pmatrix}$}} &	{\scriptsize { $\begin{pmatrix}
1740\\670
\end{pmatrix}$}}&  2.211  \\  \hline 
{ Sol 7}{\scriptsize$\begin{array}{c} e1=+1(Below)\\e2=-1(Below)\\e3=-1(Above)\\e4=+1(Above) \end{array}$}&\scriptsize{$\begin{array}{c}-1.897
\\ 0.410\end{array}$}&\scriptsize{$\begin{array}{c} 51.69 \\73.95\end{array}$} &1523 & 1640 & 1835 &
{\scriptsize { $\begin{pmatrix}
1519\\ 194
\end{pmatrix}$}} & {\scriptsize { $\begin{pmatrix}
1641\\ 207 
\end{pmatrix}$}} &	{\scriptsize { $\begin{pmatrix}
1780\\427
\end{pmatrix}$}}&  1.018 \\  \hline \hline
\end{tabular}\label{S11_BG_table}
\end{center}
\end{table*}~
\begin{table*}[!h]
\caption{\footnotesize The extracted $P_{11}$ partial wave T-matrix poles. Additional background poles below and above physical region are added.}
\begin{center}
\begin{tabular}{|l|cc|cccc|cccc|c|}
\hline \hline 
 &  \multicolumn{6}{c|}{ Bare poles} & \multicolumn{4}{c|}{ Dressed poles } & \\ \cline{2-11}
Solutions  &  $\begin{array}{c}\mathrm{B1}\\\mathrm{B2}\end{array}$&$\begin{array}{c}\mathrm{B3}\\\mathrm{B
4}\end{array}$&$\mathrm{W_{s_1}}$ 
& $\mathrm{W_{s_2}}$ &  $\mathrm{W_{s_3}}$   & $\mathrm{W_{s_4}}$ & {\scriptsize { $\begin{pmatrix}
\mathrm{Re W}\\ \mathrm{-2Im W} 
\end{pmatrix}$}} &{\scriptsize { $\begin{pmatrix}
\mathrm{Re W}\\ \mathrm{-2Im W} 
\end{pmatrix}$}}  & {\scriptsize { $\begin{pmatrix}
\mathrm{Re W}\\ \mathrm{-2Im W} 
\end{pmatrix}$}} & {\scriptsize { $\begin{pmatrix}
\mathrm{Re W}\\ \mathrm{-2Im W} 
\end{pmatrix}$}} & $\mathrm{\chi_R^2}$  \\ 
  & $\mathrm{GeV^2}$&$\mathrm{GeV^2}$   &  & \multicolumn{2}{c}{$\mathrm{MeV}$}  &  & & \multicolumn{2}{c}{$\mathrm{MeV}$}  & & \\ \hline \hline
Standard fit & $\begin{array}{c}0.814\\0.9886\end{array}$&&1607 & 1772 & 2182 & 2841&{\scriptsize { $\begin{pmatrix}
1365\\ 157
\end{pmatrix}$}} & {\scriptsize { $\begin{pmatrix}
1708\\ 174 
\end{pmatrix}$}} &	{\scriptsize { $\begin{pmatrix}
1731\\ 136
\end{pmatrix}$}}& {\scriptsize { $\begin{pmatrix}
2117\\ 345
\end{pmatrix}$}}&  0.958  \\  \hline 

{ Sol 1}{\scriptsize$\begin{array}{c} e1=+1(Below)\\e2=-1(Below)\\e3=+1(Below) \end{array}$} &\scriptsize{$\begin{array}{c}0.749
\\ 0.945\end{array}$} &\scriptsize{0.928}&1606 & 1772 & 2181 & 2859&{\scriptsize { $\begin{pmatrix}
1364\\ 162
\end{pmatrix}$}} & {\scriptsize { $\begin{pmatrix}
1708\\ 174 
\end{pmatrix}$}} &	{\scriptsize { $\begin{pmatrix}
1731\\ 136
\end{pmatrix}$}}& {\scriptsize { $\begin{pmatrix}
2116\\ 344
\end{pmatrix}$}}&  0.964  \\  \hline 

{ Sol 2}{\scriptsize$\begin{array}{c} e1=+1(Below)\\e2=-1(Below)\\e3=-1(Below) \end{array}$} &\scriptsize{$\begin{array}{c}0.612
\\-11285\end{array}$}&\scriptsize{1.241}&1606 & 1772 & 2180 & 2816&{\scriptsize { $\begin{pmatrix}
1361\\ 161
\end{pmatrix}$}} & {\scriptsize { $\begin{pmatrix}
1709\\ 172
\end{pmatrix}$}} &	{\scriptsize { $\begin{pmatrix}
1730\\ 138
\end{pmatrix}$}}& {\scriptsize { $\begin{pmatrix}
2118\\ 345
\end{pmatrix}$}}&  0.960  \\  \hline 

{ Sol 3}{\scriptsize$\begin{array}{c} e1=+1(Below)\\e2=-1(Below)\\e3=+1(Below)\\e4=-1(Below) \end{array}$} &
\scriptsize{$\begin{array}{c}0.530\\1.145\end{array}$} &\scriptsize{$\begin{array}{c}0.668\\0.195\end{array}$}
 &1605 & 1772 & 2177 & 2859&{\scriptsize { $\begin{pmatrix}
1354\\ 169
\end{pmatrix}$}} & {\scriptsize { $\begin{pmatrix}
1710\\ 172
\end{pmatrix}$}} &	{\scriptsize { $\begin{pmatrix}
1731\\ 141
\end{pmatrix}$}}& {\scriptsize { $\begin{pmatrix}
2115\\ 340
\end{pmatrix}$}}&  0.957  \\ \hline 

\rule{0cm}{0.4cm}{ Sol 4}{\scriptsize$\begin{array}{c} e1=+1(Below)\\e2=-1(Above) \end{array}$}&
 {\scriptsize$\begin{array}{c}0.030\\95.24\end{array}$}&&1597 & 1768 & 2207 & 3508&{\scriptsize { $\begin{pmatrix}
1375\\176 
\end{pmatrix}$}} & {\scriptsize { $\begin{pmatrix}
1717\\ 133 
\end{pmatrix}$}} &	{\scriptsize { $\begin{pmatrix}
1733\\ 154
\end{pmatrix}$}}& {\scriptsize { $\begin{pmatrix}
2082\\ 351
\end{pmatrix}$}}&  2.695  \\ [1.2ex] \hline 
{ Sol 5} {\scriptsize$\begin{array}{c} e1=+1(Below)\\e2=-1(Below)\\e3=+1(Above) \end{array}$}& {\scriptsize$\begin{array}{c}0.354\\
0.351\end{array}$}&\scriptsize{$0.961\cdot 10^{10}$}&1606 & 1772 & 2182 & 2801&{\scriptsize { $\begin{pmatrix}
1364\\165 
\end{pmatrix}$}} & {\scriptsize { $\begin{pmatrix}
1710\\ 169 
\end{pmatrix}$}} &	{\scriptsize { $\begin{pmatrix}
1732\\ 137
\end{pmatrix}$}}& {\scriptsize { $\begin{pmatrix}
2112\\ 334
\end{pmatrix}$}}&  1.010  \\  \hline 
{ Sol 6}{\scriptsize$\begin{array}{c} e1=+1(Below)\\e2=-1(Below)\\e3=+1(Above) \end{array}$}&{\scriptsize$\begin{array}{c} 0.169\\
79.84\end{array}$}&\scriptsize{1746}&1606 & 1773& 2201 & 2860&{\scriptsize { $\begin{pmatrix}
1365\\168 
\end{pmatrix}$}} & {\scriptsize { $\begin{pmatrix}
1705\\ 168 
\end{pmatrix}$}} &	{\scriptsize { $\begin{pmatrix}
1733\\ 135
\end{pmatrix}$}}& {\scriptsize { $\begin{pmatrix}
2108\\ 338
\end{pmatrix}$}}&  1.019  \\  \hline 
{ Sol 7}{\scriptsize$\begin{array}{c} e1=+1(Below)\\e2=-1(Below)\\e3=-1(Above)\\e4=+1(Above) \end{array}$}& 
{\scriptsize$\begin{array}{c}0.793\\0.885\end{array}$}&{\scriptsize$\begin{array}{c}33.64\\
5954\end{array}$}&1606 & 1773& 2177 & 2901&{\scriptsize { $\begin{pmatrix}
1359\\163 
\end{pmatrix}$}} & {\scriptsize { $\begin{pmatrix}
1709\\ 175 
\end{pmatrix}$}} &	{\scriptsize { $\begin{pmatrix}
1730\\ 140
\end{pmatrix}$}}& {\scriptsize { $\begin{pmatrix}
2115\\ 331
\end{pmatrix}$}}&  0.959  \\  \hline \hline
\end{tabular}\label{P11_BG_table}
\end{center}
\end{table*}~

\begin{table*}[!h]
\caption{\footnotesize The extracted $D_{13}$ partial wave T-matrix poles. Additional background poles below and above physical region are added.
}
\begin{center}
\begin{tabular}{|l|cc|ccc|cccc|c|}
\hline \hline 
 & \multicolumn{5}{c|}{ Bare poles} & \multicolumn{4}{c|}{ Dressed poles } & \\ \cline{2-10}
Solutions  &$\begin{array}{c}\mathrm{B1}\\\mathrm{B2}\end{array}$  &$\begin{array}{c}\mathrm{B3}\\\mathrm{B4}\end{array}$&
 $\mathrm{W_{s_1}}$ & $\mathrm{W_{s_2}}$ &  $\mathrm{W_{s_3}}$ & {\scriptsize { $\begin{pmatrix}
\mathrm{Re W}\\ \mathrm{-2Im W} 
\end{pmatrix}$}} & {\scriptsize { $\begin{pmatrix}
\mathrm{Re W}\\ \mathrm{-2Im W} 
\end{pmatrix}$}} & {\scriptsize { $\begin{pmatrix}
\mathrm{Re W}\\ \mathrm{-2Im W} 
\end{pmatrix}$}}    & {\scriptsize { $\begin{pmatrix}
\mathrm{Re W}\\ \mathrm{-2Im W} 
\end{pmatrix}$}}  &$\mathrm{\chi_R^2}$  \\ 

  & $\mathrm{GeV^2}$& $\mathrm{GeV^2}$&  & $\mathrm{MeV}$ &  &&  \multicolumn{2}{c}{$\mathrm{MeV}$}  & &\\ \hline \hline
Standard fit&$\begin{array}{c}0.729\\ -14628.0\end{array}$&&1582 & 1880 & 2499 &{\scriptsize { $\begin{pmatrix}
1506\\ 121
\end{pmatrix}$}} & {\scriptsize { $\begin{pmatrix}
1807\\ 127 
\end{pmatrix}$}} &	{\scriptsize { $\begin{pmatrix}
1939\\ 485
\end{pmatrix}$}}&{\scriptsize { $\begin{pmatrix}
2691\\ 583
\end{pmatrix}$}}&  1.027  \\  \hline

{ Sol 1}{\scriptsize$\begin{array}{c} e1=+1(Below)\\e2=-1(Below)\\e3=+1(Below) \end{array}$}&\scriptsize{$\begin{array}{c}-0.254\\
-254\end{array}$}&\scriptsize{ 1.077 } &1582 & 1882 & 2500 &{\scriptsize { $\begin{pmatrix}
1506\\ 120
\end{pmatrix}$}} & {\scriptsize { $\begin{pmatrix}
1808\\ 127 
\end{pmatrix}$}} &	{\scriptsize { $\begin{pmatrix}
1937\\ 486
\end{pmatrix}$}}& {\scriptsize { $\begin{pmatrix}
2699\\ 572
\end{pmatrix}$}}& 1.019  \\  \hline 
{ Sol 2}{\scriptsize$\begin{array}{c} e1=+1(Below)\\e2=-1(Below)\\e3=-1(Below) \end{array}$}&\scriptsize{$\begin{array}{c}0.669\\
-12317\end{array}$} &\scriptsize{0.677 } &1582 & 1879 & 2498 &{\scriptsize { $\begin{pmatrix}
1506\\ 122
\end{pmatrix}$}} & {\scriptsize { $\begin{pmatrix}
1807\\ 128
\end{pmatrix}$}} &	{\scriptsize { $\begin{pmatrix}
1938\\ 495
\end{pmatrix}$}}& {\scriptsize { $\begin{pmatrix}
2698\\ 580
\end{pmatrix}$}}& 1.038  \\  \hline
{ Sol 3}{\scriptsize$\begin{array}{c} e1=+1(Below)\\e2=-1(Below)\\e3=+1(Below)\\e4=-1(Below) \end{array}$}&
\scriptsize{$\begin{array}{c}0.446\\0.310\end{array}$} &\scriptsize{$\begin{array}{c}0.201\\-109.7\end{array}$} &1583 & 1884 & 2499 &{
\scriptsize { $\begin{pmatrix}
1505\\ 120
\end{pmatrix}$}} & {\scriptsize { $\begin{pmatrix}
1808\\ 127
\end{pmatrix}$}} &	{\scriptsize { $\begin{pmatrix}
1934\\ 498
\end{pmatrix}$}}&  	{\scriptsize { $\begin{pmatrix}
2696\\ 572
\end{pmatrix}$}}& 1.034  \\  \hline 

\rule{0cm}{0.4cm}{ Sol 4}{\scriptsize {$\begin{array}{c} e1=+1(Below)\\e2=-1(Above) \end{array}$}}&\scriptsize{$\begin{array}{c}-0.89\cdot10^{7}
\\ 15.839\end{array}$} &   &1577 & 1880 & 2507 &{\scriptsize { $\begin{pmatrix}
1508\\ 118
\end{pmatrix}$}} & {\scriptsize { $\begin{pmatrix}
1829\\ 135 
\end{pmatrix}$}} &	{\scriptsize { $\begin{pmatrix}
1928\\ 564
\end{pmatrix}$}}& 	{\scriptsize { $\begin{pmatrix}
2775\\ 429
\end{pmatrix}$}}&  2.142 \\ [1.2ex] \hline 
{ Sol 5}{\scriptsize$\begin{array}{c} e1=+1(Below)\\e2=-1(Below)\\e3=+1(Above) \end{array}$}&\scriptsize{$\begin{array}{c}0.517\\
-10037 \end{array} $}&\scriptsize{$0.1\cdot10^{9}$}  &1582 & 1879 & 2498 &{\scriptsize { $\begin{pmatrix}
1506\\ 122
\end{pmatrix}$}} & {\scriptsize { $\begin{pmatrix}
1808\\ 128
\end{pmatrix}$}} &	{\scriptsize { $\begin{pmatrix}
1938\\ 496
\end{pmatrix}$}}& 	{\scriptsize { $\begin{pmatrix}
2698\\ 580
\end{pmatrix}$}}&  1.039  \\  \hline 
{ Sol 6}{\scriptsize$\begin{array}{c} e1=+1(Below)\\e2=-1(Above) \\e3=+1(Above)\end{array}$}&\scriptsize{$\begin{array}{c}1.125\\
6626\end{array}$ }&\scriptsize{ 220 } &1581 & 1880 & 2498 &{\scriptsize { $\begin{pmatrix}
1506\\ 120
\end{pmatrix}$}} & {\scriptsize { $\begin{pmatrix}
1809\\ 131 
\end{pmatrix}$}} &	{\scriptsize { $\begin{pmatrix}
1935\\ 500
\end{pmatrix}$}}& 	{\scriptsize { $\begin{pmatrix}
2700\\ 570
\end{pmatrix}$}}&  1.035  \\  \hline 
{ Sol 7}{\scriptsize$\begin{array}{c} e1=+1(Below)\\e2=-1(Below)\\e3=-1(Above)\\e4=+1(Above) \end{array}$}&
\scriptsize{$\begin{array}{c}0.564\\ 0.563\end{array}$ }&\scriptsize{$\begin{array}{c}90.68\\
 656.6\end{array}$} &1583 & 1899 & 2500 &{\scriptsize { $\begin{pmatrix}
1506\\ 119
\end{pmatrix}$}} & {\scriptsize { $\begin{pmatrix}
1808\\ 126 
\end{pmatrix}$}} &	{\scriptsize { $\begin{pmatrix}
1933\\ 497
\end{pmatrix}$}}& 	{\scriptsize { $\begin{pmatrix}
2689\\ 580
\end{pmatrix}$}}&  1.038  \\  \hline\hline 
\end{tabular}\label{D13_BG_table}
\end{center}
\end{table*}~

\vspace*{2.cm}

\begin{figure}[!ht]
\centering
\vspace*{-8.cm}
{\includegraphics[width=11cm]{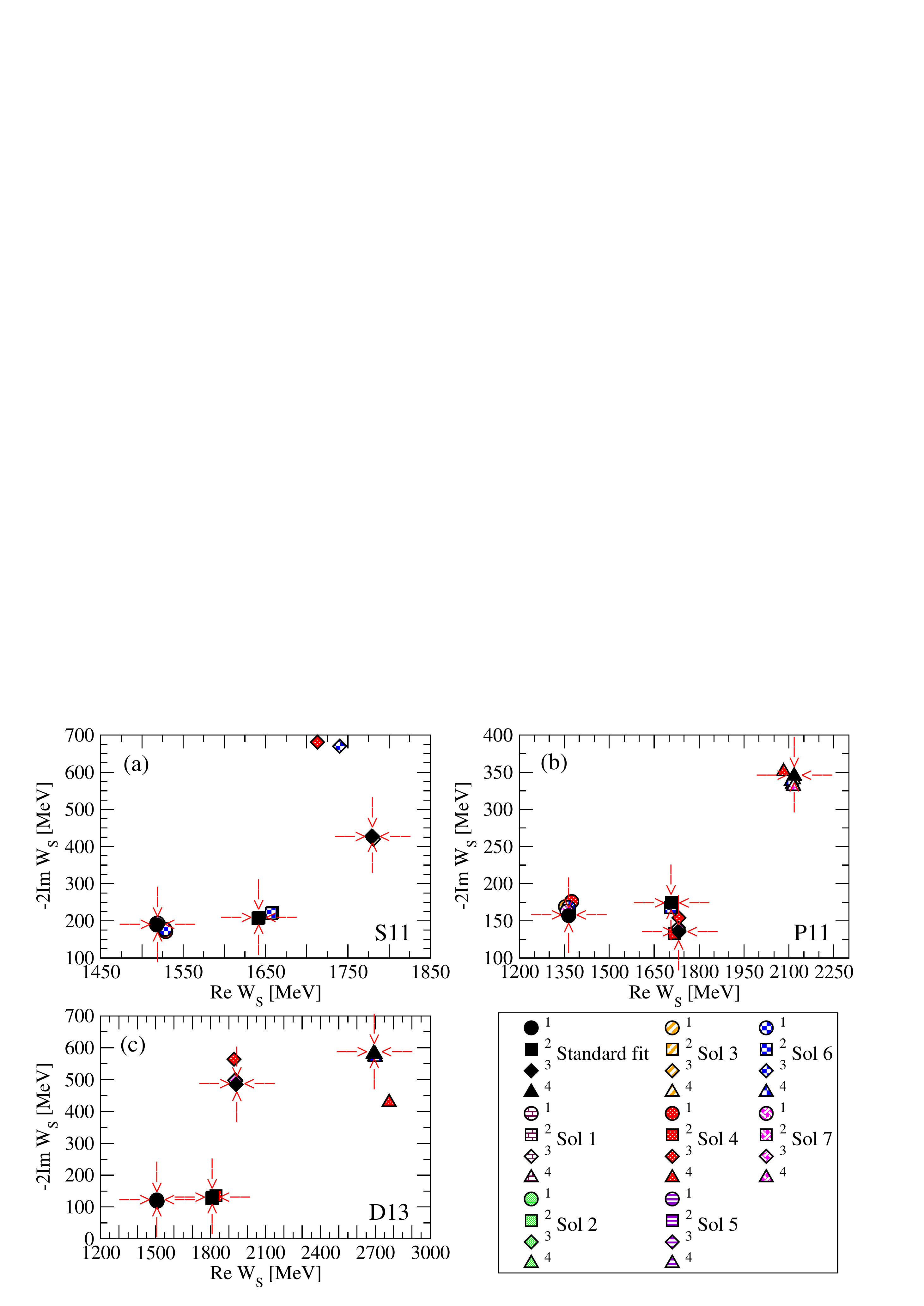}}
\caption{\footnotesize (Color online) The extracted $S_{11}$, $P_{11}$ and $D_{13}$ partial waves T-matrix poles. Additional background poles below and above
 physical region are added. { Red arrows indicate the position of Standard solution poles.}}
\label{S11_BG_poles}
\end{figure}~
\clearpage
\noindent \\ 
\underline{\textit{Conclusion:}} \\ \\
In Tables  \ref{S11_BG_table}- \ref{D13_BG_table}  in Fig. (\ref{S11_BG_poles}) we see that all extracted bare and dressed poles are practically
 identical with the exception of solutions $Sol \ 4$ and $Sol\ 6$. And these are solutions where only 1 subthreshold pole has been allowed. \\ \\ 
Therefore: 
\begin{enumerate}
   \item at least two background poles are needed to adequately represent the energy dependent background.
   \item two subthreshold poles are enough to adequately represent the background. Two pole meromorphic approximation for the background 
   representation is justified.
 \end{enumerate}
\subsection{Number of channels }
 It is not self evident that introducing new channels, i.e. new contributions to the self energy term in the denominator of  Eq. (\ref{eq:final}), 
 does not influence pole positions at all. Namely, self energy matrix is of higher order, and self energy is distributed in a different way. 
 To test this assumption we have introduced the fourth channel, and fitted the existing data base with this channel. \\

In addition to the existing $\pi$N elastic and $\pi \! N \! \rightarrow \! \eta N \:$ processes we have introduced a third channel opening at 
 1612 MeV (K$\Lambda$ channel).  The mass of the effective channel has not been altered, as the effective channel is primarily dominated by the 
 $\pi\pi$N channel. 
The results are given in Tables \ref{S11_4th_ch} - \ref{D13_4th_ch}.

\begin{table*}[!h]
\caption{\footnotesize Extracted $S_{11}$ partial wave T-matrix poles obtained by including three and four channels into a model.}
\begin{center}
\begin{tabular}{|l|ccc|ccc|c|}
\hline \hline 
 & \multicolumn{3}{c|}{ Bare poles} & \multicolumn{3}{c|}{ Dressed poles } & \\ \cline{2-7}
\rule{0cm}{0.45cm}Solutions  & $\mathrm{W_{s_1}}$ & $\mathrm{W_{s_2}}$ &  $\mathrm{W_{s_3}}$ & {\scriptsize { $\begin{pmatrix}
\mathrm{Re W}\\ \mathrm{-2Im W} 
\end{pmatrix}$}} & {\scriptsize { $\begin{pmatrix}
\mathrm{Re W}\\ \mathrm{-2Im W} 
\end{pmatrix}$}} & {\scriptsize { $\begin{pmatrix}
\mathrm{Re W}\\ \mathrm{-2Im W} 
\end{pmatrix}$}}   & $\mathrm{\chi_R^2}$  \\ 
     & & $\mathrm{MeV}$ & &  & $\mathrm{MeV}$ & &  \\ \hline \hline
\rule{0cm}{0.4cm}Sol 3CH &1523 & 1640 & 1834 &{\scriptsize { $\begin{pmatrix}
1518\\ 189
\end{pmatrix}$}} & {\scriptsize { $\begin{pmatrix}
1642\\ 208 
\end{pmatrix}$}} &	{\scriptsize { $\begin{pmatrix}
1779\\ 427
\end{pmatrix}$}}&  1.016  \\[1.2ex]  \hline 
\rule{0cm}{0.4cm}Sol 4CH &1523& 1640 & 1836 &{\scriptsize { $\begin{pmatrix}
1516\\ 190
\end{pmatrix}$}} & {\scriptsize { $\begin{pmatrix}
1642\\ 211 
\end{pmatrix}$}} &	{\scriptsize { $\begin{pmatrix}
1776\\ 418
\end{pmatrix}$}}&  1.013 \\[1.2ex]  \hline \hline
\end{tabular}
\end{center}
\label{S11_4th_ch}
\end{table*}

\begin{table*}[!h]
\caption{\footnotesize Extracted $P_{11}$ partial wave T-matrix poles obtained by including three and four channels into a model.}
\begin{center}
\begin{tabular}{|l|cccc|cccc|c|}
\hline \hline 
 &  \multicolumn{4}{c|}{ Bare poles} & \multicolumn{4}{c|}{ Dressed poles } & \\ \cline{2-9}
\rule{0cm}{0.45cm}Solutions  &$\mathrm{W_{s_1}}$ & $\mathrm{W_{s_2}}$ &  $\mathrm{W_{s_3}}$   & $\mathrm{W_{s_4}}$ & 
{\scriptsize { $\begin{pmatrix}
\mathrm{Re W}\\ \mathrm{-2Im W} 
\end{pmatrix}$}} & {\scriptsize { $\begin{pmatrix}
\mathrm{Re W}\\ \mathrm{-2Im W} 
\end{pmatrix}$}}& {\scriptsize { $\begin{pmatrix}
\mathrm{Re W}\\ \mathrm{-2Im W} 
\end{pmatrix}$}} & {\scriptsize { $\begin{pmatrix}
\mathrm{Re W}\\ \mathrm{-2Im W} 
\end{pmatrix}$}} & $\mathrm{\chi_R^2}$  \\ 
  & &  \multicolumn{2}{c}{$\mathrm{MeV}$}    &  &&  \multicolumn{2}{c}{$\mathrm{MeV}$}   &&  \\ \hline \hline
\rule{0cm}{0.4cm}Sol 3CH &1607 & 1772 & 2182 & 2841&{\scriptsize { $\begin{pmatrix}
1365\\ 157
\end{pmatrix}$}} & {\scriptsize { $\begin{pmatrix}
1708\\ 174 
\end{pmatrix}$}} &	{\scriptsize { $\begin{pmatrix}
1731\\ 136
\end{pmatrix}$}}& {\scriptsize { $\begin{pmatrix}
2117\\ 345
\end{pmatrix}$}}&  0.958  \\[1.2ex]  \hline 
\rule{0cm}{0.4cm}Sol 4CH &1607 & 1772 & 2182 & 2841&{\scriptsize { $\begin{pmatrix}
1363\\ 161
\end{pmatrix}$}} & {\scriptsize { $\begin{pmatrix}
1708\\ 174 
\end{pmatrix}$}} &	{\scriptsize { $\begin{pmatrix}
1731\\ 136
\end{pmatrix}$}}& {\scriptsize { $\begin{pmatrix}
2117\\ 343
\end{pmatrix}$}}&  0.970  \\ [1.2ex] \hline \hline
\end{tabular}
\end{center}
\label{P11_4th_ch}
\end{table*}~

\begin{table*}[!h]
\caption{\footnotesize Extracted $D_{13}$ partial wave T-matrix poles obtained by including three and four channels into a model.}
\begin{center}
\begin{tabular}{|l|ccc|cccc|c|}
\hline \hline 
 & \multicolumn{3}{c|}{ Bare poles} & \multicolumn{4}{c|}{ Dressed poles } & \\ \cline{2-8}
\rule{0cm}{0.45cm}Solutions  & $\mathrm{W_{s_1}}$ & $\mathrm{W_{s_2}}$ &  $\mathrm{W_{s_3}}$ & {\scriptsize { $\begin{pmatrix}
\mathrm{Re W}\\ \mathrm{-2Im W} 
\end{pmatrix}$}} & {\scriptsize { $\begin{pmatrix}
\mathrm{Re W}\\ \mathrm{-2Im W} 
\end{pmatrix}$}} & {\scriptsize { $\begin{pmatrix}
\mathrm{Re W}\\ \mathrm{-2Im W} 
\end{pmatrix}$}}   &{\scriptsize { $\begin{pmatrix}
\mathrm{Re W}\\ \mathrm{-2Im W} 
\end{pmatrix}$}}&   $\mathrm{\chi_R^2}$  \\ 
     & & $\mathrm{MeV}$ &&  &\multicolumn{2}{c}{$\mathrm{MeV}$} && \\ \hline \hline
\rule{0cm}{0.4cm}Sol 3CH &1582 & 1880 & 2499 &{\scriptsize { $\begin{pmatrix}
1506\\ 121
\end{pmatrix}$}} & {\scriptsize { $\begin{pmatrix}
1807\\ 127 
\end{pmatrix}$}} &	{\scriptsize { $\begin{pmatrix}
1939\\ 485
\end{pmatrix}$}}& {\scriptsize { $\begin{pmatrix}
2691\\ 583 
\end{pmatrix}$}}   &1.027  \\ [1.2ex] \hline 
\rule{0cm}{0.4cm}Sol 4CH & 1581 & 1890 & 2499 &{\scriptsize { $\begin{pmatrix}
1500\\ 169
\end{pmatrix}$}} & {\scriptsize { $\begin{pmatrix}
1808\\ 131
\end{pmatrix}$}} &	{\scriptsize { $\begin{pmatrix}
1944\\ 492
\end{pmatrix}$}}&  {\scriptsize { $\begin{pmatrix}
2689\\ 586 
\end{pmatrix}$}} & 1.035 \\[1.2ex]  \hline \hline 
\end{tabular}
\end{center}
\label{D13_4th_ch}
\end{table*}

\noindent
\underline{\textit{Conclusion:}} \\ \\
Pole position for the analyzed partial waves are extremely stable with respect to the opening of a new channel. 
\clearpage
\subsection{Mass of the effective channel}
However, in spite of looking identical to all other tests, choice of the effective cannel mass undergoes strong physical constraints:
\begin{enumerate}
   { 
   \item The effective channel represents all inelastic two and three body channels as only one,  stable, two body channel. This is a serious assumption because some channels, the 
dominant three body channel like $\pi \pi N$ and  unstable ones like $\sigma N$  and $\rho N$ introduce  a significantly different analytic structure, so this might influence the 
conclusion on existence of some resonant states. An indication that it might be actually happening is already given for the $\rho N$ channel where its complex branchpoint for the 
P$_{11}$ partial wave lies dangerously close to the P$_{11}$(1710) resonance - see ref. \cite{Ceci2011}. Genuine three body effects for the dominant $\pi \pi N$ channel might also 
cause serious problem, This particularities  should be analyzed as they are detected, but the method in general can enable statistical analyses of most of other resonant states. }
   \item As the effective channel represents the loss of flux to all inelastic channels, its mass has to be neither to high nor too low. 
   If it is too high, the loss of flux to inelastic channels imposed by the data at energies below effective channel threshold will not 
   be compensated by the third channel, so the T-matrix values will simply be too high to fit the data. If the mass is too low, we shall
    artificially open the new degrees of freedom where they actually do not exist. So, the mass of the effective channel will be fairly
	 precisely defined by all inelastic channels. 
  \item When the effective channel mass is changed, the effective channel T-matrix threshold is also shifted. Therefore, the code will
   unsuccessfully try to reproduce the inelastic channel data which are generated by the different threshold, and move other resonances.
\end{enumerate}

Therefore, we have decided to show both aspects of the problem for the $S_{11}$, $P_{11}$ and $D_{13}$ partial waves: In Fig (\ref{Fig:S11massEF}) 
and Tables \ref{S11_massEF_table} - \ref{D13_massEF_table} we illustrate what is happening to the poles when the mass of the effective channel is 
changed \vspace*{1.cm} $\pm$ 40\%.

\begin{figure}[!ht]
\centering
\vspace*{-6.cm}
{\includegraphics[width=12cm]{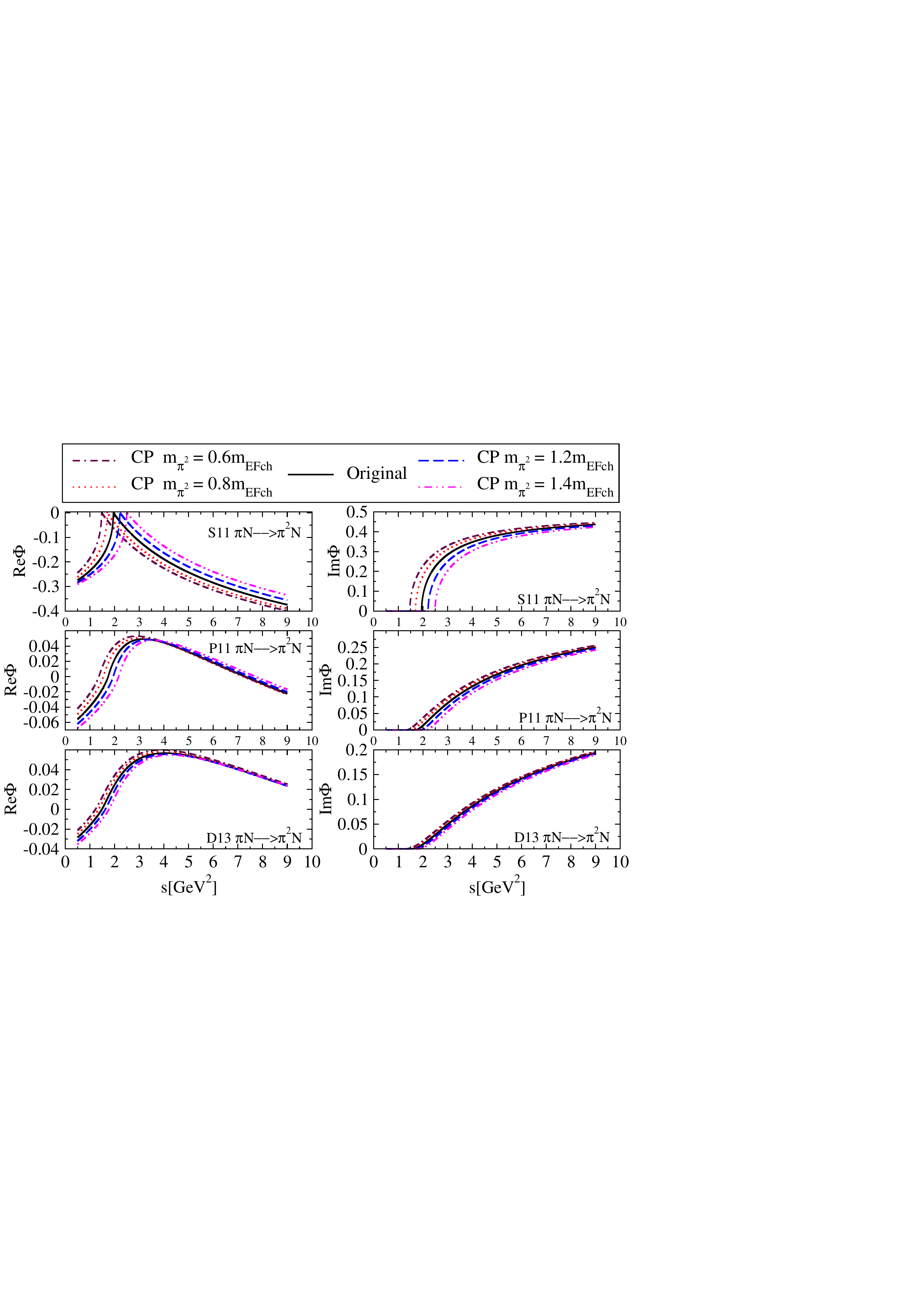}}
\vspace*{-5.cm}
\caption[]{\footnotesize (Color online) Real and imaginary parts  of $S_{11}$,  $P_{11}$ and $D_{13}$ effective channel propagator.}
\label{Fig:S11massEF}
\end{figure}

\noindent
\underline{\textit{Result:}} \\ \\ \noindent
The shift in pole positions for partial waves S$_{11}$, P$_{11}$ and D$_{13}$ are presented in Tables \ref{S11_massEF_table}- 
\ref{D13_massEF_table} and depicted in Fig. (\ref{MassEF_poles}).

\begin{table*}[!ht]
\caption{\footnotesize{The extracted $S_{11}$ partial wave T-matrix poles using different mass of effective channel}}
\begin{center}

\begin{tabular}{|l|ccc|ccc|c|}
\hline \hline 
\rule{0cm}{0.4cm}
 & \multicolumn{3}{c|}{ Bare poles} & \multicolumn{3}{c|}{ Dressed poles } & \\[1ex] \cline{2-7}
\rule{0cm}{0.45cm}Solutions  & $\mathrm{W_{s_1}}$ & $\mathrm{W_{s_2}}$ &  $\mathrm{W_{s_3}}$ & {\scriptsize { $\begin{pmatrix}
\mathrm{Re W}\\ \mathrm{-2Im W} 
\end{pmatrix}$}} & {\scriptsize { $\begin{pmatrix}
\mathrm{Re W}\\ \mathrm{-2Im W} 
\end{pmatrix}$}} & {\scriptsize { $\begin{pmatrix}
\mathrm{Re W}\\ \mathrm{-2Im W} 
\end{pmatrix}$}}   & $\mathrm{\chi_R^2}$  \\ 

  &   & $\mathrm{MeV}$ & &  & $\mathrm{MeV}$ & &  \\ \hline \hline
\rule{0cm}{0.4cm}Sol $0.6m_{EF}$  &1522 & 1637 & 1823 &{\scriptsize { $\begin{pmatrix}
1520\\ 184
\end{pmatrix}$}} & {\scriptsize { $\begin{pmatrix}
1644\\ 210 
\end{pmatrix}$}} &	{\scriptsize { $\begin{pmatrix}
1776\\ 430
\end{pmatrix}$}}&  1.028 \\[1.2ex]  

Sol $0.7m_{EF}$  & 1523& 1638 & 1825 &{\scriptsize { $\begin{pmatrix}
1520\\ 184
\end{pmatrix}$}} & {\scriptsize { $\begin{pmatrix}
1644\\ 209 
\end{pmatrix}$}} &	{\scriptsize { $\begin{pmatrix}
1777\\ 430
\end{pmatrix}$}}&  1.024 \\  [1.2ex]

Sol $0.8m_{EF}$ &1523 & 1638 & 1827 &{\scriptsize { $\begin{pmatrix}
1520\\ 185
\end{pmatrix}$}} & {\scriptsize { $\begin{pmatrix}
1644\\ 210 
\end{pmatrix}$}} &	{\scriptsize { $\begin{pmatrix}
1776\\ 429
\end{pmatrix}$}}&  1.020 \\  [1.2ex]
Sol $0.9m_{EF}$  &1523 & 1639 & 1830 &{\scriptsize { $\begin{pmatrix}
1519\\ 186
\end{pmatrix}$}} & {\scriptsize { $\begin{pmatrix}
1643\\ 209 
\end{pmatrix}$}} &	{\scriptsize { $\begin{pmatrix}
1779\\ 426
\end{pmatrix}$}}&  1.014 \\ [1ex] \hline 

\rule{0cm}{0.4cm}Standard fit&1523& 1640 & 1834 &{\scriptsize { $\begin{pmatrix}
1518\\ 189
\end{pmatrix}$}} & {\scriptsize { $\begin{pmatrix}
1642\\ 208
\end{pmatrix}$}} &	{\scriptsize { $\begin{pmatrix}
1779\\ 427
\end{pmatrix}$}}&  1.016  \\ [1ex] \hline 
\rule{0cm}{0.4cm}Sol $1.1m_{EF}$  &1524 & 1641 & 1837 &{\scriptsize { $\begin{pmatrix}
1518\\ 188
\end{pmatrix}$}} & {\scriptsize { $\begin{pmatrix}
1642\\ 207 
\end{pmatrix}$}} &	{\scriptsize { $\begin{pmatrix}
1778\\ 427
\end{pmatrix}$}}&  1.003 \\  [1.2ex]
Sol $1.2m_{EF}$  &1524 & 1643 & 1842 &{\scriptsize { $\begin{pmatrix}
1516\\ 188
\end{pmatrix}$}} & {\scriptsize { $\begin{pmatrix}
1642\\ 206 
\end{pmatrix}$}} &	{\scriptsize { $\begin{pmatrix}
1781\\ 427
\end{pmatrix}$}}&  1.016 \\  [1.2ex]
Sol $1.3m_{EF}$  &1523 & 1643 & 1846 &{\scriptsize { $\begin{pmatrix}
1511\\ 189
\end{pmatrix}$}} & {\scriptsize { $\begin{pmatrix}
1640\\ 202 
\end{pmatrix}$}} &	{\scriptsize { $\begin{pmatrix}
1786\\ 437
\end{pmatrix}$}}&  1.062 \\ [1.2ex] 
Sol $1.4 m_{EF}$ &1522 & 1644 & 1850 &{\scriptsize { $\begin{pmatrix}
1499\\ 185
\end{pmatrix}$}} & {\scriptsize { $\begin{pmatrix}
1638\\ 199 
\end{pmatrix}$}} &	{\scriptsize { $\begin{pmatrix}
1791\\ 445
\end{pmatrix}$}}&  1.172 \\ [1ex] \hline  \hline
\end{tabular}\label{S11_massEF_table}
\end{center}
\end{table*}~

\begin{table*}[!ht]
\caption{\footnotesize{The extracted $P_{11}$ partial wave T-matrix poles using different mass of the effective channel.}}
\begin{center}
\begin{tabular}{|l|cccc|cccc|c|}
\hline \hline 
\rule{0cm}{0.4cm} &  \multicolumn{4}{c|}{ Bare poles} & \multicolumn{4}{c|}{ Dressed poles } & \\[1ex] \cline{2-9}
\rule{0cm}{0.45cm}Solutions  &$\mathrm{W_{s_1}}$ & $\mathrm{W_{s_2}}$ &  $\mathrm{W_{s_3}}$   & $\mathrm{W_{s_4}}$ & 
{\scriptsize { $\begin{pmatrix}
\mathrm{Re W}\\ \mathrm{-2Im W} 
\end{pmatrix}$}} & {\scriptsize { $\begin{pmatrix}
\mathrm{Re W}\\ \mathrm{-2Im W} 
\end{pmatrix}$}} & {\scriptsize { $\begin{pmatrix}
\mathrm{Re W}\\ \mathrm{-2Im W} 
\end{pmatrix}$}} & {\scriptsize { $\begin{pmatrix}
\mathrm{Re W}\\ \mathrm{-2Im W} 
\end{pmatrix}$}}  & $\chi_R^2$  \\ [1ex]
  & &  \multicolumn{2}{c}{$\mathrm{MeV}$} & &&  \multicolumn{2}{c}{$\mathrm{MeV}$}  & &  \\ \hline \hline
\rule{0cm}{0.4cm}Sol $0.6 m_{EF}$  & 1574 & 1775 & 2204 & 2826&{\scriptsize { $\begin{pmatrix}
1348\\ 220
\end{pmatrix}$}} & {\scriptsize { $\begin{pmatrix}
1725\\ 235
\end{pmatrix}$}} &	{\scriptsize { $\begin{pmatrix}
1729\\ 148
\end{pmatrix}$}}& {\scriptsize { $\begin{pmatrix}
2131\\ 383
\end{pmatrix}$}}&  1.827  \\  [1.2ex]
Sol $0.7 m_{EF}$ & 1582 & 1774 & 2199 & 2811&{\scriptsize { $\begin{pmatrix}
1347\\ 215
\end{pmatrix}$}} & {\scriptsize { $\begin{pmatrix}
1722\\ 218
\end{pmatrix}$}} &	{\scriptsize { $\begin{pmatrix}
1728\\ 144
\end{pmatrix}$}}& {\scriptsize { $\begin{pmatrix}
2129\\ 376
\end{pmatrix}$}}&  1.587  \\ [1.2ex]

Sol $0.8 m_{EF}$ & 1595 & 1772 & 2182 & 2773&{\scriptsize { $\begin{pmatrix}
1344\\ 216
\end{pmatrix}$}} & {\scriptsize { $\begin{pmatrix}
1718\\ 183
\end{pmatrix}$}} &	{\scriptsize { $\begin{pmatrix}
1725\\ 145
\end{pmatrix}$}}& {\scriptsize { $\begin{pmatrix}
2130\\ 347
\end{pmatrix}$}}&  1.272  \\ [1.2ex]
Sol $0.9 m_{EF}$  & 1600 & 1772 & 2178 & 2799&{\scriptsize { $\begin{pmatrix}
1353\\ 192
\end{pmatrix}$}} & {\scriptsize { $\begin{pmatrix}
1714\\ 169
\end{pmatrix}$}} &	{\scriptsize { $\begin{pmatrix}
1729\\ 145
\end{pmatrix}$}}& {\scriptsize { $\begin{pmatrix}
2115\\ 344
\end{pmatrix}$}}&  1.002  \\[1ex]  \hline 
\rule{0cm}{0.4cm}Standard fit &1607 & 1772 & 2182 & 2841&{\scriptsize { $\begin{pmatrix}
1365\\ 157
\end{pmatrix}$}} & {\scriptsize { $\begin{pmatrix}
1708\\ 174 
\end{pmatrix}$}} &	{\scriptsize { $\begin{pmatrix}
1731\\ 136
\end{pmatrix}$}}& {\scriptsize { $\begin{pmatrix}
2117\\ 345
\end{pmatrix}$}}&  0.958  \\ [1ex] \hline 
\rule{0cm}{0.4cm}Sol $1.1 m_{EF}$  & 1614 & 1773 & 2184 & 2811&{\scriptsize { $\begin{pmatrix}
1377\\ 132
\end{pmatrix}$}} & {\scriptsize { $\begin{pmatrix}
1707\\ 163
\end{pmatrix}$}} &	{\scriptsize { $\begin{pmatrix}
1732\\ 132
\end{pmatrix}$}}& {\scriptsize { $\begin{pmatrix}
2109\\ 328
\end{pmatrix}$}}&  1.182  \\ [1.2ex]
Sol $1.2 m_{EF}$  & 1615 & 1775 & 2182 & 2809&{\scriptsize { $\begin{pmatrix}
1399\\ 120
\end{pmatrix}$}} & {\scriptsize { $\begin{pmatrix}
1705\\ 156
\end{pmatrix}$}} &	{\scriptsize { $\begin{pmatrix}
1735\\ 132
\end{pmatrix}$}}& {\scriptsize { $\begin{pmatrix}
2103\\ 320
\end{pmatrix}$}}&  2.215  \\  [1.2ex]
Sol $1.3 m_{EF}$  & 1608 & 1777 & 2175 & 2832&{\scriptsize { $\begin{pmatrix}
1423\\ 114
\end{pmatrix}$}} & {\scriptsize { $\begin{pmatrix}
1704\\ 158
\end{pmatrix}$}} &	{\scriptsize { $\begin{pmatrix}
1735\\ 132
\end{pmatrix}$}}& {\scriptsize { $\begin{pmatrix}
2103\\ 320
\end{pmatrix}$}}&  4.166  \\  [1.2ex]
Sol $1.4 m_{EF}$  & 1592 & 1779 & 2166 & 2883&{\scriptsize { $\begin{pmatrix}
1447\\ 119
\end{pmatrix}$}} & {\scriptsize { $\begin{pmatrix}
1702\\ 162
\end{pmatrix}$}} &	{\scriptsize { $\begin{pmatrix}
1747\\ 137
\end{pmatrix}$}}& {\scriptsize { $\begin{pmatrix}
2092\\ 330
\end{pmatrix}$}}&  7.021 \\ [1ex] \hline \hline 
\end{tabular}\label{P11_massEF_table}
\end{center}
\end{table*}~

\begin{table*}[!ht]
\caption{\footnotesize{The extracted $D_{13}$ partial wave T-matrix poles using different mass of the effective channel}
\label{tab:D13_massEF_table}}

\begin{center}
\begin{tabular}{|l|ccc|cccc|c|}
\hline \hline 
\rule{0cm}{0.4cm} & \multicolumn{3}{c|}{ Bare poles} & \multicolumn{4}{c|}{ Dressed poles } & \\ [1ex]\cline{2-8}
\rule{0cm}{0.45cm}Solutions  &  $\mathrm{W_{s_1}}$ & $\mathrm{W_{s_2}}$ &  $\mathrm{W_{s_3}}$ & {\scriptsize { $\begin{pmatrix}
\mathrm{Re W}\\ \mathrm{-2Im W} 
\end{pmatrix}$}} & {\scriptsize { $\begin{pmatrix}
\mathrm{Re W}\\ \mathrm{-2Im W} 
\end{pmatrix}$}} & {\scriptsize { $\begin{pmatrix}
\mathrm{Re W}\\ \mathrm{-2Im W} 
\end{pmatrix}$}}   & {\scriptsize { $\begin{pmatrix}
\mathrm{Re W}\\ \mathrm{-2Im W} 
\end{pmatrix}$}} & $\mathrm{\chi_R^2}$  \\ 

  &    &$\mathrm{MeV}$ & & & \multicolumn{2}{c}{$\mathrm{MeV}$}   &  &\\ \hline \hline
\rule{0cm}{0.4cm}Sol $0.6 m_{EF}$  & 1580 & 1878 & 2496 &{\scriptsize { $\begin{pmatrix}
1509\\ 129
\end{pmatrix}$}} & {\scriptsize { $\begin{pmatrix}
1808\\ 129
\end{pmatrix}$}} &	{\scriptsize { $\begin{pmatrix}
1945\\ 526
\end{pmatrix}$}}&  {\scriptsize { $\begin{pmatrix}
2687\\588 
\end{pmatrix}$}}&1.096 \\ [1.2ex]
Sol $0.7 m_{EF}$ & 1580 & 1878 & 2496 &{\scriptsize { $\begin{pmatrix}
1508\\ 129
\end{pmatrix}$}} & {\scriptsize { $\begin{pmatrix}
1808\\ 128
\end{pmatrix}$}} &	{\scriptsize { $\begin{pmatrix}
1944\\ 520
\end{pmatrix}$}}&{\scriptsize { $\begin{pmatrix}
2687\\ 589
\end{pmatrix}$}}&  1.069 \\ [1.2ex]
Sol $0.8 m_{EF}$  & 1580 & 1879 & 2499 &{\scriptsize { $\begin{pmatrix}
1507\\ 127
\end{pmatrix}$}} & {\scriptsize { $\begin{pmatrix}
1807\\ 128
\end{pmatrix}$}} &	{\scriptsize { $\begin{pmatrix}
1941\\ 512
\end{pmatrix}$}}&  {\scriptsize { $\begin{pmatrix}
2688\\ 589
\end{pmatrix}$}}&1.041 \\   [1.2ex]

Sol $0.9 m_{EF}$  & 1582 & 1880 & 2499 &{\scriptsize { $\begin{pmatrix}
1507\\ 124
\end{pmatrix}$}} & {\scriptsize { $\begin{pmatrix}
1807\\ 127
\end{pmatrix}$}} &	{\scriptsize { $\begin{pmatrix}
1940\\ 488
\end{pmatrix}$}}& {\scriptsize { $\begin{pmatrix}
2691\\ 578
\end{pmatrix}$}}& 1.044 \\[1ex]  \hline 
\rule{0cm}{0.4cm}Standard fit &1582 & 1880 & 2499 &{\scriptsize { $\begin{pmatrix}
1506\\ 121
\end{pmatrix}$}} & {\scriptsize { $\begin{pmatrix}
1807\\ 127 
\end{pmatrix}$}} &	{\scriptsize { $\begin{pmatrix}
1939\\ 485
\end{pmatrix}$}}&{\scriptsize { $\begin{pmatrix}
2691\\ 583
\end{pmatrix}$}}&  1.027  \\ [1ex] \hline 

\rule{0cm}{0.4cm}Sol $1.1 m_{EF}$ & 1583 & 1880 & 2499 &{\scriptsize { $\begin{pmatrix}
1505\\ 118
\end{pmatrix}$}} & {\scriptsize { $\begin{pmatrix}
1807\\ 127
\end{pmatrix}$}} &	{\scriptsize { $\begin{pmatrix}
1936\\ 472
\end{pmatrix}$}}&{\scriptsize { $\begin{pmatrix}
2694\\ 572
\end{pmatrix}$}}&  0.996 \\  [1.2ex]
Sol $1.2 m_{EF}$ & 1584 & 1881 & 2501 &{\scriptsize { $\begin{pmatrix}
1503\\ 113
\end{pmatrix}$}} & {\scriptsize { $\begin{pmatrix}
1806\\ 126
\end{pmatrix}$}} &	{\scriptsize { $\begin{pmatrix}
1935\\ 452
\end{pmatrix}$}}& {\scriptsize { $\begin{pmatrix}
2690\\ 570
\end{pmatrix}$}}& 1.031 \\  [1.2ex]
Sol $1.3 m_{EF}$  & 1586& 1882 & 2504 &{\scriptsize { $\begin{pmatrix}
1502\\ 108
\end{pmatrix}$}} & {\scriptsize { $\begin{pmatrix}
1805\\ 123
\end{pmatrix}$}} &	{\scriptsize { $\begin{pmatrix}
1935\\ 426
\end{pmatrix}$}}&{\scriptsize { $\begin{pmatrix}
2681\\ 571
\end{pmatrix}$}}&  1.127 \\  [1.2ex]

Sol $1.4 m_{EF}$ & 1589 & 1883 & 2507 &{\scriptsize { $\begin{pmatrix}
1501\\ 101
\end{pmatrix}$}} & {\scriptsize { $\begin{pmatrix}
1804\\ 120
\end{pmatrix}$}} &	{\scriptsize { $\begin{pmatrix}
1934\\ 397
\end{pmatrix}$}}&{\scriptsize { $\begin{pmatrix}
2670\\ 573
\end{pmatrix}$}}&  1.325 \\  [1ex]\hline \hline
\end{tabular}\label{D13_massEF_table}
\end{center}
\end{table*}~
\begin{figure}[!ht]
\vspace{-6.cm}
 \includegraphics[width=13cm]{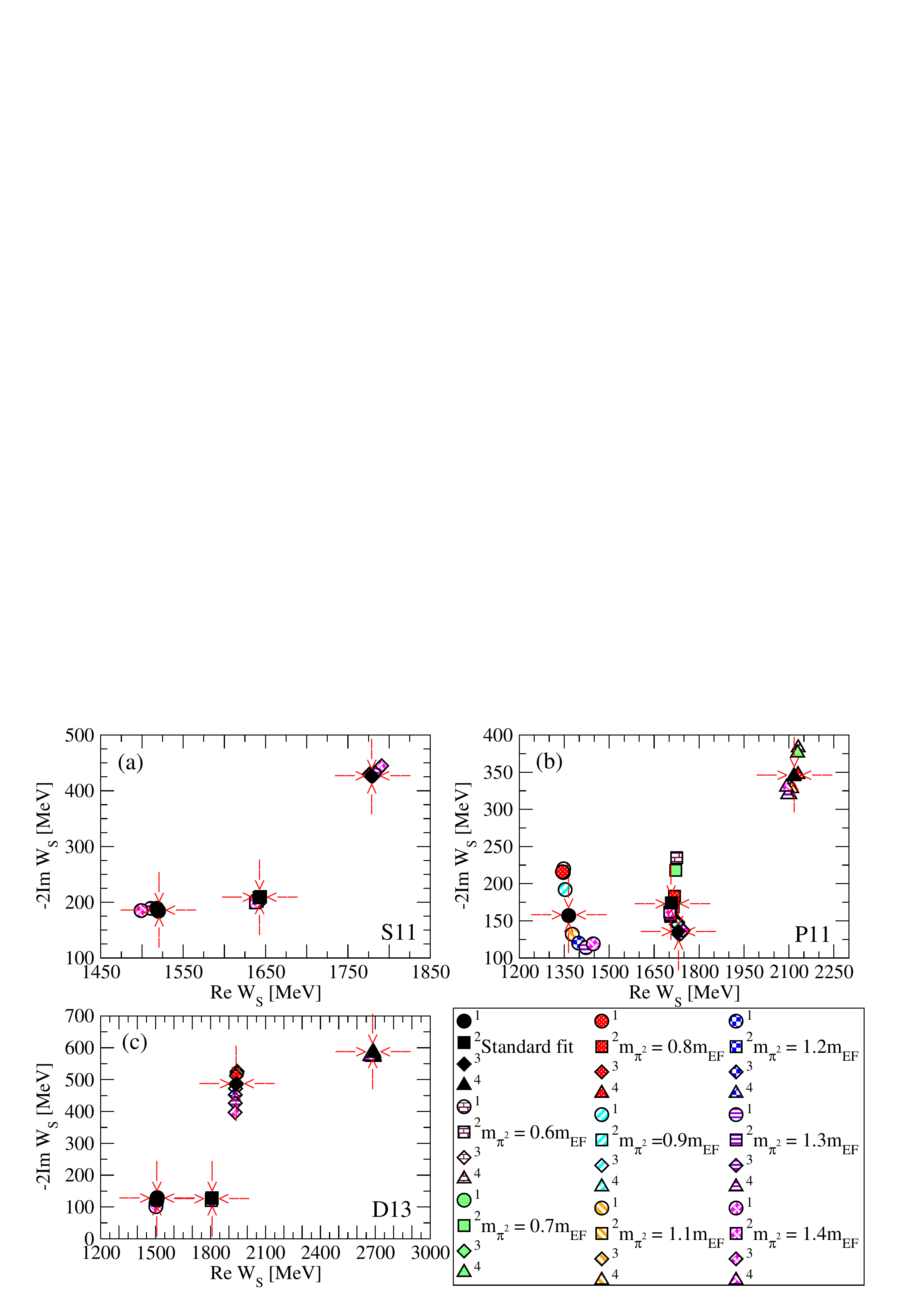}
\caption{\footnotesize{(Color online) The extracted $S_{11}$, $P_{11}$ and $D_{13}$ partial waves T-matrix poles using different mass of the effective
 channel.}}\label{MassEF_poles}
\end{figure}

\vbox{
\noindent
\underline{\textit{Conclusion:}} \\ 

We see that increasing the mass of the effective channel too much above the threshold of the first inelastic channel contributing to the particular
 partial wave is producing unrealistic shifts of the pole positions. \\

Reducing the mass for up to 40\% is producing some shift, but not significant. And again we see that the T-matrix poles form two classes: 
the first class--almost stable poles  S$_{11}$ (1518), S$_{11}$ (1642), D$_{13}$ (1506), D$_{13}$ (1807), and D$_{13}$ (2691) and the second
 class-- more sensitive poles S$_{11}$ (1779), P$_{11}$ (1506), { P$_{11}$ (1708)}, { P$_{11}$ (1731)}, P$_{11}$ (2117), and D$_{13}$ (1939). 
The pole is not sensitive to the model parameters if it is dominantly determined by the bare singularity
 (small self energy correction), and it is moderately or strongly sensitive upon the model choices if the pole is created more dynamically 
 (self energy contribution is strong).\\}

 \vbox{
From Tables \ref{S11_massEF_table}-\ref{D13_massEF_table} it is seen that reduced $\chi^2$ significantly raises when we increase the mass 
of the effective channel for up to 40\%. This is, however,  not so when the mass of effective channel is decreased for the same amount. 
{ We can easily understand the effect. Namely, the effective channel serves for ensuring unitarity for all open inelastic channels. However, 
if we increase the mass of the inelastic channel too much, then the fit, which is manifestly unitary, will not have an extra channel at its 
disposal to get rid of unwanted flux, and the code will be forced to fit the input data with amplitudes which are much more ``elastic" then in
 reality. Simply, there will be no room for the loss of flux to inelastic channels which are open at these energies. Therefore, the fit will produce 
 unrealistic results. On the other hand, if we lower  the mass of the effective channel for 40\%, we shall allow the fit to send the flux to inelastic
  channel if needed. However, the fit declines this possibility because the input data are dominantly elastic. }}

\section{Conclusions}

{  In our model} we find two classes of poles: ones strongly determined by the bare poles, and ones which are more influenced by the self energy correction. \\

Poles  S$_{11}$ (1518), S$_{11}$ (1642), P$_{11}$ (1731), D$_{13}$ (1506) and D$_{13}$ (1807)  belong to the first class, and  S$_{11}$ (1779), 
{ $P_{11}(1365)$}, P$_{11}$ (1708), P$_{11}$ (2117), D$_{13}$ (1939) and D$_{13}$ (2691) to the second. \\
				         
{  We show that the} position of bare poles is model dependent. {  It} is strongly influenced by dispersion relation subtraction constant and the channel propagator threshold 
behavior. As we believe that the threshold behavior is fairly well determined by the $q^{2L+1}$ law, the major model freedom of  bare poles is originating from the choice of 
dispersion relation subtraction constant.  \\ 

{  The dressed pole positions are much better fixed.  Poles for which the self energy correction is small are  stable, and ones {  for which} the self energy correction is 
stronger are only moderately running.  
The ``complex plane distance" between bare and dressed poles representing the size of the self energy contributions can be taken as a measure how model dependent a certain pole 
is. The smaller the difference is, the more confident the pole extraction is.} \\ \\ 

{  We observe that} the reduced $\chi^2_R$, the quantity representing the measure of ability of our model to reproduce the input data, is systematically smallest for the 
non-modified solution. In other words, our model is unable to correctly fit the input data whenever any significant change to the model ingredients is introduced. {  We conclude 
that} our model is sensitive to the analytic structure embedded into the input data by the way they are generated, because  the underlying analytical structure of the Zagreb model 
is transferred to the data when we  created the input data set by distributing Zagreb theoretical curves. If our analytic structure is wrong, and it certainly has to be up to a 
certain level, the real data coming from the experiment {  are described by the different analytic function and} will not be ideally fitted by our curves. So, pole positions will 
be shifted. \\  

{\vbox{
However, we could at least in principle, use this feature as to test the shape of the model ingredients determining the self energy corrections.  We propose the following 
procedure to improve the analytic structure of Zagreb CMB model with respect to any input data, either originating from different PWA or preferably coming directly from experiment 
in the form of partial wave data:  
\indent

\begin{enumerate}
\item We fit the obtained input set and  get the pole positions with the standard channel propagator, and note the belonging $\chi^2_R$.
\item We change the form of the channel propagator, re do  the fit, and compare the new $\chi^2_R$ with the old value. 
\item If the $\chi^2_R$ is better, we keep on changing the form of the channel propagator because we infer that our imposed analytic structure 
is not adequate, and that we are finding the better one.
\item When we achieve the best $\chi^2_R$, we pronounce that we have found the optimal value of the channel propagator for our formalism, and 
declare the found pole positions a correct ones.
\item We should, in principle, be able to follow which poles are changed, and which remain stable, and on the basis of that infer the error. 
\end{enumerate} 
}
However, the afore hinted procedure is not within the scope of this paper.\\ \\ 
\noindent 
In conclusion:\\ \\ \indent
Zagreb CMB model can be used for extracting {  dressed} pole positions from either partial wave data or partial wave amplitudes.  The bare pole masses are strongly model 
dependent, while  the dressed pole positions are much better fixed.   {  The poles which are strongly influenced by the self energy corrections, however, show  more model 
dependence then the ones which are dominated by the bare poles.  At this point it is important to mention that our findings about the relative size of bare vs. self energy 
contributions  of a certain pole are not eternal truth for all models. As bare mass is a movable quantity, it is easy to imagine that for some other model the shift of the bare 
pole position in proper direction might result in reducing the importance of self energy corrections. So, poles which are dominated by self energy corrections in our model might 
be dominated by the bare pole in others.  This allows us to speculate that changing the subtraction constant that is a completely free parameter of our model may shift the results 
of Zagreb CMB fit in the direction of other models with respect to the value of bare poles.} \\ 

There are other interesting questions like the role of physical requirements on the S-matrix that are missing in the Zagreb CMB model. Those are, e.g., sub-threshold cuts 
(circular, short nucleon, left-hand cuts as we  have a trivial left-hand cut from kinematics, but that is, of course, not the full one...), or true and not effective three-body 
intermediate states with the corresponding additional analytic structures in the complex plane. Those are the really questions of exceptional interest which are really hard to 
answer, and will not be answered yet in this paper.


\begin{thebibliography}{12} 

\bibitem{Mirza2011} M. Had\v{z}imehmedovi\'{c}, S. Ceci, A. \v{S}varc,  H. Osmanovi\'{c},   and J. Stahov, arXiv:1103.2653, submitted for publication to Phys. Rev. C. 
{
\bibitem{KH80} G. H\"{o}hler, in {\em Pion-Nucleon Scattering}, Landolt-B\"{o}rnstein, Vol {\bf I/9b2} (Springer-Verlag, Berlin, 1983);   G. H\"{o}hler and  A. Schulte, $\pi N$ 
Newsletter, {\bf 7} (1992) 407.
\bibitem{GWUWEB} \emph{http://gwdac.phys.gwu.edu/analysis/pin\_analysis.html}.
\bibitem{Arn04}R. A. Arndt, W.J. Briscoe, I.I. Strakovsky, R.L. Workman, and M.M. Pavan, Phys. Rev. {\bf C69}, 035213 (2004), R. A. Arndt, W. J. Briscoe, I. I. Strakovsky, and R. 
L. Workman, Phys. Rev. \textbf{C 74}, 045205 (2006).
\bibitem{Diaz07} B. Juli\'{a} - D\'{i}az, T.-S. H. Lee, A. Matsuyama, and T. Sato, Phys. Rev. \textbf{C 76}, 065201 (2007).
\bibitem{Dur08} J. Durand, B. Juli\'{a} - D\'{i}az, T.-S. H. Lee, B. Saghai, and T. Sato, Rev. \textbf{ C 78}, 025204 (2008).
\bibitem{Matsuyama2007}A. Matsuyama, T. Sato, and T.-S.H. Lee, Physics Reports \textbf{439}, 193 (2007).
\bibitem{Juelich} M. D\"{o}ring, C. Hanhart,, F. Huang, S. Krewald, and U.-G. Meissner, Nucl.Phys,  \textbf{A 829}, 170 (2009); C. Sch\"{u}tz, J. Haidenbauer, J. Speth, and J. W. 
Durso, Phys. Rev. \textbf{C 57} 1464 (1998); O. Krehl, C. Hanhart, S. Krewald, and J. Speth, Phys. Rev. \textbf{C 62} 025207 (2000); A. M. Gasparyan, J. Haidenbauer, C. Hanhart, 
and J. Speth, Phys. Rev. \textbf{C 68} 045207 (2003).
\bibitem{Che03} Guan-Yeu Chen, S.S. Kamalov, Shin Nan Yang, D. Drechsel, and L. Tiator, Nuclear Physics \textbf{A 723},  447  (2003).
\bibitem{Che07} Guan Yeu Chen, S. S. Kamalov, Shin Nan Yang, D. Drechsel and L. Tiator, Phys. Rev. \textbf{C76}, 035206 (2007).
\bibitem{Giessen} V. Shklyar, H. Lenske, and U. Mosel, Phys. Rev. \textbf{C 72},  015210 (2005), and private communication.
\bibitem{Cut79} R. E. Cutkosky,  C.P. Forsyth, R.E. Hendrick, and R.L. Kelly, Phys. Rev. {\bf D 20}, 2839 (1979).
\bibitem{Vrana2000} T. P. Vrana, S. A. Dytman, T. -S. H. Lee, Phys. Rep. \textbf{328},181 (2000).
\bibitem{BonnGatchina} A.V. Anisovich, A.V. Sarantsev,  Eur.Phys.J. \textbf{A30 }, 427 (2006),  and \\
\emph{http://pwa.hiskp.uni-bonn.de/}.
}
\bibitem{Hoe2001} G. H\"{o}hler, in NSTAR2001, Proceedings of the Workshop on The Physics of Excited Nucleons, ed. D. Drechsel and L. Tiator,
 World Scientific 2001, 185.  
\bibitem{Chew1976} G.F. Chew, Resonances, Particles, and Poles from the Experimenter�s Point of View. Berkeley UCRL-16983 (1966).
\bibitem{Hoe2000} G. H\"{o}hler in  D.E. Groom et al., Particle Data Group, Eur. Phys. Jour. \textbf{C 15}, 1 (2000).
\bibitem{Breit-Wigner} W. N. Cottingham, D. A. Greenwood, ``An Introduction to Nuclear Physics",
Cambridge University Press, Cambridge 1986 and 2001. 
\bibitem{DalitzMoorhouse} R. H. Dalitz and R. G. Moorhouse, What is Resonance?
Proc. R. Soc. Lond. \textbf{A 318}, 279-298 (1970).
\bibitem{Dalitz61} R. H. Dalitz, Ann. Rev. Nucl. Science \textbf{13}, 471 (1961).
\bibitem{Neumann-Wigner29} J. von Neumann and E. P. Wigner, Z. Physik \ {\bf 30}, 467 (1929).
\bibitem{Bat98} M. Batini\'{c}, I. \v{S}laus, A. \v{S}varc, and B.M.K. Nefkens, Phys. Rev. {\bf C 51}, 2310 (1995); M. Batini\'{c}, I. Dadi\'{c}, I. \v{S}laus, A. \v{S}varc, 
B.M.K. Nefkens, and T.-S.H. Lee, Physica Scripta {\bf 58}, 15, (1998).
\bibitem{Batinic2010} M. Batini\'{c}, S. Ceci, A. \v{S}varc, and B. Zauner, Phys. Rev. {\bf C 82}, 038203 (2010).
\bibitem{Ceci2011} S. Ceci, M. D\"{o}ring, C. Hanhart, S. Krewald, U.-G. Meissner, A. \v{S}varc, Phys. Rev. {\bf C84}, 015205 (2011). 
\bibitem{Pietarinen} E. Pietarinen, Nuovo Cimento Soc. Ital. Fis. \textbf{12A}, 522 (1972).
\bibitem{Ciu62} I. Ciulli, S. Ciulli, and J. Fisher, Il Nuovo Cimento 23, 1129 (1962).
\bibitem{Ceci06} S. Ceci, A. \v{S}varc, and B. Zauner, Phys. Rev. Lett. \textbf{97}, 062002 (2006).
\bibitem{Analyticity-piecewise} P.M. Morse and H. Fesbach, Methods of Theoretical Physics, McGraW- Hill, New York, 1953.
\bibitem{CapstickEPJA}  S. Capstick, A. \v{S}varc, L. Tiator, J. Gegelia, M.M. Giannini, E. Santopinto,
C. Hanhart, S. Scherer, T.-S.H. Lee, T. Sato and N. Suzuki, Eur. Phys. J. \textbf{A 35}, 253 (2008), and references therein. 





\end{thebibliography}
\end{document}